\documentclass[manuscript,iop, numberedappendix]{emulateapj}
\usepackage{apjfonts}
\usepackage{graphicx}
\usepackage{epstopdf}
\usepackage{epsf}
\usepackage{amsmath, amsthm, amssymb} 
\usepackage{threeparttable}
\usepackage{layouts}
\usepackage{rotating}

\usepackage{color}
\definecolor{citeRGB}{rgb}{0,0.1,0.7}
\usepackage[hyperfootnotes=true,naturalnames=true,letterpaper,pdfstartview=FitH,pdfpagemode=UseNone,colorlinks=true,citecolor=citeRGB]{hyperref}

\def\la{\mathrel{\hbox{\rlap{\hbox{\lower4pt\hbox{$\sim$}}}\hbox{$<$}}}}
\def\gt{\mathrel{\hbox{$>$}}}

\def\ga{\mathrel{\hbox{\rlap{\hbox{\lower4pt\hbox{$\sim$}}}\hbox{$>$}}}}
\def\lt{\mathrel{\hbox{$<$}}}

\shortauthors{Skelton et al.}
\shorttitle{3D-HST Photometric Catalogs}

\slugcomment{Submitted to The Astrophysical Journal Supplement Series}

\begin{document}

\title{3D-HST WFC3-selected Photometric Catalogs in the Five CANDELS/3D-HST Fields: Photometry, Photometric Redshifts and Stellar Masses}
\author{Rosalind E. Skelton,\altaffilmark{1,2} Katherine E. Whitaker,\altaffilmark{3} Ivelina G. Momcheva,\altaffilmark{2} Gabriel B. Brammer,\altaffilmark{4} Pieter G. van Dokkum,\altaffilmark{2}
\\
{\footnotesize{AND}}
\\
Ivo Labb\'e,\altaffilmark{5} Marijn Franx,\altaffilmark{5}
Arjen van der Wel,\altaffilmark{6} 
Rachel Bezanson,\altaffilmark{7,2} Elisabete Da Cunha,\altaffilmark{6}  Mattia Fumagalli,\altaffilmark{5}  Natascha F\"orster Schreiber,\altaffilmark{8} 
 Mariska Kriek,\altaffilmark{9} Joel Leja,\altaffilmark{2} Britt F. Lundgren,\altaffilmark{10} Daniel Magee,\altaffilmark{11} Danilo Marchesini,\altaffilmark{12} Michael V. Maseda,\altaffilmark{6} Erica J. Nelson,\altaffilmark{2}  Pascal Oesch,\altaffilmark{2} Camilla Pacifici,\altaffilmark{13} Shannon G. Patel,\altaffilmark{14} Sedona Price,\altaffilmark{9} Hans-Walter Rix,\altaffilmark{6} Tomer Tal,\altaffilmark{11}  David A. Wake,\altaffilmark{10,15} Stijn Wuyts\altaffilmark{8}}
\altaffiltext{1}{South African Astronomical Observatory, PO Box 9, Observatory, Cape Town, 7935, South Africa}
\altaffiltext{2}{Department of Astronomy, Yale University, 260 Whitney Avenue, New Haven, CT 06511, USA}
\altaffiltext{3}{Astrophysics Science Division, Goddard Space Center, Greenbelt, MD 20771, USA}
\altaffiltext{4}{Space Telescope Science Institute, 3700 San Martin Drive, Baltimore, MD 21218, USA}
\altaffiltext{5}{Leiden Observatory, Leiden University, Leiden, The Netherlands}
\altaffiltext{6}{Max Planck Institute for Astronomy (MPIA), K\"onigstuhl 17, D-69117, Heidelberg, Germany}
\altaffiltext{7}{Steward Observatory, University of Arizona, 933 North Cherry Avenue, Tucson, AZ 85721}
\altaffiltext{8}{Max-Planck-Institut f\"ur extraterrestrische Physik, Giessenbachstrasse, D-85748 Garching, Germany}
\altaffiltext{9}{Astronomy Department, University of California, Berkeley, CA 94720, USA}
\altaffiltext{10}{Department of Astronomy, University of Wisconsin-Madison, 475 N. Charter Street, Madison, WI 53706, USA}
\altaffiltext{11}{Department of Astronomy \& Astrophysics, University of California, Santa Cruz, CA, USA}
\altaffiltext{12}{Department of Physics and Astronomy, Tufts University, Medford, MA 02155, USA}
\altaffiltext{13}{Yonsei University Observatory, Yonsei University, Seoul 120-749, Republic of Korea}
\altaffiltext{14}{Carnegie Observatories, Pasadena, CA 91101, USA}
\altaffiltext{15}{Department of Physical Sciences, The Open University, Milton Keynes, MK7 6AA, UK}
\altaffiltext{16}{\url{http://3dhst.research.yale.edu}}

\email{ros@saao.ac.za}

\begin{abstract}
The 3D-HST and CANDELS programs have provided WFC3 and
ACS spectroscopy and photometry over $\approx 900$ arcmin$^2$ in five fields: AEGIS, COSMOS, GOODS-North, GOODS-South, and the UKIDSS UDS field.  All these fields have a wealth of publicly available imaging datasets in addition to the \textit{HST} data, which makes it possible to construct the spectral energy distributions (SEDs) of objects over a wide wavelength range. In this paper we describe a photometric analysis of the CANDELS and 3D-HST \textit{HST} imaging and the ancillary imaging data at wavelengths $0.3\,\mu$m -- $8\,\mu$m. Objects were selected in the WFC3 near-IR bands, and their SEDs were determined by carefully taking the effects of the point spread function in each observation into account. A total of 147 distinct imaging datasets were used in the analysis. The photometry is made available in the form of six catalogs: one for each field, as well as a master catalog containing all objects in the entire survey. We also provide derived data products: photometric redshifts, determined with the EAZY code, and stellar population parameters determined with the FAST code.  We make all the imaging data that were used in the analysis available, including our reductions of the WFC3 imaging in all five fields. 3D-HST is a spectroscopic survey with the WFC3 and ACS grisms, and the photometric catalogs presented here constitute a necessary first step in the analysis of these grism data. All the data presented in this paper are available through the 3D-HST website.\altaffilmark{16}
\end{abstract}

\section{Introduction}
\label{sec:intro}

Large multi-wavelength photometric surveys have made it possible to study galaxy populations over most of cosmic history. Near-infrared selected samples have been used to trace the evolution of the stellar mass function \citep[e.g.,][]{Marchesini09, Perez-Gonzalez08}, the star formation--mass relation (e.g., \citealt{Whitaker12}), and the structural evolution of galaxies \citep[e.g.,][]{Franx08, Bell12, Wuyts12, vanderWel12}. Until recently most of these surveys relied on deep, wide-field imaging from ground-based telescopes (e.g., \citealt{Muzzin13a, Williams09}). The WFC3 camera on the \textit{Hubble Space Telescope} (\textit{HST}) has opened up the possibility to select and study galaxies at near-infrared wavelengths with excellent sensitivity and spatial resolution. Furthermore, the WFC3 grisms enable space-based near-IR slitless spectroscopy of all objects in the camera's field-of-view (see, e.g., \citealt{vanDokkumBrammer10}). 

The largest area WFC3 imaging survey done to date is the Cosmic Assembly Near-infrared Deep Extragalactic Legacy Survey (CANDELS, \citealt{Grogin11, Koekemoer11}), a 912-orbit Multi-Cycle Treasury imaging program (PIs: S.~Faber
and H.~Ferguson). This survey encompasses five well-studied extragalactic fields: the All-wavelength Extended Groth Strip International Survey (AEGIS) field, the Cosmic Evolution Survey (COSMOS) field, the Great Observatories Origins Survey (GOODS) Northern and Southern fields (GOODS-North and GOODS-South) and the UKIRT InfraRed Deep Sky Surveys (UKIDSS) Ultra Deep Field (UDS). The coordinates of the five fields are given in Table~\ref{table:fields}. As these fields have been observed extensively over the past decade, the CANDELS imaging builds on a vast array of publicly available photometry at other wavelengths, ranging from the near-UV to the far-IR (see \citealt{Grogin11}). 

\begin{figure*}[!ht]
\centering
\includegraphics[width = \textwidth]{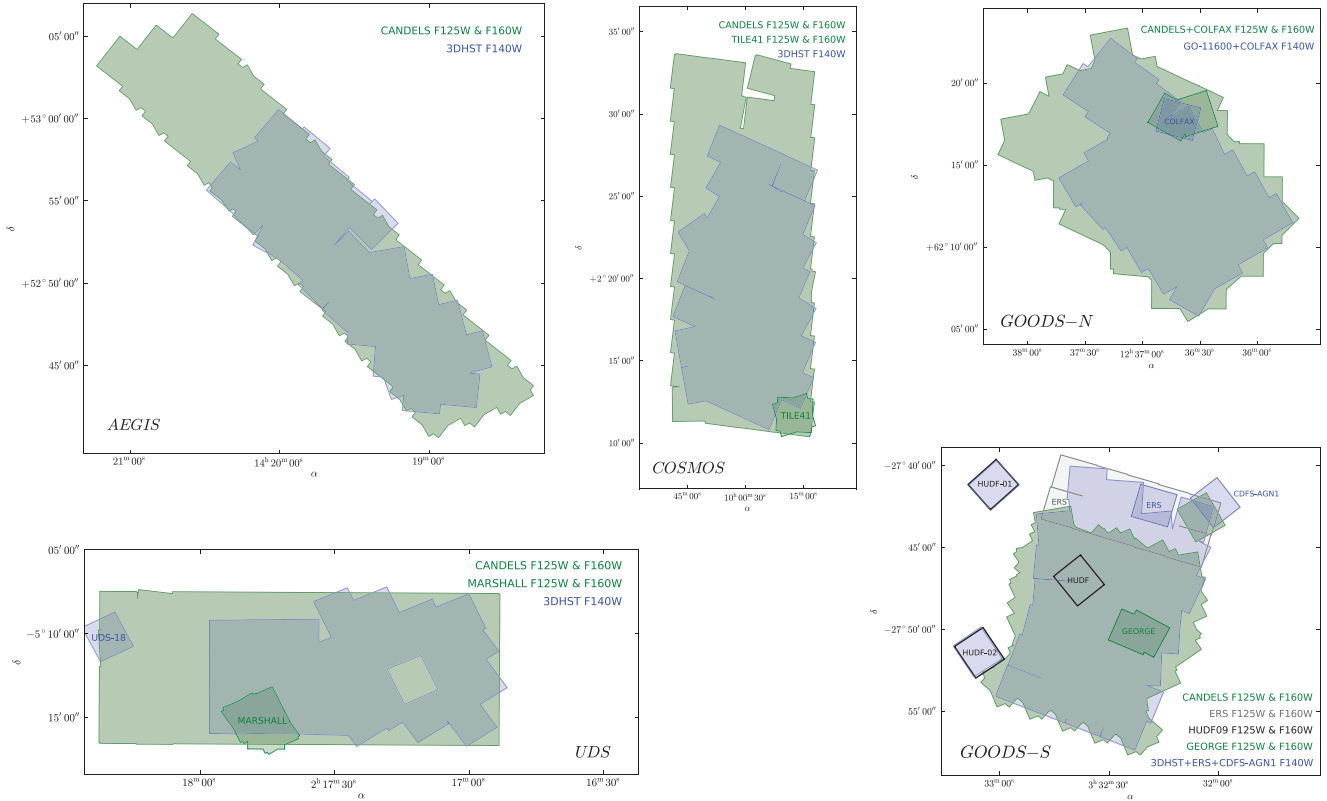}
\caption{Layout of the WFC3 observations used in this paper. The catalogs presented in this paper cover the entire area that
is covered by either F125W, F140W, or F160W, in each of the five fields. Table \ref{table:hstdata} lists the programs and PIs for all the HST/WFC3 observations that were used in our work.  North is up and East is to the left.}\label{fig:fieldlayout}
\end{figure*}

Building, in turn, on the CANDELS survey, we have undertaken a WFC3 spectroscopic survey in these same fields.
3D-HST is a 248-orbit \textit{HST} Treasury program (Programs 12177 and 12328; PI: P.\ van Dokkum) that uses the WFC3 G141 grism for slitless spectroscopy across $\sim700$ arcmin$^2$ of the sky, approximately 75\% of the CANDELS area (see Figure \ref{fig:fieldlayout}). This rich dataset is providing excellent redshifts and spatially-resolved spectral lines for thousands of galaxies in the key epoch $1 < z < 3$ (e.g., \citealt{Whitaker13, Nelson12, Brammer12b}).
The survey is described in \citet{Brammer12a}. We targeted four of the five CANDELS fields: AEGIS, COSMOS, GOODS-South, and UDS. WFC3/G141 grism data in GOODS-North were already available from program GO-11600 (PI: B.\ Weiner); these data are incorporated in the 3D-HST analysis and data releases. The 3D-HST observations yield the following four types of data: WFC3 G141 grism observations;  WFC3 F140W imaging for wavelength calibration of the spectra; parallel ACS G800L grism spectroscopy; and parallel F814W imaging. 

The scientific returns from CANDELS, 3D-HST and all other surveys in these five fields are maximized when the various datasets are combined in a homogeneous way, as it is often the combination of different kinds of data that provides new insight. To give just one example, \citet{Wuyts12} study the structure of galaxies (determined from \textit{HST} imaging) as a function of photometric redshift (determined from fits to multi-wavelength, broad-band spectral energy distributions, SEDs) and star formation rate (determined from SEDs and space-based infrared photometry). The interpretation of the data is also made easier when all information is used: it is much easier to correctly identify an emission line in a grism spectrum when the redshift range of the object is constrained by the available photometric information.

The 3D-HST project has the aim of providing this homogenous combination of datasets in the five CANDELS/3D-HST fields. This undertaking has several linked aspects:  
\begin{enumerate}
\item{We obtained and reduced the available HST/WFC3 imaging in the fields, using the same pixel scale and tangent point as those used by the CANDELS team. The WFC3 imaging includes the CANDELS data and also the Early Release Science data in GOODS-South and various other programs such as the HUDF09 Ultra Deep Field campaign.}
\item{Source catalogs are created with SExtractor \citep{Bertin96}, detecting objects in deep combined F125W + F140W + F160W images.}
\item{These source catalogs, along with the detection images, associated segmentation maps and PSFs,  are used as the basis to measure photometric fluxes at wavelengths $0.3\,\mu$m -- $8\,\mu$m from a large array of publicly available imaging datasets. The resulting SEDs are of very high quality, particularly in fields with extensive optical and near-IR medium band photometry.}
\item{Photometric redshifts, and redshift probability distributions, are estimated from the SEDs.}
\item{Stellar population parameters are determined by fitting stellar population synthesis models to the SEDs, using the photometric redshifts as input.}
\item{Mid- and far-IR photometry is obtained from Spitzer/MIPS and Herschel imaging. These data, combined with rest-frame UV emission measurements from the SEDs, are used to determine star formation rates of the galaxies.}
\item{The set of images, PSFs, and catalogs is used to measure structural parameters of the objects in the WFC3 and ACS bands, following the methodology of \citet{vanderWel12}.}
\item{The coordinates in the catalogs and segmentation maps are mapped back to the original (interlaced) coordinate system of the WFC3 and ACS grism data, and spectra are extracted for each object in the photometric catalog that is covered by the grism. No source matching is required, since each extracted spectrum is associated with a particular object in the photometric catalog. The photometric SED can  be combined directly with the grism spectroscopy of each object for further analysis.}
\item{The spectra and SEDs are fitted simultaneously, to measure redshifts and emission line fluxes.}
\item{Parameters measured in steps 5, 6, and 7 are re-measured using the updated redshifts.}
\end{enumerate}
In this paper we describe steps 1--5 of the 3D-HST project; steps 6--10 will be described in future papers. As outlined above the photometric catalogs ultimately serve as input to the fits of the grism spectroscopy, but as we show here they constitute a formidable dataset in their own right. Furthermore,  the majority of objects in the photometric catalogs are so faint that the grism does not provide useful additional information. We provide the homogenized set of imaging datasets that are used in this paper to the community, as well as the photometric catalogs and the EAZY and FAST fits to the photometry. The structure of this paper is as follows. In Section~\ref{sec:wfc3im} we describe the data reduction and mosaicking of the WFC3 detection images.  Section~\ref{sec:otherdata} details the additional multi-wavelength data available for each field. Section~\ref{sec:phot} describes our photometric methods, accounting for differences in the depth and resolution of the data in different bands. We discuss the survey completeness in Section~\ref{sec:completeness}. We verify the quality and consistency of the catalogs in Section~\ref{sec:tests}. In Section~\ref{sec:eazy}, Section~\ref{sec:rfcols} and Section~\ref{sec:fast} we describe the photometric redshift, rest-frame color and stellar population parameter fits to the SEDs.  Additional information on the PSFs and  zero point offsets applied to the catalogs are provided in Appendix~\ref{app:psfs} and \ref{app:zps}. We present a comparison of our photometry with other available catalogs for each of the five fields in Appendix~\ref{app:comparisons}.

We use the AB magnitude system throughout \citep{Oke71} and where necessary, a $\Lambda$CDM cosmology with $\Omega_M$=0.3, $\Omega_{\Lambda}$ = 0.7 and $H_0 = 70~$km s$^{-1}$ Mpc$^{-1}$.

\begin{table*}[ht]
\centering
\caption{3D-HST Fields}\label{table:fields}
\begin{tabular}{lllll}
\hline \hline
\noalign{\smallskip}
 & RA & Dec & Total area & Science Area \\
Field& (h m s) & (d m s) & (arcmin$^2$) & (arcmin$^2$) \\
\noalign{\smallskip}
\hline
\noalign{\smallskip}
AEGIS &14 18 36.00 &  +52 39 0.00 & 201 & 192.4 \\
COSMOS & 10 00 31.00 & +02 24 0.00 & 199 & 183.9 \\
GOODS-North & 12 35 54.98 & +62 11 51.3 & 164 &  157.8  \\
GOODS-South & 03 32 30.00 & -27 47 19.00  & 177 & 171.0 \\
UDS & 02 17 49.00 & -05 12 2.00 &201 & 191.2 \\
\noalign{\smallskip}
\hline
\noalign{\smallskip}
\end{tabular}
\end{table*}

\section{Datasets}

\begin{figure*}[!ht]
\centering
\includegraphics[width = \textwidth]{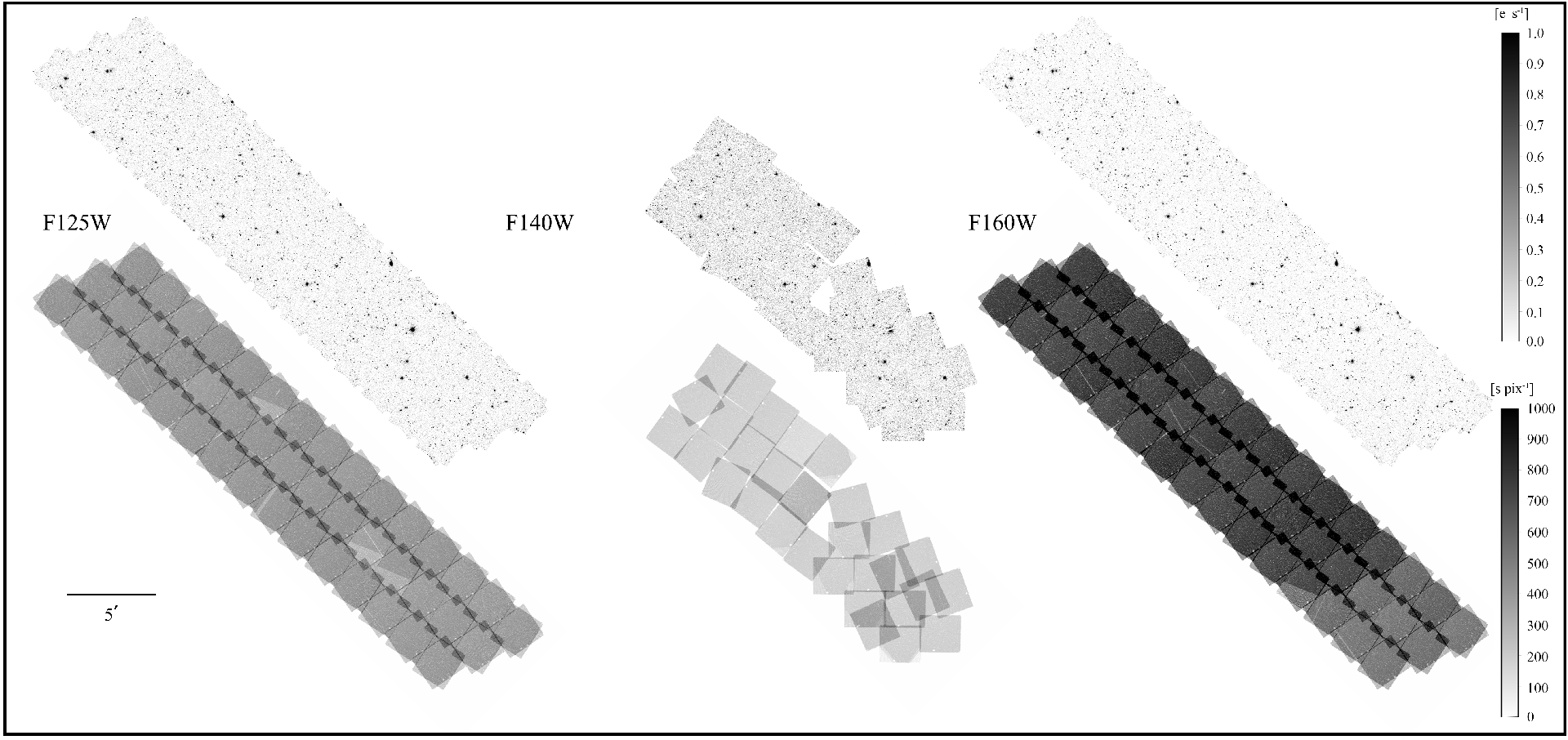}
\caption{WFC3 mosaic science images and exposure maps for F125W, F140W and F160W in the AEGIS field. North is up, East is to the left. The science image is in units of electrons per second per pixel. The exposure map is in units of seconds per pixel. See the text for descriptions of how these mosaics were created. }\label{fig:AEGIS_sci_wht}
\end{figure*}

\begin{figure*}
\centering
\includegraphics[width = \textwidth]{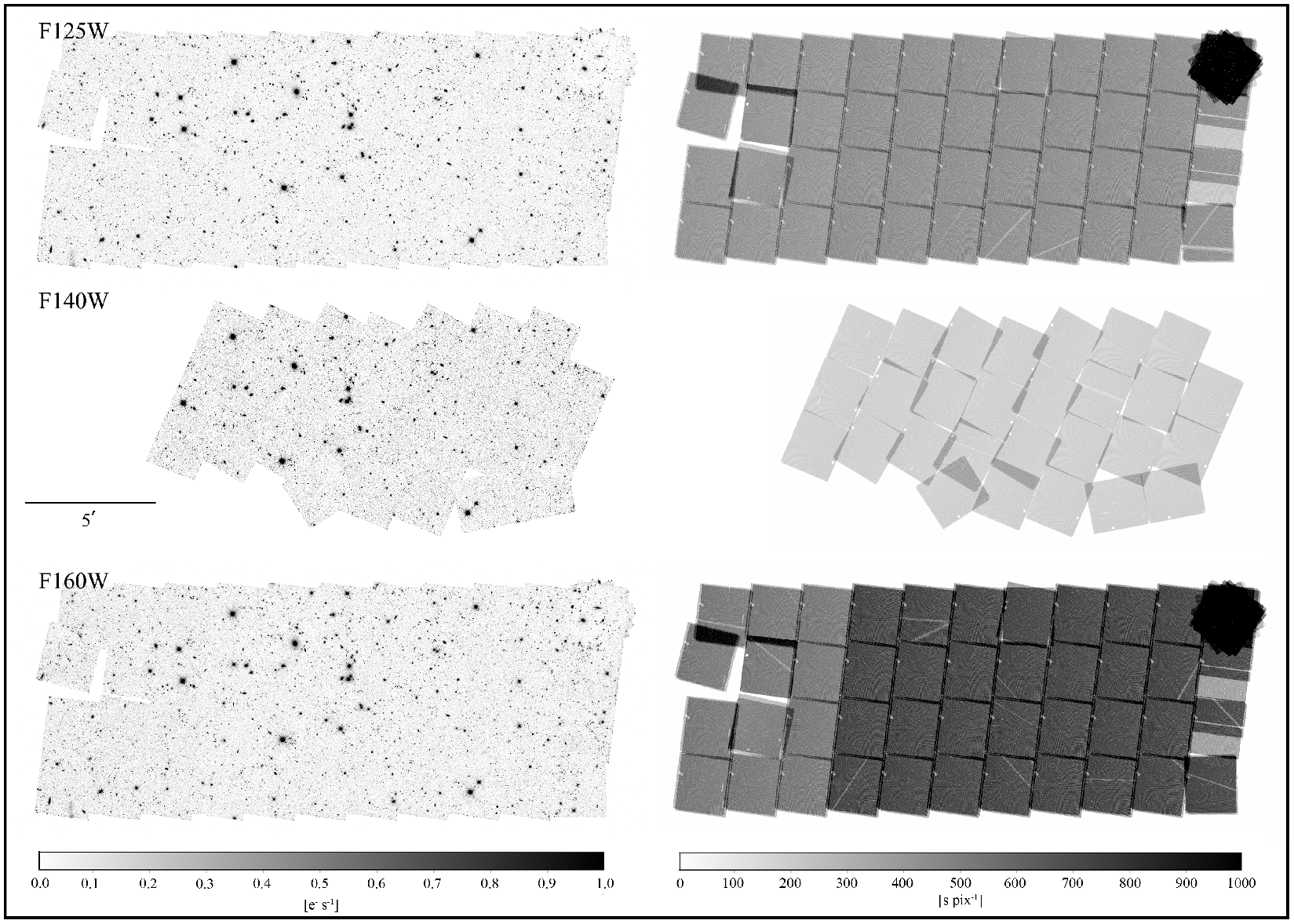}
\caption{Same as Figure \ref{fig:AEGIS_sci_wht}, for the COSMOS field. North is to the left, East is to the bottom of the page. Note that the deep region of the image saturates on the scale shown for the exposure map.}\label{fig:COSMOS_sci_wht}
\end{figure*}

\begin{figure*}[!ht]
\centering
\includegraphics[width = \textwidth]{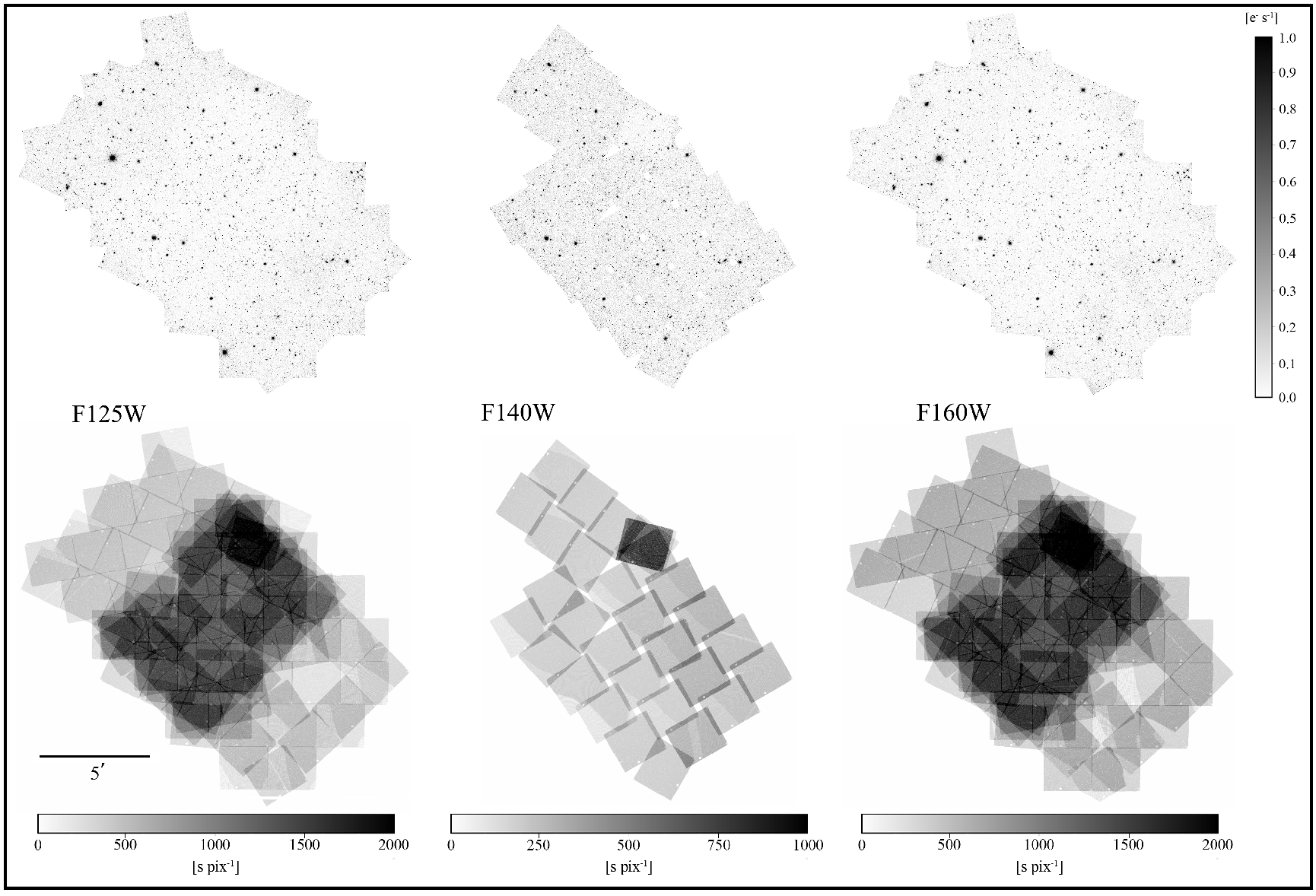}
\caption{Same as Figure \ref{fig:AEGIS_sci_wht}, for the GOODS-N field. North is up, East is to the left. Note that the scale of the F140W weight image differs from that of F125W/F160W.}\label{fig:GOODS-N_sci_wht}
\end{figure*}

\begin{figure*}[!ht]
\centering
\includegraphics[width = \textwidth]{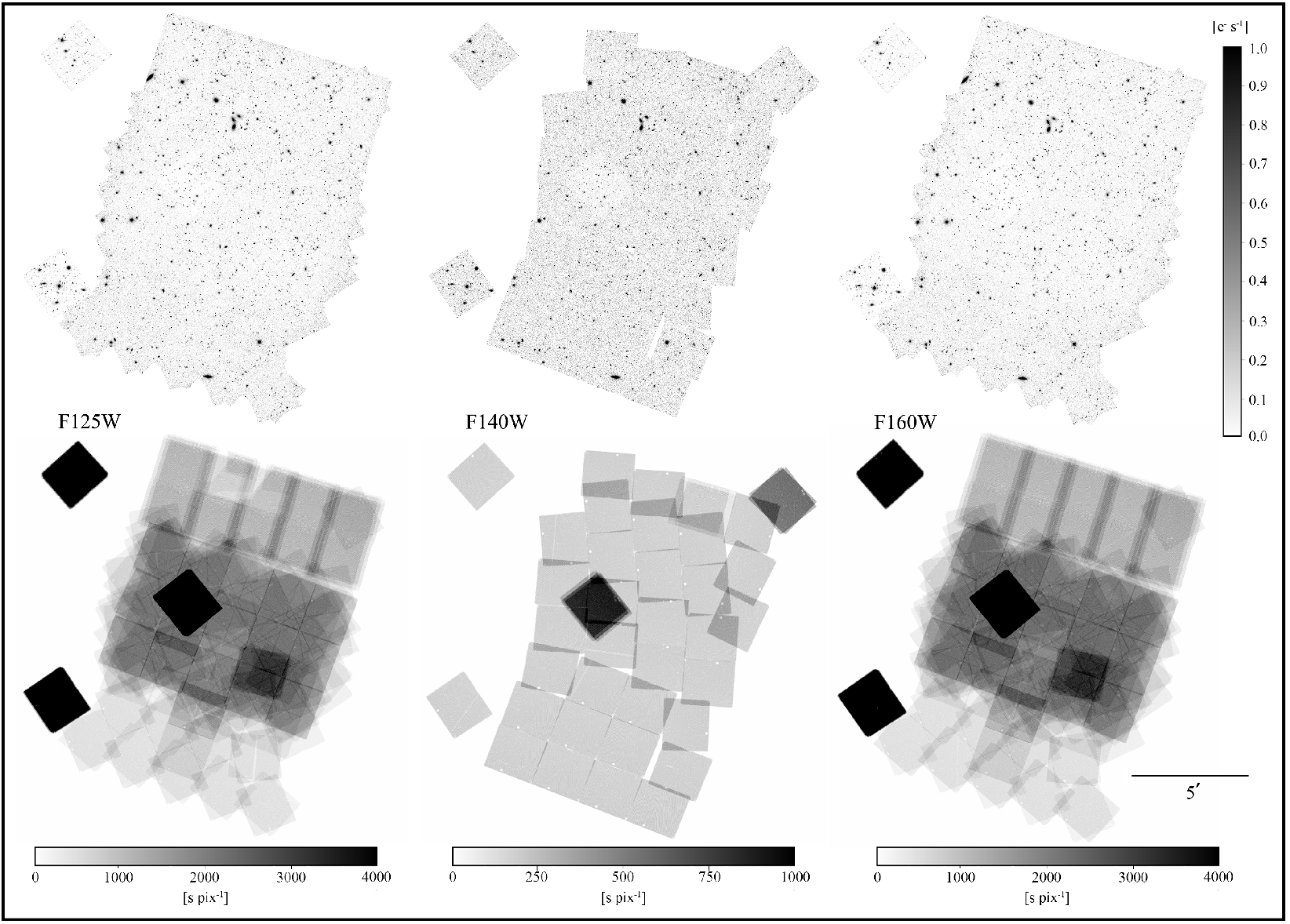}
\caption{Same as Figure \ref{fig:AEGIS_sci_wht}, for the GOODS-S field. North is up, East is to the left.   Note that the scale of the F140W weight image differs from that of F125W/F160W and that the deep HUDF regions are saturated on the exposure map scale shown here.}\label{fig:GOODS-S_sci_wht}
\end{figure*}

\begin{figure*}[!ht]
\centering
\includegraphics[width = \textwidth]{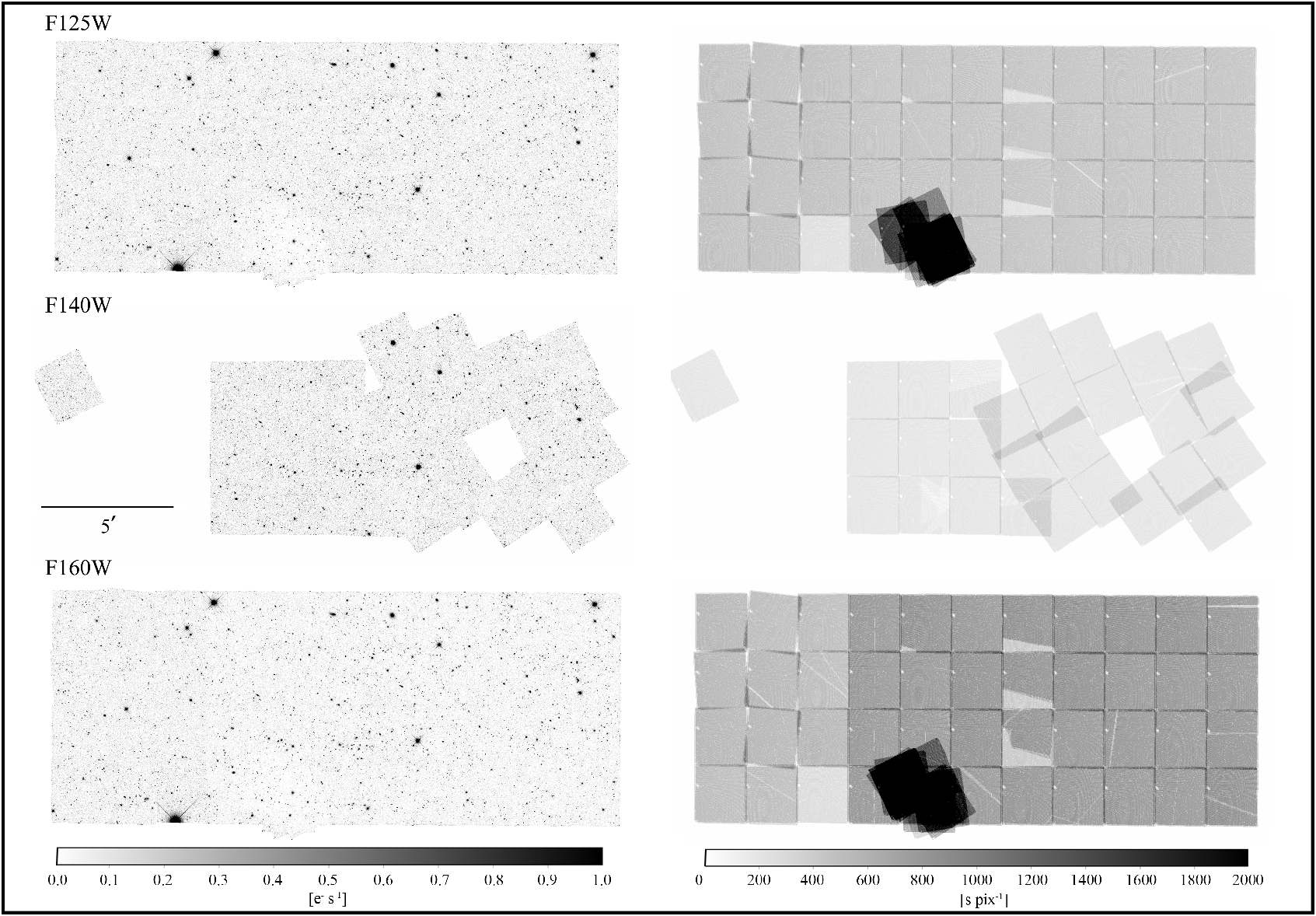}
\caption{Same as Figure \ref{fig:AEGIS_sci_wht} for the UDS field. North is up, East is to the left. Note that the deep region of the image saturates on the scale shown for the exposure map.}\label{fig:UDS_sci_wht}
\end{figure*}

The five CANDELS/3D-HST fields have been observed with HST/WFC3, HST/ACS, Spitzer, and many ground-based telescopes. In each field the heart of the data consists of the WFC3 F125W, F140W, and F160W images obtained by the CANDELS and 3D-HST Treasury programs. In this Section we describe our reductions of these data, and briefly discuss all other space- and ground-based data that are used to construct the SEDs. 

The photometric catalogs make use of some 150 different image mosaics.
As part of the analysis we
projected these data onto our astrometric grid and pixel scale (which
is identical to that used by the CANDELS team) and, in some cases,
process the images to remove artifacts.
For convenience, all images used in our work and the derived photometry are made available for
download on the 3D-HST website. The images are also available through the Mikulski Archive for Space Telescopes (MAST).
The majority of these data
have been made public previously, but this is the first time all of
them are offered as one comprehensive dataset.

\subsection{WFC3 Imaging}
\label{sec:wfc3im}
\subsubsection{Sources of Data}

The majority of HST/WFC3 imaging comes from the 3D-HST and CANDELS surveys which, jointly, have covered $\sim$940 arcmin$^2$ in three infrared filters: F125W, F140W and F160W. The coordinates of the five fields and the areas covered by the WFC3 imaging are given in Table~\ref{table:fields}. The ``Total area" column indicates the total area for which there is data in F125W, F140W or F160W, while the "Science area" column indicates the useful area after accounting for bright stars and regions without sufficient coverage in one of the CANDELS bands (see \S\,\ref{sec:numbercounts}). Other \textit{HST} programs have carried out observations of portions of these fields with combinations of the three filters. In order to increase the depth of the data and maximize
the footprint of the mosaics we have incorporated many of these additional  datasets. Table~\ref{table:hstdata} lists all HST/WFC3 datasets used in our work as well as the \textit{HST} proposal ID which requested the observations, the proposal PI and the total number of orbits. In total, we utilize 1160 orbits of HST/WFC3 imaging observations. Figure~\ref{fig:fieldlayout} illustrates the layout of the WFC3 observations. We summarize the relevant observational details for each field below, focusing primarily on the CANDELS and 3D-HST data. Details for the remaining programs can be found on the MAST archive\footnote{\url{http://archive.stsci.edu/hst/search.php}}. 

All near-infrared \textit{HST} observations were obtained using the Wide Field Camera 3 IR detector (WFC3/IR) which has a $1024\times1024$ HgCdTe array. The usable portion of the detector is $1014\times1014$ pixels, covering a region of $136\arcsec\times123\arcsec$ across with a native pixel scale of $0\farcs128$ pixel$^{-1}$ (at the central reference pixel). The majority of currently-available observations in the five deep fields are done in three wide filters: F125W, F140W and F160W, which cover the wavelength ranges of $\sim1.1\mu m - 1.4\mu m$, $\sim1.2\mu m - 1.6 \mu m$ and $1.4\mu m - 1.7 \mu m$, respectively. The F125W filter is slightly wider than the standard ground-based $J$-band, while the F160W is slightly narrower to better match the detector QE and to limit the effects of the thermal background. The F140W filter covers the gap between the $J$ and $H$ bands which is inaccessible from the ground. The standard designations for the three filters are $J_{F125W}$, $JH_{F140W}$ and $H_{F160W}$, however throughout this paper we will refer to them by the filter name to avoid confusion with ground-based $J$ and $H$ bandpasses. The WFC3/IR PSF has a FWHM between $0\farcs13$ and $0\farcs15$ over this wavelength range (1.02 to 1.18 native pixels). 

3D-HST is primarily a spectroscopic survey, with most of the $2\times 248$ primary and parallel orbits devoted to grism observations \citep[see, e.g.,][]{Brammer12a}. In addition to the grism exposures we obtain direct images in broad band filters, as required for wavelength calibration of the spectra and for associating spectra with objects \citep{Kuemmel09}. These direct images are obtained in the F814W filter for ACS and in the F140W filter for WFC3. The F140W filter is broad and overlaps with most of the wavelength range of the G141 grism.  In the context of the available imaging in the CANDELS fields the F140W data offer an important photometric datapoint between the CANDELS F125W and F160W imaging described below.  For the current photometric catalogs, we do not make use of the direct images taken with the F814W filter for 3D-HST, but rather use the deeper, publicly available CANDELS mosaics in this band. The majority of the 3D-HST data were obtained between October 2010 and May 2012, with two pointings in the AEGIS field obtained in December 2012. This paper makes use of all 124 pointings, as well as 28 pointings in the GOODS-N field, which was observed between September 2009 and April 2011 in the program GO-11600 (PI: B.\ Weiner). Each pointing was observed for two orbits, with $\sim800$ s of direct imaging in the F140W filter and 4511--5111 s with the G141 grism per orbit (amounting to $\sim 0.3$ orbits of imaging data per pointing in total). The average 5$\sigma$ depth of the F140W images is $JH_{F140W} \sim 25.8$~mag within a 1\arcsec aperture.  The point-source depth is 0.05~mag brighter than this, after correcting the depth for the flux outside of the 1\arcsec aperture using the growth curve; see \S\,\ref{sec:psfmatching}.

The HST/WFC3 observations for the CANDELS survey cover all five fields and have a two-tiered depth structure. The ``wide'' observations cover $\sim800$ arcmin$^2$ combined over the five fields to $2/3$-orbit depth in F125W and $4/3$-orbit depth in F160W. The F160W median 5$\sigma$ depths in a $1\arcsec$ aperture are 26.4, 26.4, 26.2, 26.6, 26.9, and 26.4 in the AEGIS, COSMOS, GOODS-N, GOODS-S and UDS fields, respectively. The CANDELS ``deep'' observations cover a smaller $\sim125$ arcmin$^2$ area in GOODS-N and GOODS-S with 4 orbits in F125W and 6 orbits in F160W. A third tier in depth is added by the even-deeper HUDF area in GOODS-S where the CANDELS, HUDF09 (GO: 11563; PI: G.\ Illingworth) and HUDF12 (GO: 12498; PI: R.\ Ellis) observations add up to 38 orbits in F125W, 33 orbits in F140W and 85 orbits in F160W.  The CANDELS data we use in this paper were taken between August 2010 and May 26, 2013. The final GOODS-N epoch, which was observed in August 2013, is not included in our current mosaics. Adding this epoch was not a priority for the current release as it does not provide additional coverage but only additional depth in the ``deep'' portion of the field. These data will be included in future reductions. We summarize the relevant details here \citep[for a detailed description of the CANDELS observations we refer the reader to][]{Koekemoer11}. 

{\it AEGIS: }  (a.k.a. EGS) The CANDELS footprint is a rectangular region of $3\times15$ pointings or $\sim6.5\arcmin\times32.5\arcmin$ (Figure \ref{fig:AEGIS_sci_wht}). Observations in F125W and F160W were carried out in two epochs at different roll angles.  The 3D-HST F140W observations in AEGIS comprise 30 pointings. 15 of them are arranged in a regular $3\times5$ pattern covering the north-western area of the field, minimizing overlap and maximizing coverage. Scheduling constraints limited the range of available roll angles for the remaining 15 pointings which led to gaps in the mosaic and more substantial overlap between pointings. The 3D-HST footprint in this field is smaller than the CANDELS one: approximately 2/3 of the CANDELS footprint also has 3D-HST F140W coverage. 

{\it COSMOS: } The CANDELS mosaic is a rectangular region of $4\times11$ pointings or $\sim8.6\arcmin\times23.8\arcmin$ (Figure~\ref{fig:COSMOS_sci_wht}). Observations in F125W and F160W were carried out in two epochs at the same roll angle. Deeper F125W and F160W observations of the TILE41 supernova (GO: 12461; PI: A.\ Riess) were added to our mosaics to aid in the reduction of the supernova grism observations. The 3D-HST F140W observations in COSMOS constitute 28 pointings, most of them arranged in a $3\times8$ pattern. The 3D-HST footprint covers $\sim2/3$ of the CANDELS footprint in this field.

{\it GOODS-N: } The CANDELS observations in this field are two-tiered. The ``deep'' area consists of  a rectangular grid of $3\times5$ pointings in F125W and F160W. The observations were done over 10 epochs (9 of which are used here) and roll angles vary by $\sim45-50$ degrees. The remaining southern and northern areas of the field are part of the shallower tier, each covered with $\sim2\times4$ pointings in both filters. The areas of the "deep" and "wide" coverage are distinctly visible in Figure~\ref{fig:GOODS-N_sci_wht}. The F140W observations were taken by GO:11600\footnote{\url{http://mingus.as.arizona.edu/~bjw/aghast/}} (PI: B.\ Weiner ) using a strategy identical to the one described above for the 3D-HST survey.  The field is covered with 30 pointings arranged in a $4\times6$ grid with 4 additional pointings covering the north-east edge of the field. There is no F140W imaging (or grism spectra) in the north-western edge of the field.  Additional images in F125W, F140W and F160W were added over the field of the COLFAX supernova (GO:12461; PI: A.\ Riess). 

{\it GOODS-S: } The CANDELS observations in GOODS-S are also two-tiered. The ``deep'' area observations mirror those in GOODS-N: they cover an area of $3\times5$ pointings in the F125W and F160W filters, obtained over 10 epochs. The observations of the southern portion of the field are in the shallow tier over an area of $\sim2\times4$ pointings. The northern portion of the field is covered by the 
WFC3/IR Early Release Science-2  (ERS) observations \citep[GO/DD: 11359,11360; PI: R.\ O'Connell; ][]{Windhorst2011} over an area of $2\times4$ pointings with two orbits in each of the F125W and F160W filters. We have further incorporated the observations form the Hubble Ultra Deep Field 2009 (HUDF09\footnote{\url{http://archive.stsci.edu/prepds/hudf09/}}) program (GO: 11563; PI: G.\ Illingworth). HUDF09 provides 34 orbits of F125W observations and 53 orbits of F160W observations in a single pointing in the center of the GOODS-S ``deep'' area. In addition, observations are carried out in two flanking fields HUDF09-1 and HUDF09-2, which coincide with prior ACS coverage. The depth in HUDF09-1 is 12 orbits in F125W and 13 orbits in F160W. The depth in HUDF09-2 is 18 and 19 orbits in the two filters, respectively. The F125W and F160W  mosaics also include the observations of the supernova GEORGE (GO: 12099; PI: A.\ Riess). The 3D-HST F140W coverage in this field is broken into 38 individual pointings. Of these, 32 cover a rectangular region $\sim8.6\arcmin\times17.3\arcmin$ in area. Two more pointings, GOODS-S-1 and GOODS-S-28, cover the flanking fields. The final four pointings overlap on the HUDF area to provide deep G141 grism spectra. The F140W mosaics also include data from the ERS2 program in a single pointing which fills a gap in the 3D-HST mosaic. Finally, we added the F140W direct observations for CDFS-AGN1 from GO:12190 (PI: A.\ Koekemoer), slightly extending the footprint of the mosaic. Figure~\ref{fig:fieldlayout} indicates the areas covered by different programs and Figure~\ref{fig:GOODS-S_sci_wht} indicates the depths across the field in each of the WFC3 bands.
  
{\it UDS: } The CANDELS mosaic is a rectangular region identical to the one in COSMOS: $4\times11$ pointings or $\sim8.6\arcmin\times23.8\arcmin$ (Figure \ref{fig:UDS_sci_wht})  with observations in both F125W and F160W taken over two epochs at the same roll angle. The pointings are arranged in a tight mosaic which maximizes contiguous coverage. Observations of the MARSHALL supernova in the F125W and F160W filters (GO: 12099; PI: A.\ Riess) are added to the mosaics. The 3D-HST F140W observations consist of 28 pointings. Ten of the 28 pointings form a regular grid which covers the central portion of the CANDELS footprint, matching the F125W/F160W roll angle. Due to scheduling constraints, the remaining 18 pointings are rotated by $\sim45$ degrees, 17 of them providing a more uneven coverage of the east portion of the field. The F140W coverage has a hole because the final pointing, UDS-18, was moved to the western-most edge of the CANDELS coverage to carry out G141 observations of the IRC0812A $z=1.62$ cluster \citep{Papovich2010}. Figure \ref{fig:fieldlayout} illustrates the position of  UDS-18 relative to the full mosaic. 

\begin{table*}[ht]
\centering
\caption{HST Observations}\label{table:hstdata}
\begin{tabular}{lllrccll}
\hline
\hline
\noalign{\smallskip}
Field & Instrument & Filters & N$_{orbits}$
&Proposal ID & \textit{HST} Cycle & Survey
& PI \\
\noalign{\smallskip}
\hline
\noalign{\smallskip}
AEGIS & WFC3
&F125W, F160W & 90 &12063  & 18
&CANDELS & Faber \\
& WFC3
&F140W & 8$^{\dagger}$
&12177  & 18 & 3D-HST
& van Dokkum \\
& ACS
&F606W, F814W  & 90 & 12063  & 18
&CANDELS & Faber
\\ 
\noalign{\smallskip}
\hline
\noalign{\smallskip}
COSMOS & WFC3
& F125W, F160W  & 88 &12440
& 18 &CANDELS
& Faber \\
& WFC3
& F125W, F160W  & 10$^{\dagger}$ &12461 
& 19 & TILE41  & Riess \\
& WFC3
& F140W & 8$^{\dagger}$
&12328 & 18
& 3D-HST & van Dokkum
\\
&ACS
& F606W, F814W  & 88 & 12440
& 18 &CANDELS
&  Faber \\
\noalign{\smallskip}
\hline
\noalign{\smallskip}
GOODS-N& WFC3 & F125W, F160W
& 173 &12443-12445 & 19
& CANDELS & Faber
\\
& WFC3
& F140W & 7$^{\dagger}$
&11600 & 17
& AGHAST & Weiner
\\
& WFC3
& F125W, F140W, F160W
& 5$^{\dagger}$  &12461 & 19
& COLFAX & Riess\\
&ACS
& F606W, F814W,F850LP
& 219 &12442-12445 &19 & CANDELS
&  Faber \\
&ACS
& F435W,F606W,F775W,F850W
&199 & 9583 & 11
& GOODS & Giavalisco
\\
\noalign{\smallskip}
\hline
\noalign{\smallskip}
GOODS-S& WFC3 & F125W, F160W
& 173 &12061,12062 & 18
& CANDELS & Faber
\\ 
& WFC3
& F140W & 11$^{\dagger}$
&12177 & 18
& 3D-HST & van Dokkum \\
& WFC3
& F125W, F160W  & 4$^{\dagger}$ &12099 & 18
& GEORGE & Riess
\\
& WFC3
& F140W & 1$^{\dagger}$
&12190 & 18
& CDFS-AGN1 & Koekemoer
\\
& WFC3
& F125W, F140W, F160W
&12 &11359,11360 & 17
& ERS & O'Connell
\\
& WFC3
& F125W, F160W  & 149 &11563  
& 17 & HUDF09 & Illingworth
\\
&ACS
& F606W, F814W,F850LP 
& 229 & 12060-12062& 18
& CANDELS &  Faber
\\
&ACS
& F435W,F606W,F775W,F850W& 199 
& 9425 & 11
& GOODS & Giavalisco
\\
\noalign{\smallskip}
\hline
\noalign{\smallskip}
UDS & WFC3
& F125W, F160W  & 88 &12064
& 18 & CANDELS  & Faber\\
& WFC3
& F140W &8$^{\dagger}$
&12328 & 18
& 3D-HST & van Dokkum \\
& WFC3
& F125W, F160W  & 18$^{\dagger}$ & 12099
& 18 & MARSHALL  & Riess \\
&ACS
& F606W, F814W  & 88 & 12064
& 18 & CANDELS  & Faber \\
\noalign{\smallskip}
\hline
\noalign{\smallskip}
\end{tabular}
\begin{tablenotes}
\item[]${\dagger}${ For orbits which contain grism observations, the number of orbits has been determined based on the fraction of the time dedicated to direct images and rounded to the nearest full orbit. }
\end{tablenotes}
\end{table*}

\subsubsection{Data Reduction}
\label{sec:wfc3reduction}

We downloaded the calibrated images and association tables from the MAST archive between April and June 2013. These images were processed on the fly with the best available calibrations at the time by the {\it calwfc3} pipeline. We briefly summarize the calibration steps. The {\it calwfc3} task starts with the raw files ({\it *\_raw.fits}) and first populates the data quality arrays from the known bad pixel tables. It subtracts the bias for each read based on the overscan regions. It then subtracts the zeroth read to remove the bias structure across the detector, subtracts the dark current reference file based on the readout sequence and performs a non-linearity correction. Following these corrections on the individual reads, the task does an up-the-ramp fit to each pixel to maximize the dynamic range of the images and identify cosmic rays. The count rate is computed from the unflagged reads for each pixel and stored in the final calibrated exposure. The uncertainty in the slope is stored as the error array. Finally, the appropriate multiplicative corrections for the gain and the flat-field are applied. The resulting images  ({\it *\_flt.fits}) are placed on the STAGE drive and downloaded via FTP. The flat-field correction is reapplied as described below.

A number of corrections are applied to improve the data quality and produce the final data products: masking satellite trails, persistence correction, sky-subtraction, flat-field re-application, initial astrometric alignment, and additional cosmic-ray and bad pixel rejection. Some of the these steps have already been described in \citet{Brammer12a} in the context of the F140W and G141 reduction. These are briefly summarized here with more attention given to new steps.

The pipeline-reduced images occasionally contain satellite trails and other cosmetic blemishes, which we identify by visually inspecting all {\it *\_flt.fits} images.  
When necessary we create mask files which mark the positions of any cosmetic blemishes in the following manner.  Each {\it *\_flt.fits} image is displayed in DS9 and blemishes are marked with a polygon region. The coordinates of all regions are saved and then used in the reduction to set the 2048 bit for the corresponding pixels in the quality array (i.e., interpreted and ignored as a cosmic ray).  The positions of such masked blemishes can be seen in the weight images; an example is shown in Figure~\ref{fig:wht_maps}. Contamination from scattered light from the bright Earth limb has been noticed in some F140W images, especially in the GOODS-N field. 
Occasionally, a similar issue affects F125W and F160W exposures. For the current release we mitigate the problem by masking the affected exposures in the same manner as the satellite trails. In future releases, the F125W and F160W issue will be treated more carefully by removing only the affected reads (typically, at the beginning or the end of the exposure). Unfortunately, the short direct F140W observations will not benefit from the new approach as they are comprised of very few reads.

Persistence is a concern with the WFC3/IR detector, causing residual flux in an exposure
from bright targets in preceding separate observing programs or in other exposures within the same program. The 3D-HST observations interleave the direct F140W images and the G141 grism spectra and persistence from the spectra of bright stars commonly occurs in the subsequent direct images. The spatial extent of the persistent flux is larger than the dither pattern steps and they are not usually flagged by {\it Multidrizzle}. The imaging observations from the other grism programs incorporated in our mosaics suffer the same effect. We download the persistence images provided by STScI ({\it *\_persist.fits}), which provide estimates for the total persistence, for all grism programs as well as the HUDF09 program from the MAST archive\footnote{\url{http://archive.stsci.edu/prepds/persist/}}. Rather than subtract the persistence, we adopt a more conservative approach and use the persistence images to flag affected pixels. For each individual image, we create a mask of the image where the persistence flux is greater than 0.6 times the error, convolve the mask with a $3\time3$ maximum filter to grow the area slightly and then set the masked pixels in the error array to 4196. These are later treated as cosmic rays and are not used in the final mosaics. The most severe persistence in the CANDELS observations was masked by hand as described above for the satellite trails. 

Even though HST is above Earth's atmosphere the near-IR background is non-negligible, with the background arising predominantly from zodiacal light.  The background subtraction of the F140W images is described in detail by \citet{Brammer12a}. For each association table, we align the images to each other using {\it tweakshifts} and then create a combined pointing image with {\it Multidrizzle}. Objects are detected in the combined image using SExtractor and masked aggressively in the original distorted frames. The background is determined in two passes: first a median is subtracted, then a first order polynomial is fit to the background and subtracted. The same procedure is applied to the F125W and F160W images.

We find that the standard flat applied as part go the {\it calwfc3} task is insufficient to correct for the time-variable behavior of some features, namely the appearance of new ``IR blobs'' with time. We therefore create and apply time-dependent flats in the three WFC3/IR filters by splitting the CANDELS and 3D-HST observations in epochs (two for F125W and F160W, three for F140W) and creating super-sky flats from the masked science exposures themselves with a method similar to that described by \cite{Pirzkal11}. 

We use the {\it Multidrizzle} software \citep{Koekemoer02} to identify hot pixels and cosmic rays not flagged by the instrument calibration pipeline \citep{Brammer12a}. This step is applied separately for each epoch of the CANDELS observations to avoid flagging the diffraction spikes of stars (which vary between epochs due to the different rotation angles). In the ``wide'' areas of CANDELS where each epoch only has two exposures, some cosmic rays and hot pixels may be missed. Therefore, our mosaics may appear less cosmetically clean. We accept this compromise to preserve the structure of stellar PSFs, as otherwise the core of the PSF and the diffraction spikes are frequently flagged as cosmic rays.  We adopt a conservative value of the relevant {\it Multidrizzle} parameter, {\sc driz\_cr\_scale = `2.5 0.7'}. 

In order to provide sub-pixel sampling and mitigate the effects of  hot pixels and other artifacts, all observations are dithered between exposures. Both CANDELS and 3D-HST observations employ a four-point dither pattern that provides half-pixel subsampling. In 3D-HST, all four exposures are taken during a single visit \citep[see][]{Brammer12a}, while for CANDELS-Wide, two exposures are taken during each of two epochs \citep{Koekemoer11}. The largest blemish on the WFC3/IR detector is a circle of dead pixels $\sim50$ pixels in diameter, dubbed the ``Death Star''. The CANDELS epoch-to-epoch dither steps are large enough to cover the hole; however, 3D-HST only has a single epoch with small $\leq10$ pixels offsets. Therefore the ``Death Star'' is present in the F140W mosaics.

At the end of the processing steps we add back a constant background estimated by comparing the median levels in the pipeline and the processed FLT images. This is done so that the Poisson term from the background subtraction is added to the pixel errors by {\it Astrodrizzle} when creating the final mosaics.

\subsubsection{The WFC3/IR Mosaics}

\begin{figure*}
\centering
\includegraphics[width = \textwidth]{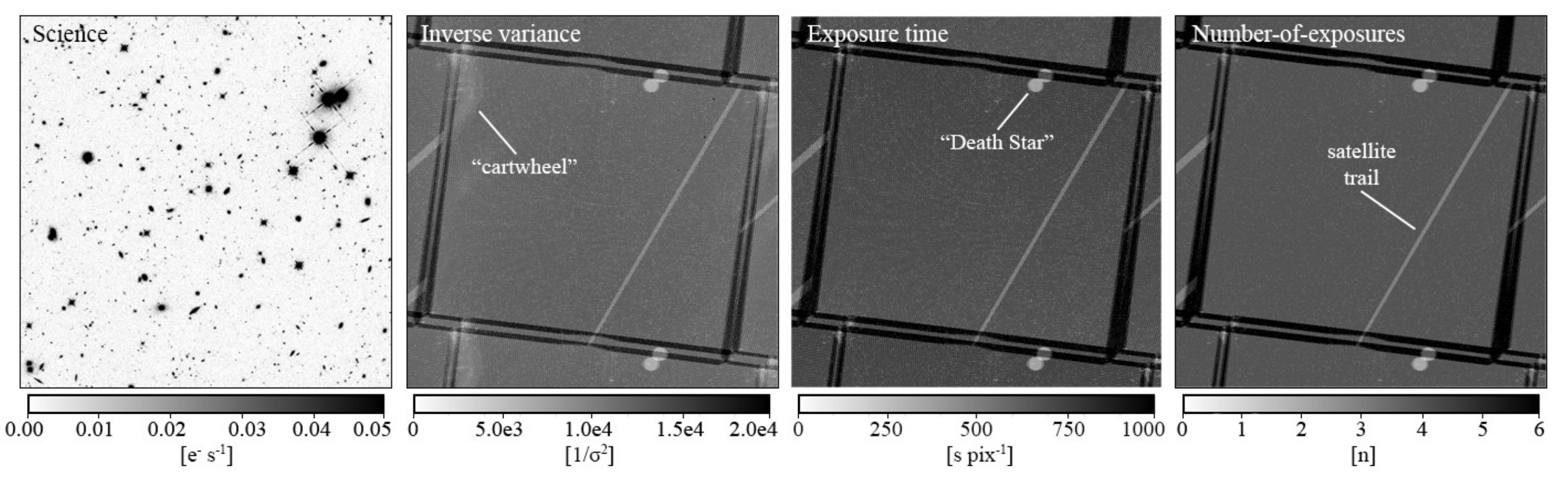}
\caption{The WFC3 science images are accompanied by three types of weight maps: inverse variance, exposure time and number-of-exposures. We show an area of the COSMOS F160W mosaic (pointing COSMOS-V24/COSMOS-V68) to illustrate the differences in the weight maps. Each cutout is $\sim2\farcm5$ on the side. A scale for each of the images is provided below the cutout. All three weight maps reflect masked and flagged pixels (satellite trails, the "Death Star"), but only the inverse variance map takes into account other error sources (flat field, background subtraction, read noise, dark current, etc.). All weight maps, as well as area maps that can be used for clustering analysis, are made available as part of the data release.}\label{fig:wht_maps}
\end{figure*}

Finally, the individual processed images are corrected for distortions
and combined into mosaics for each field and each filter. A total of 3,477 individual {\it *\_flt.fits} images containing $3.77\times10^9$ pixels are combined to produce these mosaics. Here we describe the alignment to a common world coordinate system (WCS) and the production of the final mosaics. 

Due to uncertainties of the guide star positions, small adjustments to the commanded telescope positions are needed to align the individual images with accuracy $<1$ pixel. We use {\it Tweakreg} (version 1.2.1, updated 2013-01-25) to align each image to an external reference frame. {\it Tweakreg} fits for small differences in position, orientation, and scale between adjacent \textit{HST} pointings. Measurements of positions of objects are used to determine the relative offsets; the WCS information in the image header is used as an initial estimate.  The source finding algorithm used by {\it Tweakreg} has been optimized for point sources, which are sparse in the deep extragalactic fields of 3D-HST/CANDELS. For this reason we do not rely on the automatic source finding algorithm, but supply a reference catalog to be used in place of a catalog extracted from the reference image. This procedure allows us to use all the sources in the image for the alignment, and enables a transformation to an absolute reference frame. The reference catalogs are created using SExtractor v2.8.6 \citep{Bertin96}  on the following publicly available images: AEGIS, ACS-F814W \citep{Davis07}; COSMOS, WFC3-F160W (v1.0, \citealt{Koekemoer11}); GOODS-N and GOODS-S,  ACS-F814W; and UDS, WFC3-F160W (v1.0, \citealt{Koekemoer11}). The 3D-HST F140W images are shallower than the CANDELS F125W and F160W images. For these images we used a magnitude limited catalog (F814W, F160W$\,<\,24$) to decrease the rms of the differences in matched positions. For the UDF parallel fields in GOODS-S, HUDF09-01 and HUDF09-02, we used the F850LP tiles 42 and 25 respectively from the GEMS survey \citep{Rix04}. Aside from the reference catalog input variables, we used the defaults for all other {\it Tweakreg} parameters. The typical rms in the differences of matched positions is 0.4 native pixels, corresponding to an uncertainty of $\approx 0.3$ pixels or  $\approx 0\farcs03$ in the positions of the (generally faint, spatially-extended) objects that were used in the procedure. 

The final mosaics in each filter are produced with {\it AstroDrizzle} (version 1.1.9.dev23803, 2013-02-06). We use inverse-variance weighting, a square kernel, and {\it pixfrac=0.8}. In order to exactly match the pixel scale ($0\farcs06$), the center, and the tangent point of our output mosaics to those produced by CANDELS we provide {\it AstroDrizzle} with reference images from the CANDELS public data releases (v0.5 for AEGIS and GOODS-N; v1.0 for COSMOS, GOODS-S and UDS). The publicly available GOODS-S v1.0 images from CANDELS cover an area much larger than the WFC3 data which makes the final file sizes unwieldy. Our final mosaics are matched to a cropped version of the CANDELS mosaic such that the $x$ and $y$ positions of all objects are smaller by 5000 and 11000 pixels respectively. We create an association table of all images taken with a given filter and use it as an input to {\it AstroDrizzle}. The final images have a pixel scale of $0\farcs06$/pixel. A small portion of the UDS-18 F140W 3D-HST pointing falls outside the CANDELS footprint, however all other pointings are fully contained within the mosaics. The reduction process is very similar to that used to produce the publicly available CANDELS mosaics, as described in \citet{Koekemoer11}. A key difference is that we include all the available epochs, as well as observations from other programs. Other differences include the lower threshold for cosmic ray rejection, the application of a persistence correction, the time-variable post-pipeline flat-fielding, the use of different reference images for astrometric alignment and {\it Astrodrizzle} rather than {\it MultiDrizzle} for the production of the final mosaics.

Accompanying the images are exposure maps, inverse variance weight maps and number-of-exposures maps. These outputs are used throughout the photometric analysis described below to calculate the errors, determine the depths and define source reliability flags. The inverse variance and exposure maps are standard outputs of {\it AstroDrizzle} and were created by doing consecutive runs with different {\it final\_wht\_type} selection. The exposure map is in units of seconds. In order to preserve flux, the exposure time in each original pixel is divided by the ratio in the areas of the original to the final pixels. The inverse variance weight map is based on the flat-field reference file and computed dark value from the image header and the final weight image accounts for all background noise sources (sky level, read noise, dark current, etc.) but not the Poisson noise from the objects themselves. The inverse variance weight map is used as a weight image input for {\it SExtractor}. The number-of-exposures map encodes the number of individual images that have contributed to the flux in each pixel. These are produced from the context images output by {\it AstroDrizzle}. A comparison of the three weight maps for a region of the COSMOS field is shown in Figure~\ref{fig:wht_maps}. The maps have different scales but generally reflect the same features because flagged pixels are taken into account by all of them. Features in the flat field (the "cartwheel" and the IR-blobs which are too small to be visible in the figure) are only visible in the inverse variance map. The moire patterns in the exposure and inverse variance maps are real and result from the dithering. 

The WFC3 mosaics and all weight images are made available as part of the data release. 

\subsection{Additional Data}
\label{sec:otherdata}

\begin{table*}[ht]
\centering
\caption{Image sources}\label{table:ancil_data}
\begin{tabular}{lllll}
\hline \hline
\noalign{\smallskip}
Field & Filters & Telescope/Instrument & Survey & Reference \\
\hline
\noalign{\smallskip}
AEGIS & $u$, $g$, $r$, $i$, $z$ & CFHT/MegaCam & CFHTLS & \citet{Erben09, Hildebrandt09}\\
  &F606W, F814W  & HST/ACS & CANDELS & \citet{Grogin11, Koekemoer11} \\
&$J1$, $J2$, $J3$, $H1$, $H2$, $K$ & KPNO 4m/NEWFIRM &  NMBS & \citet{Whitaker11} \\
 &$J$, $H$, $K_s$  &  CFHT/WIRCam & WIRDS & \citet{Bielby12}\\
 &F140W  & HST/WFC3 & 3D-HST & \citet{Brammer12a}\\
 &F125W, F160W  & HST/WFC3 & CANDELS  & \citet{Grogin11, Koekemoer11}\\
 &3.6, 4.5 $\mu m$ & Spitzer/IRAC & SEDS & \citet{Ashby13} \\
 & 5.8, 8 $\mu m$ & Spitzer/IRAC & EGS & \citet{Barmby08} \\
\noalign{\smallskip}
\hline
\noalign{\smallskip}
COSMOS & $u$, $g$, $r$, $i$, $z$ & CFHT/MegaCam & CFHTLS &  \citet{Erben09, Hildebrandt09}\\
 & $B$, $V$, $r'$, $i'$, $z'$, 12 medium-band optical  & Subaru/Suprime-Cam & &  \citet{Taniguchi07}\\
 &F606W, F814W   & HST/ACS & CANDELS & \citet{Grogin11, Koekemoer11}\\
 &$J1$, $J2$, $J3$, $H1$, $H2$, $K$ & KPNO 4m/NEWFIRM & NMBS & \citet{Whitaker11} \\
 &$J$, $H$, $K_s$  &  CFHT/WIRCam & WIRDS & \citet{Bielby12}\\
 &$Y$, $J$, $H$, $K_s$ & VISTA & UltraVISTA & \citet{McCracken12} \\
 &F140W & HST/WFC3 & 3D-HST & \citet{Brammer12a}\\
 &F125W, F160W & HST/WFC3 &  CANDELS & \citet{Grogin11, Koekemoer11}\\
 &3.6, 4.5 $\mu m$ & Spitzer/IRAC & SEDS & \citet{Ashby13} \\
 &5.8, 8 $\mu m$ & Spitzer/IRAC & S-COSMOS & \citet{Sanders07}\\
\noalign{\smallskip}
\hline
\noalign{\smallskip}
GOODS-N & $U$  & KPNO 4m/Mosaic & Hawaii HDFN & \citet{Capak04}\\
 & $G$, $R_s$  &  Keck/LRIS & & \citet{Steidel03}\\
 & F435W, F606W, F775W, F850LP   & HST/ACS & GOODS  &  \citet{Giavalisco04} \\
 & $B$, $V$, $R_c$, $I_c$, $z'$ & Subaru/Suprime-Cam & Hawaii HDFN  &  \citet{Capak04} \\
& F140W   & HST/WFC3 & 3D-HST &  \citet{Brammer12a}\\
 & F125W, F160W  & HST/WFC3 & CANDELS &\citet{Grogin11, Koekemoer11}\\
 & $J$, $H$, $K_s$  & Subaru/MOIRCS & MODS &  \citet{Kajisawa11}\\
  &3.6, 4.5 $\mu m$ & Spitzer/IRAC & SEDS & \citet{Ashby13} \\
&5.8, 8 $\mu m$ & Spitzer/IRAC & GOODS &  \citet{Dickinson03}\\
\noalign{\smallskip}
\hline
\noalign{\smallskip}
GOODS-S &$U$, $R$  & VLT/VIMOS & ESO/GOODS &  \citet{Nonino09}\\
& $U38$, $B$, $V$, $R_c$, $I$ & WFI 2.2m & GaBoDs & \citet{Hildebrandt06, Erben05}\\
& 14 medium bands & Subaru/Suprime-Cam & MUSYC & \citet{Cardamone10}\\
 &F435W, F606W, F775W, F850LP   & HST/ACS & GOODS &  \citet{Giavalisco04} \\
 &F606W, F814W, F850LP  & HST/ACS & CANDELS & \citet{Grogin11, Koekemoer11}\\
&F140W  & HST/WFC3 & 3D-HST &  \citet{Brammer12a} \\
& F125W, F160W  & HST/WFC3 & CANDELS & \citet{Grogin11, Koekemoer11}\\
 &$J$, $H$, $K_s$  & VLT/ISAAC & ESO/GOODS, FIREWORKS &  \citet{Retzlaff10}, \citet{Wuyts08} \\
 & $J$, $K_s$ & CFHT/WIRcam & TENIS & \citet{Hsieh12}\\
 &3.6, 4.5 $\mu m$ & Spitzer/IRAC & SEDS & \citet{Ashby13} \\
 &5.8, 8 $\mu m$ & Spitzer/IRAC & GOODS &  \citet{Dickinson03} \\
\noalign{\smallskip}
\hline
\noalign{\smallskip}
UDS &  $U$   & CFHT/MegaCam & & Almaini/Foucaud in prep.\\
 & $B$, $V$, $R_c$, $i'$, $z'$   & Subaru/Suprime-Cam & SXDS &  \citet{Furusawa08} \\
 &F606W, F814W  & HST/ACS &  CANDELS & \citet{Grogin11, Koekemoer11}\\
 &F140W  & HST/WFC3 & 3D-HST &  \citet{Brammer12a} \\
 &F125W, F160W  & HST/WFC3 & CANDELS & \citet{Grogin11, Koekemoer11} \\
 &$J$, $H$, $K_s$ &  UKIRT/WFCAM & UKIDSS DR8 & Almaini in prep. \\
  &3.6, 4.5 $\mu m$ & Spitzer/IRAC & SEDS & \citet{Ashby13} \\
&5.8, 8 $\mu m$ & Spitzer/IRAC & SpUDS & Dunlop in prep.\\
\noalign{\smallskip}
\hline
\noalign{\smallskip}
\end{tabular}
\end{table*}

\begin{figure*}[ht]
\centering
\includegraphics[height=0.93\textheight]{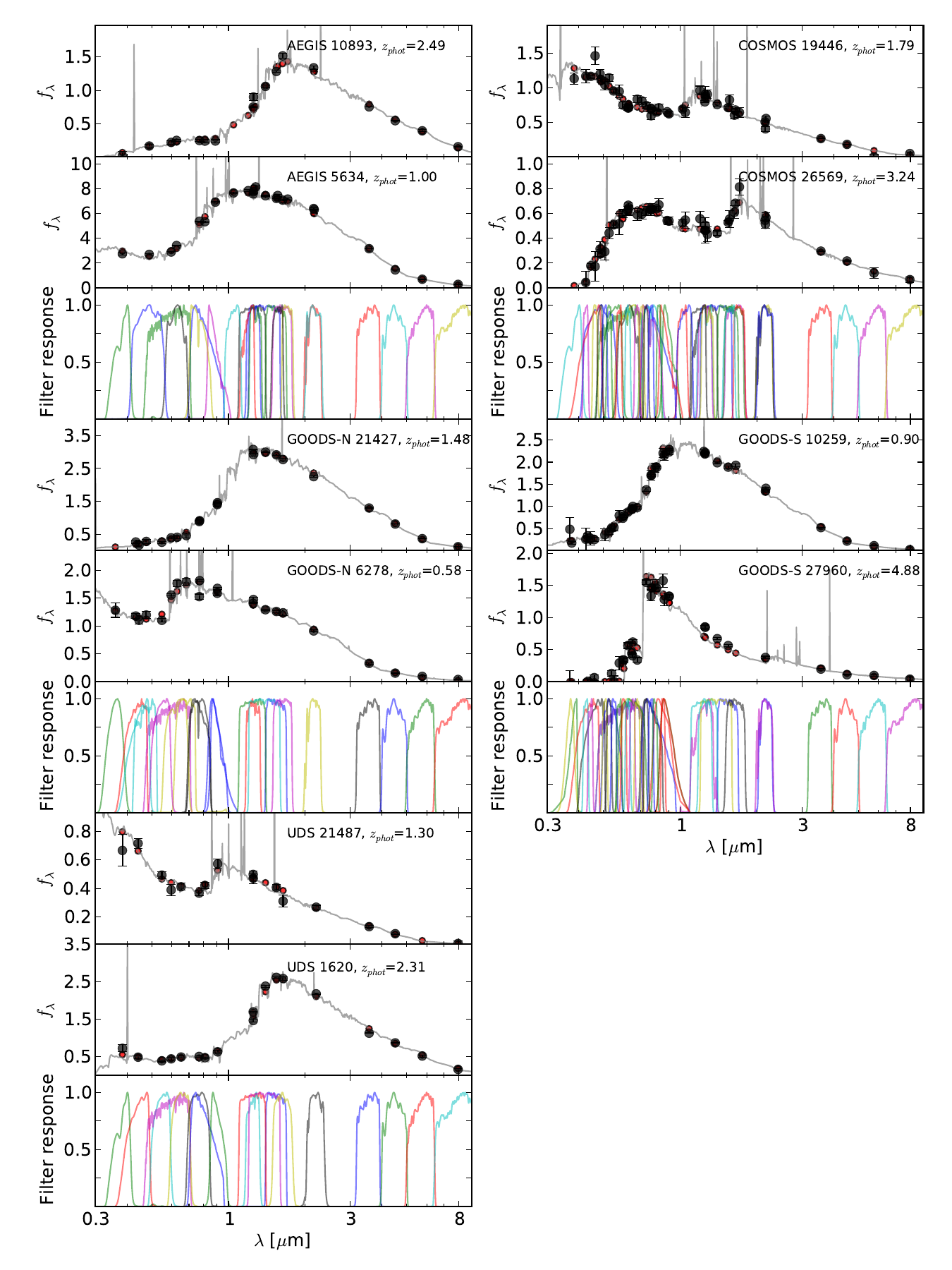}

\caption{The photometric filter set and two example SEDs for each of the five fields. The lower panels show the full set of filter curves used for the photometric catalog of each field, normalized to a maximum transmission of one. The upper two panels show the SEDs of two galaxies randomly chosen to demonstrate some of the variety of objects in the catalogs. The id and redshift of each object are shown in the top right corner. The black circles indicate the observed flux in each filter, the best-fit EAZY template spectra are shown in grey with the expected flux from the fit for each band in that catalog shown by the red circles. Some objects lie in areas of the image without coverage in one or more bands; in these cases only the predicted flux (red point) is shown. COSMOS and the GOODS-South field are particularly well sampled by the medium bands in the optical region, but there is excellent multi-wavelength coverage in all of the fields.}
\label{fig:filtercurves}
\end{figure*}

We use publicly available ancillary data from many different sources to build a comprehensive photometric catalog for each field. The image sources are listed in Table~\ref{table:ancil_data} and described in the section for each field below. Fig.~\ref{fig:filtercurves} shows the set of filter curves used for each field, together with examples of the resulting SEDs. The wavelength coverage in each field is excellent, spanning from the $U$-band through 8~$\mu$m. 
Each of the images was matched to the same pixel grid as the WFC3 images using the IRAF task {\tt{wregister}}. The GOODS-N and UDS IRAC 8 $\mu m$ images showed a clear shift with respect to the WFC3 images. In these cases the image registration was improved by using the iraf task {\tt{xregister}}, which cross-correlates source positions to determine the shifts between the images.

We obtained all \textit{HST} ACS images from the CANDELS program, and as we used the CANDELS coordinate system for our WFC3 mosaics these data are exactly aligned with ours. However, the ground-based and \textit{Spitzer} IRAC data exhibit small, position-dependent astrometric offsets, which are due to a combination of the use of slightly different absolute reference systems and small position-dependent errors in the astrometry of the various datasets.  The photometry software we use, described in \S\,\ref{sec:lowresphot}, fits not only for the position-dependence of the PSF but also for position-dependent astrometric errors. Information on astrometric offsets will be supplied with the release of the ancillary data. 

\subsubsection{AEGIS}
The 3D-HST AEGIS field lies within the larger Extended Groth Strip (EGS), which has publicly available imaging at many wavelengths from the All-wavelength Extended Groth Strip International Survey (AEGIS, \citealt{Davis07}). We incorporate imaging in 23 filters into the AEGIS photometric catalog. In the optical we use {\it ugriz} Deep Canada--France--Hawaii Telescope Legacy Survey (CFHTLS) broadband images from the CARS team \citep{Erben09, Hildebrandt09} and our own reduction of the HST/ACS F606W and F814W images from the CANDELS survey. In the NIR, we use $JHK_S$ imaging from the WIRCam Deep Survey (WIRDS;
\citealt{McCracken10, Bielby12}) and {\it J1, J2, J3, H1, H2, K} from the NEWFIRM Medium-Band Survey (NMBS, \citealt{Whitaker11}). The IRAC 3.6, 4.5$\mu m$ images are from the Spitzer Extended Deep Survey (SEDS, \citealt{Ashby13}) v1.2 data release, while the 5.8 and 8$\mu m$ images are from the EGS \citep{Barmby08}. The central wavelength of the filter, 95\% cumulative throughput width of the filter, dust attenuation from Galactic extinction, aperture used for photometry, image zero point, average FWHM and median 5$\sigma$ depth for each of the images are listed in Table~\ref{table:aegis_image_data}. Here the depths are calculated as the median of five times the total error in each band for the all the objects in the catalog, in magnitudes.

\begin{table*}[th]
\centering
\caption{AEGIS Imaging Data}
\label{table:aegis_image_data}
\begin{tabular}{lccccccc}
\hline \hline
\noalign{\smallskip}
Band & $\lambda_{\mathrm{central}}$ & Width$^a$ & $A_{\lambda}$ & Aperture & Zero point & FWHM & 5$\sigma$ Depth$^b$\\
 & ($\mu m$) &($\mu m$) & (mag) & (arcsec) &(AB) & (arcsec) & (mag) \\
\hline
\noalign{\smallskip}
  U& 0.3828& 0.0771 & 0.037& 1.2& 30.000&0.88 & 26.8 \\
  G& 0.4870& 0.1428 & 0.030& 1.2& 30.000&0.81& 27.4\\
F606W& 0.5921& 0.2225 & 0.023& 0.7& 26.491&0.13&26.8 \\
  R& 0.6245& 0.1232 &  0.020& 1.2 & 30.000&0.81&27.3 \\
  I& 0.7676& 0.1501 & 0.015& 1.2& 30.000&0.81& 27.0\\
F814W& 0.8057& 0.2358  & 0.014& 0.7& 25.943&0.11& 26.4\\
  Z& 0.8872& 0.1719 & 0.012&1.2 & 30.000&0.72& 26.2\\
 J1& 1.0460& 0.1471 & 0.010& 1.5& 23.310&1.13& 24.8\\
 J2& 1.1946& 0.1476 & 0.008& 1.5& 23.350&1.16&24.5 \\
 J3& 1.2778& 0.1394 & 0.006& 1.5& 23.370&1.08&24.4 \\
F125W& 1.2471& 0.2867 & 0.006& 0.7& 26.247&0.18& 26.3 \\
  J & 1.2530& 0.1541 & 0.006& 1.2& 30.000&0.76&  24.5\\
F140W& 1.3924& 0.3760 & 0.005& 0.7& 26.465&0.19& 25.7\\
F160W& 1.5396& 0.2744 &  0.004& 0.7& 25.956&0.19& 26.1\\
  H1& 1.5601& 0.1658 & 0.005& 1.5& 23.590&1.10&23.8 \\
  H2& 1.7064& 0.1721 & 0.004& 1.5& 23.610&1.06& 23.8\\
  H& 1.6294& 0.2766 & 0.004& 1.2& 30.000&0.68& 24.3\\
  K& 2.1684& 0.3181 & 0.003& 1.5& 23.850&1.08& 23.6\\
 Ks& 2.1574& 0.3151 & 0.003& 1.2& 30.000&0.69& 24.0\\
IRAC1& 3.5569& 0.7139 & 0.000& 3.0& 21.545& 1.7 & 25.2\\
IRAC2& 4.5020& 0.9706 & 0.000& 3.0& 21.545& 1.7 & 25.1\\
IRAC3& 5.7450& 1.3591 & 0.000& 3.0& 21.545& 1.9 &22.6 \\
IRAC4& 7.9158& 2.7893 & 0.000& 3.0& 21.545& 2.0  & 22.6\\
\noalign{\smallskip}
\hline
\noalign{\smallskip}
\multicolumn{8}{l}{$^a$ 95\% cumulative throughput width} \\
\multicolumn{8}{l}{$^b$ Median 5$\sigma$ depth calculated from the errors of objects in the final catalogs} 
\end{tabular}
\end{table*}

 \subsubsection{COSMOS}

The 3D-HST COSMOS field is within the Cosmic Evolution Survey field \citep{Scoville07a} and has a wealth of publicly available ancillary data. The catalog includes a total of 44 bands---27 broad bands and 17 medium bands that span both the optical and NIR. We use {\it ugriz} broadband images from the final release of the Deep Canada--France--Hawaii Telescope Legacy Survey (CFHTLS) \citep{Cuillandre12}, deep Subaru broadband images in the $B$, $V$ ,$ r'$, $i'$, and $z'$ bands\footnote{Note that these bands are designated $B_J$, $V_J$, $r^+$, $i^+$ and $z^+$ in some previous works on the COSMOS field \citep[e.g.][]{Whitaker11}. We use the Subaru naming convention here for consistency between the fields.} and 12 Subaru optical medium bands \citep{Taniguchi07}. We add the CANDELS ACS/F606W and ACS/F814W bands using the publicly released v1.0 mosaics.\footnote{\url{http://candels.ucolick.org/data_access/COSMOS.html}} In the NIR we use images from the UltraVISTA survey in $Y$, $J$, $H$ and $K_s$ \citep{McCracken12}, {\it J1, J2, J3, H1, H2, K} from NMBS \citep{Whitaker11} and $JHK_S$ from WIRDS \citep{McCracken10, Bielby12}. The IRAC 3.6, 4.5$\mu m$ images are from the SEDS  v1.2 data release \citep{Ashby13}, while the 5.8 and 8$\mu m$ images are from the S-COSMOS survey \citep{Sanders07}. The central wavelength of the filter, 95\% cumulative throughput width of the filter, dust attenuation from Galactic extinction, aperture used for photometry, image zero point, average FWHM and median 5$\sigma$ depth for each of the images are listed in Table~\ref{table:cosmos_image_data}.

The centers of bright stars are masked in the five optical broadband and twelve medium-band Subaru images, which results in artifacts in these localized regions after registration to the finer WFC/HST pixel scale.  To avoid these artifacts, we flag the centers of these bright stars and grow the area by a factor of ten ($0\farcs6$).  Although no weight maps were publicly released for these datasets, we generate our own maps that mark the pixels flagged during this process.

\begin{table*}[ht]
\centering
\caption{COSMOS Imaging Data}
\label{table:cosmos_image_data}
\begin{tabular}{lccccccc}
\hline \hline
\noalign{\smallskip}
Band & $\lambda_{\mathrm{central}}$ & Width$^a$ & $A_{\lambda}$ & Aperture & Zero point & FWHM & 5$\sigma$ Depth$^b$\\
 & ($\mu m$) &($\mu m$) & (mag) & (arcsec) &(AB) & (arcsec)& (mag) \\
\hline
\noalign{\smallskip}
  U& 0.3828& 0.0771 & 0.076& 1.2& 25.233&0.85 & 26.7\\
  B& 0.4448& 0.1035& 0.068& 1.2& 31.400&0.78& 27.9 \\
  G& 0.4870& 0.1428& 0.062& 1.2& 26.375&0.85& 27.2 \\
  V& 0.5470& 0.0993& 0.051& 1.2& 31.400&0.90& 27.2 \\
F606W& 0.5921& 0.2225 &  0.019& 0.7& 26.491&0.11& 26.7 \\
  R& 0.6245& 0.1232& 0.044& 1.2 & 25.926&0.78& 27.2 \\
 Rp& 0.6276& 0.1379 & 0.044& 1.2& 31.400&0.84& 27.3 \\
  I& 0.7676& 0.1501 & 0.032& 1.2& 25.703&0.81& 27.0 \\
 Ip& 0.7671& 0.1489 & 0.032& 1.2& 31.40 &0.74& 27.0 \\
F814W& 0.8057& 0.2358 &  0.029& 0.7& 25.943&0.10& 26.5 \\
  Z& 0.8872& 0.1719 &  0.025&1.2 & 24.768&0.75& 26.0 \\
 Zp& 0.9028& 0.1411 &  0.025&1.2 & 31.400&0.87& 25.8 \\
UVISTA\_Y& 1.0217 & 0.1026 & 0.020& 1.2& 30.000&0.79& 25.7 \\
 J1& 1.0460& 0.1471 & 0.019& 1.5& 23.310&1.19& 24.7 \\
 J2& 1.1946&  0.1476 &0.016& 1.5& 23.350&1.17& 24.5\\
 J3& 1.2778& 0.1394 & 0.014& 1.5& 23.370&1.12& 24.4\\
F125W& 1.2471& 0.2867 &  0.014& 0.7& 26.247&0.19&26.1 \\
  J& 1.2530& 0.1541 & 0.014& 1.2& 30.000&0.93& 23.8 \\
UVISTA\_J& 1.2527& 0.1703 & 0.017& 1.2&  30.000&0.78& 25.4 \\
F140W& 1.3924& 0.3760 & 0.012& 0.7& 26.465&0.19& 25.5 \\
F160W& 1.5396& 0.2744 & 0.010& 0.7& 25.956&0.19& 25.8\\
  H1& 1.5601&0.1658  & 0.012& 1.5& 23.590&1.03& 23.7 \\
  H2& 1.7064& 0.1721&  0.010& 1.5& 23.610&1.24& 23.6 \\
  H& 1.6294& 0.2766& 0.009& 1.2& 30.000&0.73& 24.0 \\
UVISTA\_H& 1.6433& 0.2844 & 0.009& 1.2& 30.000&0.76& 24.9 \\
  K& 2.1684&0.3181 & 0.005& 1.5& 23.850&1.08& 23.7 \\
 Ks& 2.1574& 0.3151& 0.005& 1.2& 30.000&0.68& 23.8 \\
UVISTA\_Ks& 2.1503 & 0.3109 & 0.005& 1.2& 30.000&0.74& 25.0 \\
IA427& 0.4260& 0.0223 & 0.069& 1.2& 31.400&0.81& 26.3 \\
IA464& 0.4633& 0.0238 & 0.063& 1.2& 31.400&0.91& 25.8 \\
IA484& 0.4847& 0.0250& 0.060& 1.2& 31.400&0.60& 26.4 \\
IA505& 0.5061& 0.0259& 0.058& 1.2& 31.400&0.83& 26.2 \\
IA527& 0.5259& 0.0282& 0.056& 1.2& 31.400&0.71& 26.5 \\
IA574& 0.5763& 0.0303 & 0.051& 1.5& 31.400&1.08& 25.7 \\
IA624& 0.6231& 0.0337 & 0.042& 1.2& 31.400&0.84& 26.4 \\
IA679& 0.6782&0.0372 & 0.038& 1.5& 31.400&1.12& 25.9 \\
IA709& 0.7074& 0.0358 & 0.040& 1.5& 31.400&1.05& 26.1 \\
IA738& 0.7359& 0.0355& 0.036& 1.2& 31.400&0.85& 26.0\\
IA767& 0.7680& 0.0389 & 0.032& 1.5& 31.400&1.09& 25.7 \\
IA827& 0.8247& 0.0367 & 0.028& 1.5& 31.400&1.07& 25.6\\
IRAC1& 3.5569& 0.7139 & 0.000& 3.0& 21.581& 1.7& 25.1\\
IRAC2& 4.5020& 0.9706 & 0.000& 3.0& 21.581& 1.7& 25.0\\
IRAC3& 5.7450& 1.3591& 0.000& 3.0& 21.581&  1.9& 21.6\\
IRAC4& 7.9158& 2.7893 & 0.000& 3.0& 21.581&  2.0& 21.6\\
\noalign{\smallskip}
\hline
\noalign{\smallskip}
\multicolumn{8}{l}{$^a$ 95\% cumulative throughput width} \\
\multicolumn{8}{l}{$^b$ Median 5$\sigma$ depth calculated from the errors of objects in the final catalogs} 
\end{tabular}
\end{table*}

\subsubsection{GOODS-North}
The GOODS-North field is a large field with GOODS ACS imaging centered on the Hubble Deep Field North \citep{Dickinson03}. We use images in 22 filters for the GOODS-North catalog, many of which are provided by the GOODS team. The ACS F435W, F606W, F775W, and F850LP mosaics are from the v2.0 data release (HST Cycle 11, program IDs 9425 and 9583; \citealt{Giavalisco04}). The $U$-band image was taken with the Mosaic camera on the Kitt Peak 4-m telescope by the Hawaii Hubble Deep Field North project \citep{Capak04}.\footnote{\url{http://www.astro.caltech.edu/~capak/hdf/index.html}} Broad-band optical data in the $BVRiz$ filters from Suprime-Cam on the Subaru 8.2-m is also provided by the Hawaii Hubble Deep Field North project. Optical $G$ and $R_s$ band images from LRIS on the Keck I telescope are provided by \citet{Steidel04, Reddy05}. We use the HST/WFC3 F140W images from GO: 11600 (PI: B.\ Weiner) and the F125W and F160W images from CANDELS to produce our own mosaics, as described in Section~\ref{sec:wfc3im}. Ground-based $J$, $H$, and $K_s$-band images from the Multi-Object Infrared Camera and Spectrograph (MOIRCS) on Subaru are provided by the MOIRCS Deep Survey (MODS, \citealt{Kajisawa11}).\footnote{\url{http://www.astr.tohoku.ac.jp/MODS}} We use the "convolved" mosaics provided by the MODS team, in which the four subfields making up each mosaic have been PSF-matched to the field with the worst seeing.  The IRAC 3.6, 4.5$\mu m$ images are from the SEDS v1.2 data release \citep{Ashby13}, while the 5.8 and 8$\mu m$ images are from the GOODS Spitzer $2^{nd}$ data release. There are two epochs of data for the 5.8 and 8$\mu m$ bands.  We measure fluxes independently on the two epochs and where there is overlap, combine them for the final catalog, rather than coadding the images beforehand, due to the differences in the orientation of the PSFs. The central wavelength of the filter, 95\% cumulative throughput width of the filter, dust attenuation from Galactic extinction, aperture used for photometry, image zero point, average FWHM and median 5$\sigma$ depth for each of the images are listed in Table~\ref{table:gn_image_data}.

\begin{table*}[ht]
\centering
\caption{GOODS-N Imaging Data}\label{table:gn_image_data}
\begin{tabular}{lccccccc}
\hline \hline
\noalign{\smallskip}
Band & $\lambda_{\mathrm{central}}$ & Width$^a$ & $A_{\lambda}$ & Aperture & Zero point & FWHM&  5$\sigma$ Depth$^b$\\
 & ($\mu m$) &($\mu m$) & (mag) & (arcsec) &(AB) & (arcsec)& (mag)\\
\hline
\noalign{\smallskip}
U& 0.3593 & 0.0721 &  0.052& 1.5& 31.369 & 1.26&26.4 \\
F435W& 0.4318 & 0.0993 & 0.044& 0.7& 25.689  & 0.10& 27.1\\
HDF.B& 0.4448 & 0.1035 & 0.042& 1.0& 31.136 & 0.71 &  26.7\\
G& 0.4751 & 0.0940 & 0.039& 1.2& 35.250 & 1.07&  26.3\\
HDF.V0201& 0.5470 & 0.0993 & 0.033& 1.0& 34.707 & 0.71& 27.0\\
F606W& 0.5919 & 0.2225 &  0.030& 0.7& 26.511 & 0.10& 27.4\\
HDF.R& 0.6276 & 0.1379 & 0.027& 1.5& 34.676 & 1.11& 26.2\\
Rs& 0.6819 & 0.1461 & 0.023& 1.2& 35.250 & 1.02&25.6 \\
HDF.I& 0.7671 &  0.1489 & 0.020& 1.0& 34.481 & 0.72 & 25.8\\
F775W& 0.7693 & 0.1491  & 0.020& 0.7& 25.671 & 0.11&26.9\\
HDF.Z& 0.9028 & 0.1411 &  0.015& 1.0& 33.946 & 0.67& 25.5\\
F850LP& 0.9036 &  0.2092 & 0.015& 0.7& 24.871 & 0.11& 26.7\\
F125W& 1.2471 & 0.2867 & 0.009& 0.7& 26.230 & 0.18& 26.7\\
J& 1.2517 &  0.1571 & 0.009& 1.0& 26.000 & 0.60&  25.0\\
F140W& 1.3924 & 0.3760 & 0.007& 0.7& 26.452 & 0.18& 25.9\\
F160W& 1.5396 &  0.2744 & 0.006& 0.7& 25.946 & 0.19& 26.1\\
H& 1.6347 & 0.2686 & 0.005 & 1.0& 26.000 & 0.60&  24.3\\
Ks& 2.1577 & 0.3044 & 0.004& 1.0& 26.000 & 0.60& 24.7\\
IRAC1& 3.5569 & 0.7139 & 0.000& 3.0& 21.581 & 1.7 & 24.5\\  
IRAC2& 4.5020& 0.9706 & 0.000& 3.0& 21.581 & 1.7& 24.6\\ 
IRAC3& 5.7450& 1.3591& 0.000& 3.0& 20.603 & 1.9 & 22.8\\ 
IRAC4& 7.9158& 2.7893 & 0.000& 3.0& 21.781 & 2.0 & 22.7\\ 
\noalign{\smallskip}
\hline
\noalign{\smallskip}
\multicolumn{8}{l}{$^a$ 95\% cumulative throughput width}\\
\multicolumn{8}{l}{$^b$ Median 5$\sigma$ depth calculated from the errors of objects in the final catalogs} 
\end{tabular}
\end{table*}

\subsubsection{GOODS-South}

The 3D-HST GOODS-South field is centered on the Chandra Deep Field South and contains the WFC3 Early Release Science (ERS) field and the Hubble Ultra Deep Field (HUDF) within its boundaries. In the GOODS-South field we make use of images in 26 broad bands and 14 medium bands. As for GOODS-N, much of the data is publicly available from the GOODS team. Here we make use of the ACS v2.0 data as described above. In addition, we include the publicly-released v1.0 CANDELS ACS/F606W, ACS/F814W and ACS/F850LP mosaics.\footnote{\url{http://candels.ucolick.org/data_access/GOODS-S.html}} $U$ and $R$-band images from VIMOS on the Very Large Telescope (VLT) are from the European Southern Observatory (ESO) v1.0 data release \citep{Nonino09}. We use the ESO WFI U38, B, V, Rc and I images reduced by the Garching-Bonn Deep Survey (GaBoDS) consortium \citep{Hildebrandt06, Erben05}. We incorporate 14 Subaru medium bands from the Multiwavelength Survey by Yale-Chile (MUSYC, \citealt{Cardamone10, Gawiser06}).\footnote{\url{http://www.astro.yale.edu/MUSYC/}} Four medium bands (three of which have FWHM $>1\farcs5$) were excluded from the analysis due to large zero point uncertainties. We use our own reductions of the CANDELS data in the WFC3/F125W and WFC3/F160W bands and the 3D-HST F140W image, as described in Section~\ref{sec:wfc3im}. We use the $J$, $H$ and $K_s$-bands mosaics from the FIREWORKS survey, kindly provided by S. Wuyts. The mosaics were constructed by convolving the individual ESO GOODS survey images from ISAAC on the VLT (v1.5 data release, \citealt{Retzlaff10}) to a uniform PSF (see \citealt{Wuyts08} for details).  We also include deep $J$ and $K_s$-band images from the Taiwan Extended Chandra Deep Field South Near-Infrared Survey (TENIS, \citealt{Hsieh12}). We use IRAC 3.6, 4.5$\mu m$ data from SEDS v1.2 data release \citep{Ashby13} and 5.8 and 8$\mu m$ images from the GOODS Spitzer $3^\mathrm{rd}$ data release. As with GOODS-N, there are two epochs of data for both the 5.8 and 8$\mu m$ bands, which we treat separately, combining the two flux measurements for objects in the overlapping region for the final catalog. The central wavelength of the filter, 95\% cumulative throughput width of the filter, dust attenuation from Galactic extinction, aperture used for photometry, image zero point, average FWHM and median 5$\sigma$ depth for each of the images are listed in Table~\ref{table:gs_image_data}.

\begin{table*}[th]
\centering
\caption{GOODS-S Imaging Data}\label{table:gs_image_data}
\begin{tabular}{lccccccc}
\hline \hline
\noalign{\smallskip}
Band & $\lambda_{\mathrm{central}}$ & Width$^a$ & $A_{\lambda}$ & Aperture & Zero point & FWHM& 5$\sigma$ Depth$^b$\\
 & ($\mu m$) &($\mu m$) & (mag) & (arcsec) &(AB) & (arcsec)& (mag)\\
\hline
\noalign{\smallskip}
U38& 0.3637& 0.0475 &  0.033& 1.2& 21.910&0.98& 25.7\\
  B& 0.4563&  0.0975 & 0.026& 1.2& 24.379&1.01& 26.9\\
  V& 0.5396& 0.0920 &0.021& 1.2& 24.096&0.94& 26.6\\
 Rc& 0.6517& 0.1600 &0.016& 1.2& 24.651&0.83& 26.6\\
  I& 0.7838& 0.2459 &0.012& 1.2& 23.640&0.96& 24.7\\
  U& 0.3750& 0.0591 &0.032& 1.2& 26.150& 0.8 & 27.9\\
  R& 0.6443& 0.1333 &0.017& 1.2& 27.490& 0.7 & 27.5\\
F435W& 0.4318&  0.0993&0.028& 0.7& 25.690& 0.11 & 27.3\\
F606Wcand& 0.5921& 0.2225 &0.019& 0.7& 26.493& 0.11 & 27.2\\
F606W& 0.5919& 0.2225 &0.019& 0.7& 26.511& 0.11 & 27.4\\
F775W& 0.7693& 0.1491 & 0.013& 0.7& 25.671& 0.10 & 26.9\\
F814W& 0.8057& 0.2358 & 0.012& 0.7& 25.947& 0.10 & 27.2\\
F850LPcand& 0.9033& 0.2092 &0.010& 0.7& 24.857& 0.11& 25.5\\
F850LP& 0.9036& 0.2092 &0.010& 0.7& 24.871& 0.11 & 26.5\\
  J& 1.2356& 0.2668 &0.005& 1.0& 26.000& 0.65 & 25.1\\
TENIS J& 1.2530& 0.1540 &0.005& 1.0& 23.900& 0.93& 25.0\\
F125W& 1.2471& 0.2867 & 0.005& 0.7& 26.230& 0.18& 26.1\\
F140W& 1.3924&0.3760 & 0.004& 0.7& 26.452& 0.18& 25.6\\
F160W& 1.5396&  0.2744 & 0.003& 0.7& 25.946& 0.19& 26.4\\
  H & 1.6496& 0.2832 &0.003& 1.0& 26.000 & 0.65 & 24.5\\
TENIS Ks& 2.1574& 0.3151 & 0.002& 1.0& 23.900 & 0.83& 24.5\\
 Ks & 2.1667& 0.2686 &0.002& 1.0& 26.000 & 0.65 & 24.4\\
IA427& 0.4260& 0.0223 &0.028& 1.5& 25.100&1.01& 25.4\\
IA445& 0.4443& 0.0219 &0.027& 1.5& 25.070&1.23& 25.7\\
IA505& 0.5061&  0.0259 &0.023& 1.2& 25.340&0.94 & 25.7\\
IA527& 0.5259& 0.0282 & 0.022& 1.2& 25.720&0.83 & 26.4\\
IA550& 0.5495&  0.0305 &0.021& 1.5& 25.880&1.13& 25.9\\
IA574& 0.5763& 0.0303 &0.020& 1.2& 25.710&0.95& 25.5\\
IA598& 0.6007& 0.0331 &0.019& 1.2& 26.020&0.63& 26.5\\
IA624& 0.6231&  0.0337 &0.018& 1.2& 25.890&0.61& 26.4\\
IA651& 0.6498&  0.0360 &0.017& 1.2& 26.150&0.6& 26.6\\
IA679& 0.6782& 0.0372 &0.015& 1.2& 26.200&0.8& 26.4\\
IA738& 0.7359& 0.0355 &0.013& 1.2& 26.020&0.77& 26.2\\
IA767& 0.7680& 0.0389 & 0.013& 1.2& 26.040&0.7& 25.1\\
IA797& 0.7966& 0.0404 & 0.012& 1.2& 26.020&0.68& 25.0\\
IA856& 0.8565& 0.0379 &0.011& 1.2& 25.730&0.67& 24.6\\
IRAC1& 3.5569& 0.7139 & 0.000& 3.0& 21.581& 1.7  & 24.8\\ 
IRAC2& 4.5020& 0.9706 & 0.000& 3.0& 21.581&  1.7& 24.8\\ 
IRAC3& 5.7450& 1.3591& 0.000& 3.0& 20.603& 1.9& 23.0\\ 
IRAC4& 7.9158& 2.7893 & 0.000& 3.0& 21.781& 2.0& 23.0\\ 
\noalign{\smallskip}
\hline
\noalign{\smallskip}
\multicolumn{8}{l}{$^a$ 95\% cumulative throughput width} \\
\multicolumn{8}{l}{$^b$ Median 5$\sigma$ depth calculated from the errors of objects in the final catalogs} 
\end{tabular}
\end{table*}

\subsubsection{UDS}

The 3D-HST UDS field covers part of the region observed by the UKIDSS UDS. The catalog incorporates photometry in 18 filters. In addition to the WFC3/F140W data from 3D-HST, we use our own reductions of the CANDELS WFC3/F125W and WFC3/F160W data and the ACS/F606W and ACS/F814W data from the CANDELS v1.0 data release.\footnote{\url{http://candels.ucolick.org/data_access/UDS.html}} This is supplemented by optical ground-based data from the Subaru/XMM-Newton Deep Survey (SXDS, \citealt{Furusawa08}) in the $B$, $V$, $R$, $i$ and $z$-bands.\footnote{\url{http://www.naoj.org/Science/SubaruProject/SXDS/}} We use the mosaics provided by M. Cirasuolo \citep{Cirasuolo10}.\footnote{\url{http://www.roe.ac.uk/~ciras/Scientific_Research.html}} We include a $u'$-band image reduced from CFHT archival data by R. Williams and R. Quadri (private communication). We have used the UKIDSS UDS NIR imaging data from the 8$^\mathrm{th}$ data release.\footnote{\url{http://surveys.roe.ac.uk/wsa/dr8plus_release.html}}  The UKIDSS project is defined in \citet{Lawrence07}. Further details on the UDS can be found in Almaini et al. (in prep). UKIDSS uses the UKIRT Wide Field Camera (WFCAM; \citealt{Casali07}. The photometric system is described in \citet{Hewett06}, and the calibration is described in \citet{Hodgkin09}. The pipeline processing and science archive are described in Irwin et al. (in prep.) and \citet{Hambly08}. We include the IRAC data from the Spitzer Public Legacy Survey of the UKIDSS UDS (SpUDS, PI: J.\ Dunlop) in the 5.8 and 8$\mu m$ pass bands and the SEDS v1.2 data release for the 3.6 and 4.5$\mu m$ pass bands \citep{Ashby13}. The central wavelength of the filter, 95\% cumulative throughput width of the filter, dust attenuation from Galactic extinction, aperture used for photometry, image zero point, average FWHM and median 5$\sigma$ depth for each of the images are listed inTable~\ref{table:uds_image_data}.

\begin{table*}[ht]
\centering
\caption{UDS Imaging Data}\label{table:uds_image_data}
\begin{tabular}{lccccccc}
\hline \hline
\noalign{\smallskip}
Band & $\lambda_{\mathrm{central}}$ & Width$^a$ & $A_{\lambda}$ & Aperture & Zero point & FWHM& 5$\sigma$ Depth$^b$\\
 & ($\mu m$) &($\mu m$) & (mag) & (arcsec) &(AB) & (arcsec)& (mag)\\
\hline
\noalign{\smallskip}
u& 0.3828& 0.0771 & 0.092& 1.5& 25.350& 0.90 & 26.4\\
B& 0.4408& 0.1084 & 0.081& 1.2& 34.706& 0.78& 27.4\\
V& 0.5470& 0.0993 & 0.061& 1.2& 33.602& 0.83& 27.2\\
F606W& 0.5921& 0.2225 &0.056& 0.7& 26.491& 0.11&  26.8\\
R& 0.6508& 0.1194 & 0.048& 1.2& 34.260& 0.79& 26.9\\
i& 0.7655&0.1524 & 0.037& 1.2& 34.055& 0.81&26.7\\
F814W& 0.8057& 0.2358 & 0.034& 0.7& 25.943& 0.11& 26.8\\
z& 0.9060& 0.1402 & 0.028& 1.2& 32.743& 0.80& 25.9\\
F125W& 1.2471& 0.2867 & 0.016& 0.7& 26.230& 0.18 &25.8\\
J& 1.2502& 0.1599 & 0.016& 1.0& 30.931& 0.73& 25.1\\
F140W& 1.3924& 0.3760 &0.013& 0.7& 26.452& 0.18& 25.2\\
H& 1.6360&0.2972 &  0.010& 1.0& 31.379& 0.76 & 24.3\\
F160W& 1.5396& 0.2744 &  0.011& 0.7& 26.452& 0.19 & 25.9\\
K& 2.2060& 0.3581 & 0.007& 1.0& 31.893& 0.70 & 24.9\\
IRAC1& 3.5569& 0.7139 & 0.000& 3.0& 21.581 & 1.7  &24.6\\ 
IRAC2& 4.5020& 0.9706 & 0.000& 3.0& 21.581 & 1.7&24.4 \\ 
IRAC3& 5.7450& 1.3591& 0.000& 3.0& 21.581 &  1.9&21.7\\ 
 IRAC4& 7.9158& 2.7893 & 0.000& 3.0& 21.581 & 2.0 & 21.5\\ 

\noalign{\smallskip}
\hline
\noalign{\smallskip}
\multicolumn{8}{l}{$^a$ 95\% cumulative throughput width}\\
\multicolumn{8}{l}{$^b$ Median 5$\sigma$ depth calculated from the errors of objects in the final catalogs} 
\end{tabular}
\end{table*}

\section{Photometry}
\label{sec:phot}

We construct a WFC3-selected photometric catalog for each field as detailed below. This is done in the same manner and consistently across all five fields. Briefly, we use a noise-equalized combination of the three WFC3 bands (F125W, F140W and F160W)  for detection. We convolve each of the \textit{HST} images to the same point-spread function (PSF) in order to measure consistent colors across multiple passbands. Aperture photometry was performed on the PSF-matched images in an aperture of $0\farcs7$ using SExtractor v2.8.6 \citep{Bertin96} in dual-image mode. For the ground-based and Spitzer data we used the photometry code Multi-resolution Object PHotometry oN Galaxy Observations (MOPHONGO) described in \citet{Labbe06, Wuyts07, Whitaker11} to take into account the large difference in PSF size and confusion due to neighboring sources. A combination of the PSF-matched WFC3 images is used as a high resolution prior. We measure the aperture flux in each band, with an aperture size that depends on the PSF full-width at half maximum (FWHM) for each filter, and make a correction to total flux based on the F160W growth curve. The methods we use are described below; they are similar to those described in \citet{Whitaker11}. 

\subsection{Background Subtraction}
We determine the background level as a function of position across each image. For the \textit{HST} images, we found that an initial run of SExtractor with the AUTO background subtraction using a background mesh size of 64 and filter size of 5 produced a reasonably flat background-subtracted image. We note that careful background subtraction for individual WFC3 pointings was already done as part of the reduction process described above; this additional step serves to remove any large scale gradients over the entire mosaic. For the ground-based and IRAC data, we applied a custom-developed code that uses a similar algorithm to mask sources and fit the background level in a variable-sized mesh, using sigma clipping to reduce the impact of bright sources.  We verified that the flux distribution in empty regions of the image was centered on zero after the background subtraction.

\subsection{Source Detection}
\label{sec:detection}

For each field we create a noise-equalized version of the mosaic in each of the three WFC3 bands F125W, F140W and F160W by multiplying the science image by the square root of the inverse variance map. The three noise-equalized images are then coadded to form a deep detection image. As the variable weight across the mosaic is already taken into account with this method, we do not input a weight map to SExtractor when detecting sources on these images. We use a detection and analysis threshold of 1.8$\sigma$ and require a minimum area of 14 pixels for detection. The deblending threshold is set to 32, with a minimum contrast parameter of 0.005 for all fields except GOODS-S, where we use a minimum contrast parameter of 0.0001 to improve the detection of sources within the wings of bright objects, particularly affecting the deep HUDF area. A Gaussian filter of 4 pixels is used to smooth the images before detection. The detection parameters were chosen as a compromise between deblending neighboring galaxies and splitting large objects into multiple components. An alternative approach would be to run SExtractor in a ``hot'' and ``cold'' mode, as was been done by the GEMS survey \citep{Rix04}, and more recently, CANDELS \citep{Guo13, Galametz13}. 

\subsection{HST PSF-Matching}
\label{sec:psfmatching}

\begin{figure*}[ht]
\centering
\begin{minipage}[b]{0.48\linewidth}{
\includegraphics[width=\textwidth]{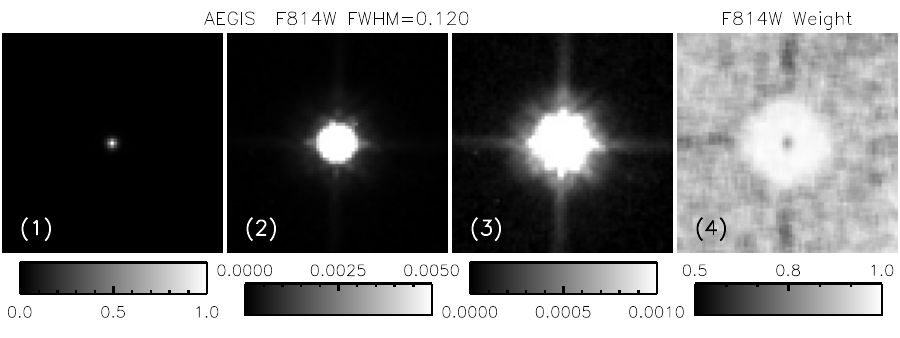}
\label{fig:subfig1}}
\end{minipage}
\begin{minipage}[b]{0.48\linewidth}{
\includegraphics[width=\textwidth]{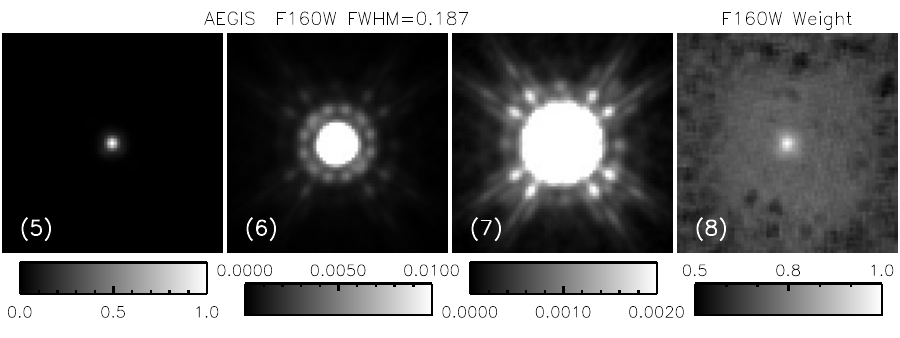}
\label{fig:subfig2}}
\end{minipage}
\caption{Point-spread functions (PSFs) for the ACS/F814W band and the WFC3/F160W band in the AEGIS field. The construction of the PSFs is described in \S\ref{sec:psfmatching}. For each filter we show three stretch levels (panels 1 to 3 and 5 to 7) to expose the structure of the PSF: the core, the first Airy ring and the diffraction spikes. The images are normalized to a maximum value of one. The grayscale bars show the stretch for each panel. These are slightly different for ACS and WFC3 as a result of the different FWHMs (listed above the images). We also show the combined weight images for each PSF. The weight is largest in the center and lower at larger radii due to masking of neighboring objects. The ACS PSFs have lower weights in the central pixels because of cosmic ray rejection flagging the centers of stars. PSFs for all other bands and fields are shown in the Appendix.}\label{fig:psfs}
\end{figure*}

We PSF-match all the \textit{HST} ACS and WFC3 images to the F160W image, which has the PSF with the largest FWHM, before performing aperture photometry. An empirical PSF was created for each \textit{HST} image by stacking isolated unsaturated stars from across the mosaic.  As discussed in \S\ref{sec:wfc3reduction}, we chose drizzle parameters that avoid clipping the centers of bright stars. 

To create PSFs for each of the \textit{HST} mosaics we made a careful selection of stars based on the tight stellar sequence in the ratio of the flux in a small aperture ($0\farcs5$) to that in a large aperture ($2^{\prime\prime}$), adjusting the selection criteria appropriately for each of the ACS and WFC3 bands. The number of stars selected varied between 35 for the F606W band in GOODS-S, which has limited coverage, to $\sim$200 in F814W in COSMOS.  We mask neighboring objects within a postage stamp cut-out of 84 pixels ($\sim5^{\prime\prime}$) around each star. The postage stamps are recentered, normalized and then averaged to determine the PSF. Stamps where half-integer or large shifts are required to recenter the star are excluded from the final stack. Finally, we subtract a background correction based on the level measured in the outskirts ($r > 67$ pixels, corresponding to $\sim4^{\prime\prime}$) of the stacked PSF stamp. We have not attempted to take variation with chip position into account, as the mosaics are made up of multiple pointings with different orientations and overlap, but such differences are likely to be small.

The PSF stamps for the ACS F814W band and the WFC3 F160W band in AEGIS are shown in Fig.~\ref{fig:psfs}. PSFs for all the other fields and bands are shown in Appendix~\ref{app:psfs}.  The PSFs in the Figure are shown at three contrast levels for each of the ACS and WFC3 filters to expose the different level of structure: the core of the PSF, the first Airy ring ($\sim0.5\%$) and the diffraction spikes ($\sim0.1\%$). For a single orientation the \textit{HST} PSF has four diffraction spikes, caused by the vanes of the secondary mirror assembly. The large number of diffraction spikes in the PSFs of Fig.~\ref{fig:psfs} (especially in the WFC3 PSFs)  is due to the varying orientations of the data that went into the mosaics. We also show weight images, which encode the sum of the weights that went into each pixel. 

\begin{figure}[!ht]
\includegraphics[width=0.48\textwidth]{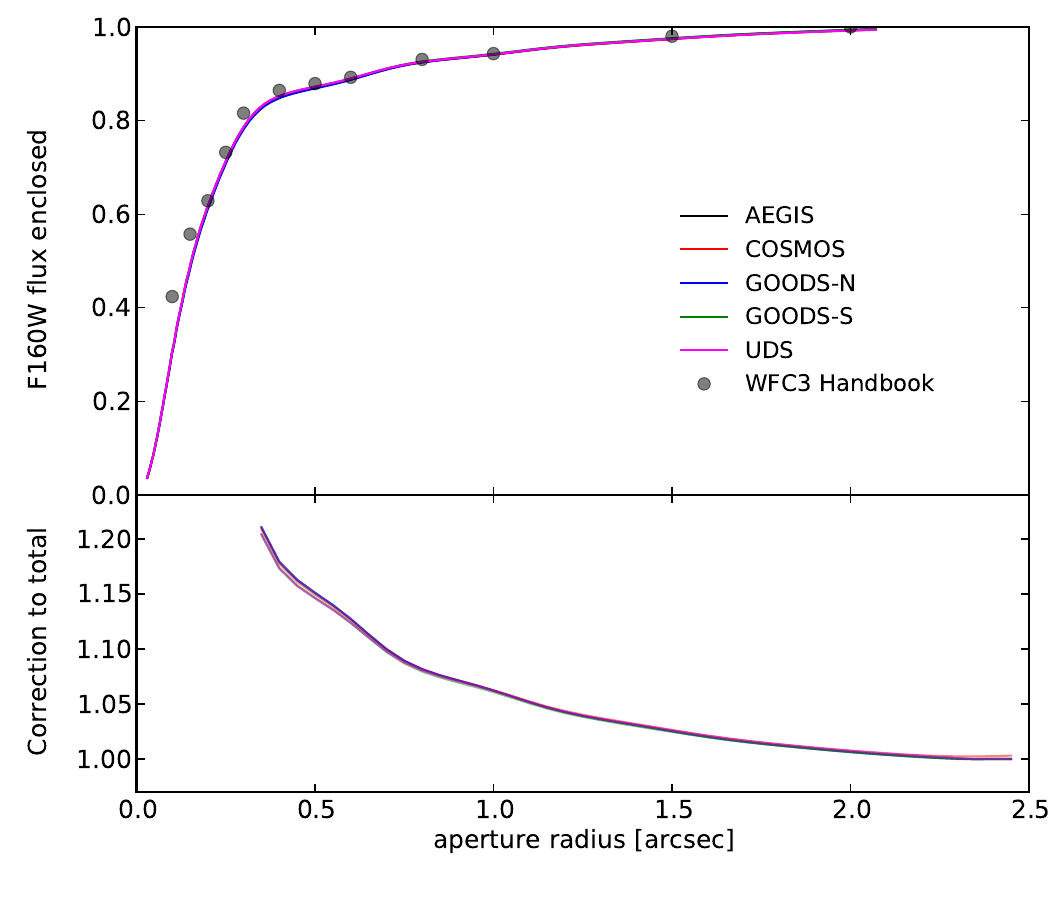}
\caption{F160W growth curves. Upper panel: The fraction of light enclosed as a function of radius relative to the total light within $2^{\prime\prime}$, $f(r)/f(2^{\prime\prime})$, from the F160W PSF stamp of each field. The PSFs of the five fields are very consistent with each other. The grey points show the encircled energy as a function of aperture size, also normalized to $2^{\prime\prime}$, from the WFC3 handbook. The empirical growth curves agree well with the theoretical expectation. Lower panel: The correction to total flux for a point source with a circularized Kron radius equal to the aperture radius on the x-axis, derived as the inverse of the growth curves in the upper panel ($f(2^{\prime\prime})/f(r)$). The minimum Kron radius is set to the aperture radius in which we measure colors, $0\farcs35$, giving rise to a maximum correction of $\sim$1.21.}
\label{fig:f160growth}
\end{figure}

The curve of growth (fraction of light enclosed as a function of aperture size) for each of the F160W PSFs, normalized at $2^{\prime\prime}$, is shown in the top panel of Fig.~\ref{fig:f160growth}. The PSFs for the five fields are very consistent with each other, with almost indistinguishable growth curves on this scale. The agreement with the encircled energy as a function of aperture provided in the WFC3 handbook (normalized to the same maximum radius of $2^{\prime\prime}$) is also excellent. We show a comparison of the growth curves for the PSFs used in this paper to those generated from the CANDELS v1.0 mosaic and the "hybrid" PSF used for the morphological analysis in \citet{vanderWel12} in Appendix~\ref{app:psfs}. 

\begin{figure*}[ht]
\centering
\begin{minipage}[b]{0.48\linewidth}{
\includegraphics[width=\textwidth]{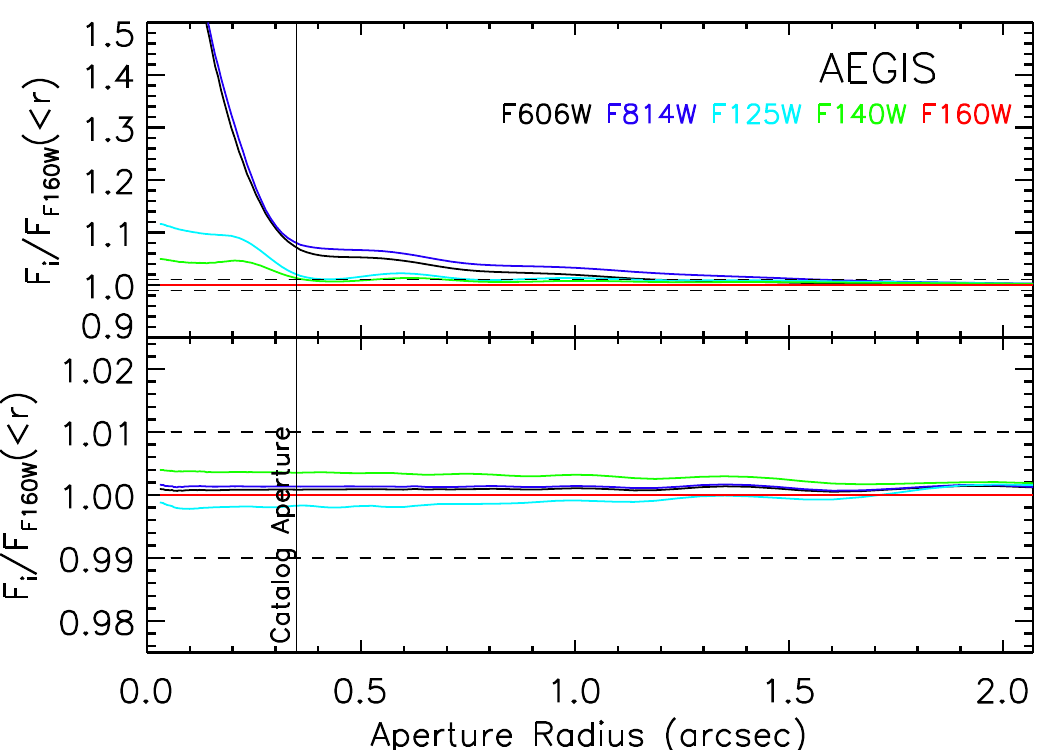}
\label{fig:subfig1}}
\end{minipage}
\begin{minipage}[b]{0.48\linewidth}{
\includegraphics[width=\textwidth]{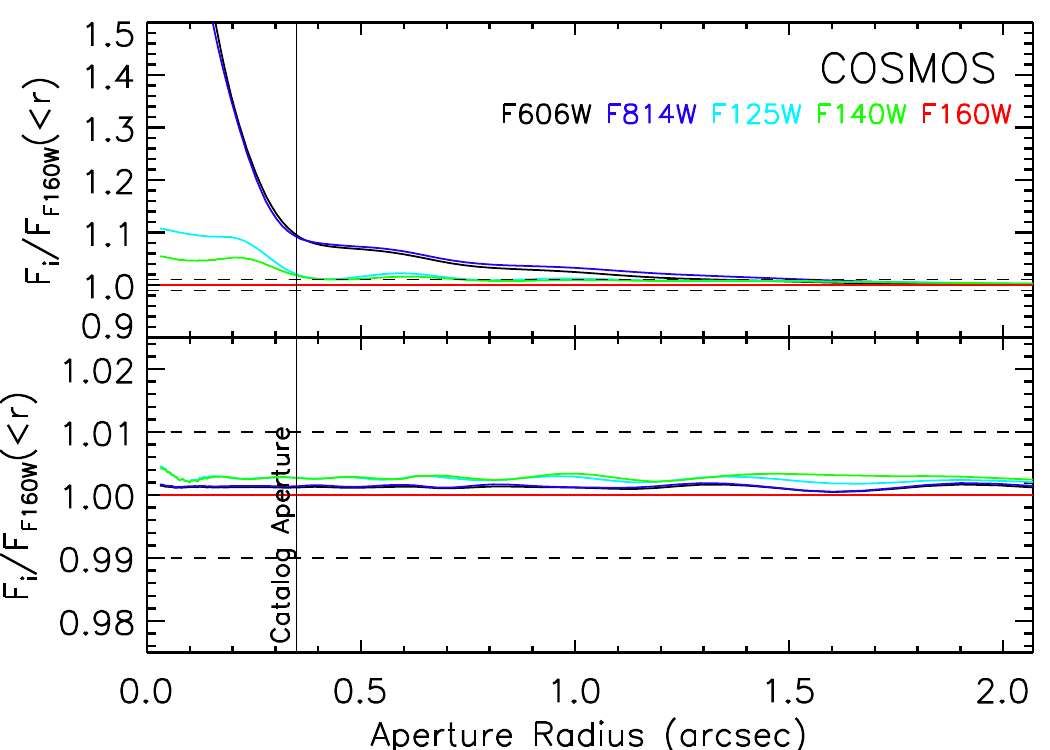}
\label{fig:subfig2}}
\end{minipage}
\begin{minipage}[b]{0.48\linewidth}{
\includegraphics[width=\textwidth]{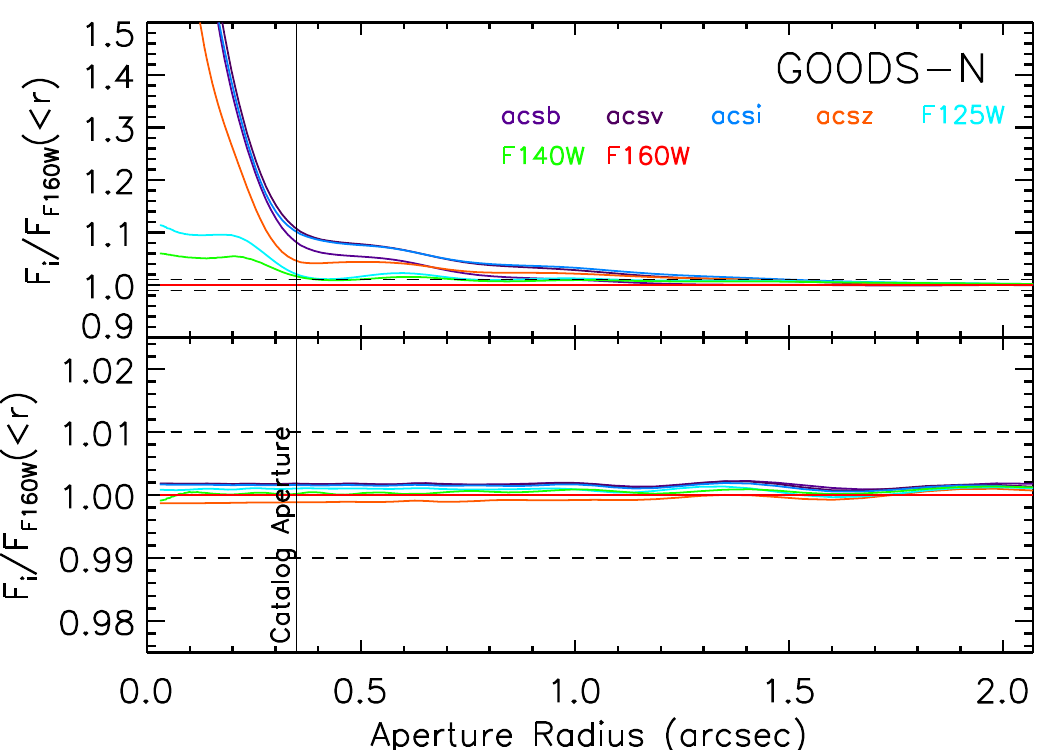}
\label{fig:subfig3}}
\end{minipage}
\begin{minipage}[b]{0.48\linewidth}{
\includegraphics[width=\textwidth]{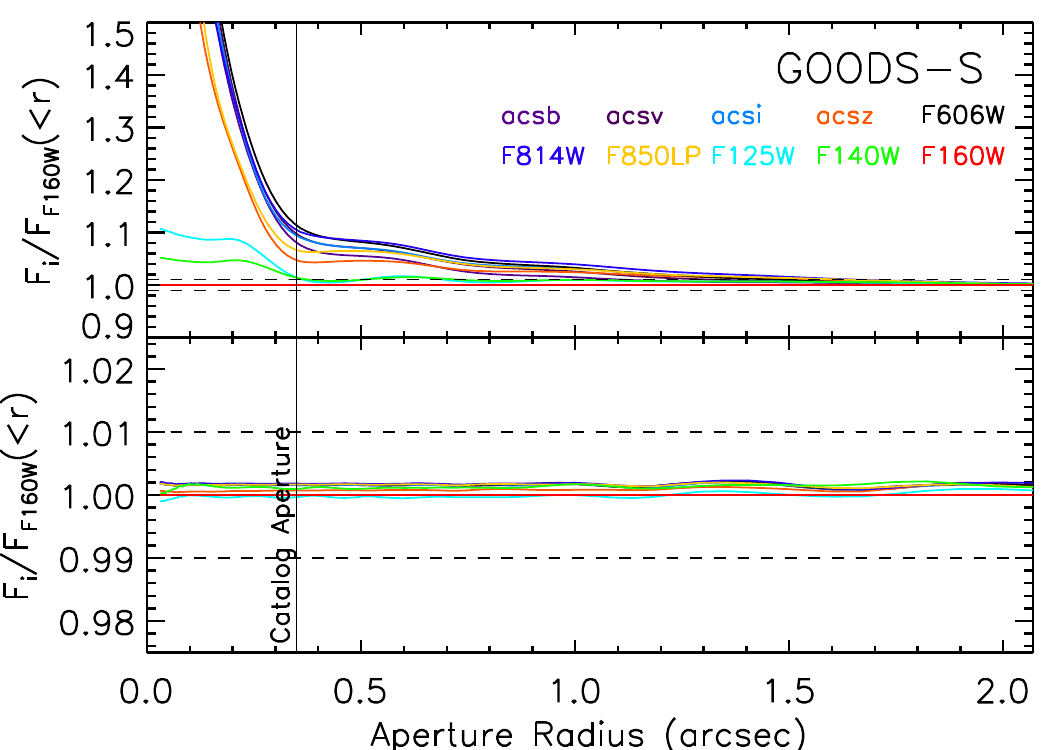}
\label{fig:subfig4}}
\end{minipage}
\begin{minipage}[b]{0.96\linewidth}{
\includegraphics[width=0.5\textwidth]{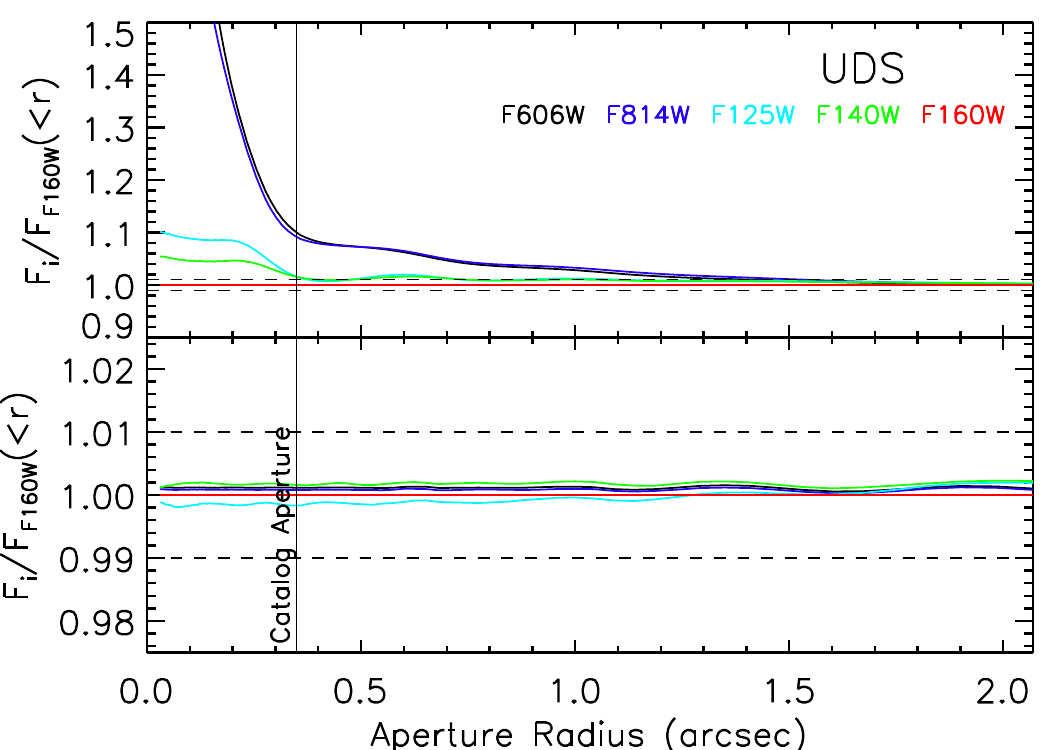}
\label{fig:subfig4}}
\end{minipage}
\caption{Growth curves showing the fraction of light enclosed as a function of radius for each \textit{HST} filter relative to the F160W growth curve in each of the five 3D-HST fields. The upper and lower panels show the growth curves before and after convolution to match the F160W PSF, respectively. Note the change in scale between the upper and the lower panels. The dashed lines in both panels represents a 1\% difference. After PSF-matching, the resulting growth curves in all bands are consistent with the F160W PSF to well within 1\% in each of the fields.}
\label{fig:growthcurves}
\end{figure*}

We use a deconvolution code developed by I. Labb{\'e}, which fits a series of Gaussian-weighted Hermite polynomials to the Fourier transform of the stacked stars, to find the kernel that convolves each PSF to match the F160W PSF. We found residuals a factor of 5 -- 10 times lower using this method than with the standard maximum entropy-type algorithms (e.g. the task \textit{lucy} in iraf). In Fig.~\ref{fig:growthcurves} we show the ratio of the growth curve in each band to that of the F160W growth curve, before and after the convolution. The PSF-matching is accurate to 1\% within a $0\farcs7$ aperture for all \textit{HST} bands and fields.

\subsection{\textit{HST} Photometry}
\label{sec:hstphot}

We ran SExtractor in dual-mode to obtain aperture flux measurements for each \textit{HST} band, using the detection images described above, the PSF-matched \textit{HST} image and the corresponding weight map. We did not do a further background subtraction. Aperture photometry was done in an aperture of diameter $0\farcs7$ in all the \textit{HST} bands. Additional measurements in apertures of 1$^{\prime\prime}$, $1\farcs2$, $1\farcs5$ and 3$^{\prime\prime}$ were done on the F140W and F160W images in order to correct the flux measurements from the ground and Spitzer bands to an equivalent color aperture, as described below. 

The total flux in the reference band, which is chosen to be F160W where there is F160W coverage (99.7\% of objects) and F140W otherwise, is determined by correcting the SExtractor AUTO flux for the approximate amount of light that falls outside the AUTO aperture for a point source. This amount was calculated from the growth curves described in the previous Section.\footnote{As is well known, these corrections slightly underestimate the required correction as galaxies have more extended profiles than point sources outside of the AUTO aperture. For our data these errors are small, as the total GALFIT-derived magnitudes of galaxies correspond very well to our total magnitudes (see \S\,\ref{sec:totalfluxes}).} The AUTO flux is measured within a flexible elliptical aperture, known as the Kron radius \citep{Kron80}, that typically encloses 90-95\% of the total light. The correction is the inverse of the fraction of light within a circular aperture enclosing the same area as the Kron aperture (the circularized Kron radius), determined from the empirical growth curve for F160W. 
\begin{equation}
f_{\mathrm{WFC3,tot}} = f_{\mathrm{WFC3, AUTO}} \frac{f (r_{tot})}{f(r<r_{K})}, \label{eqn:totcor}
\end{equation}
 where $f (r<r_{K})$ is the flux enclosed within the circularized Kron radius and $f (r_{tot})$ is the total flux for a point source in F160W. The growth curve is normalized at $r_{tot} = 2^{\prime\prime}$. We enforce a minimum radius of $0\farcs35$, corresponding to the color aperture used for the \textit{HST} photometry. The AUTO flux for objects with radii smaller than this is taken to be the flux measured in the color aperture. 
  
The errors returned by SExtractor are known to be underestimated due to the correlations between pixels introduced by the drizzling process \citep{Casertano00}. Rather than applying a single correction factor based on the drizzle parameters to all the errors, we determine the error in total magnitude and the errors on the \textit{HST} aperture magnitudes by measuring the flux within empty apertures, as described by \citet{Labbe05, Quadri07, Whitaker11}. In some of the \textit{HST} bands, the depth varies dramatically across the image, so we cannot use an average measurement to represent the error at every point. To take variable depth into account, we measure the empty aperture fluxes on the noise-equalized mosaics. Each pixel in the noise-equalized mosaic is essentially weighted by depth, bringing the noise to a standard level. We divide the resulting empty aperture errors by the square root of the weight at the position of each object for each band to obtain the errors provided in the catalogs.

\begin{figure*}[!ht]
\begin{minipage}[b]{0.48\linewidth}{
\includegraphics[width=\textwidth]{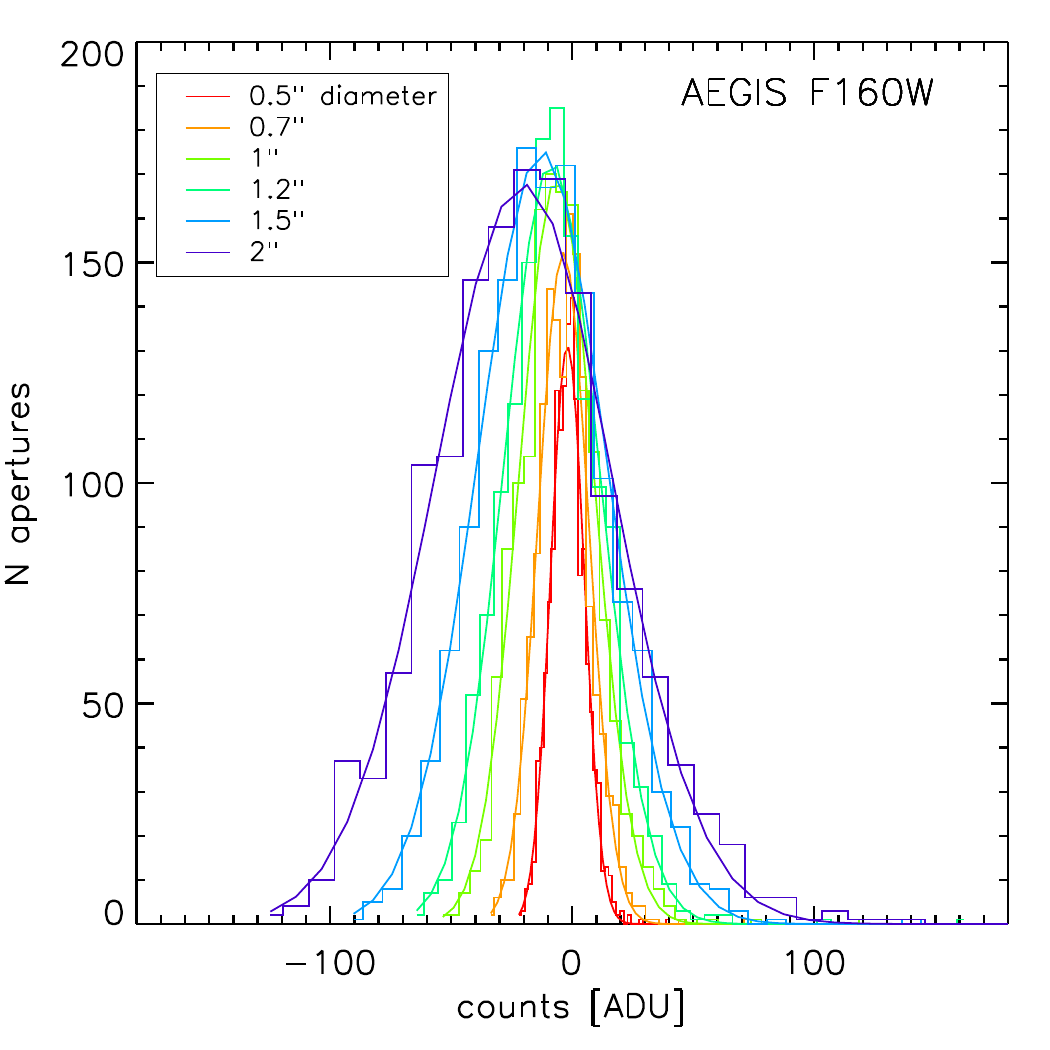}
\label{fig:subfig1}}
\end{minipage}
\begin{minipage}[b]{0.48\linewidth}{
\includegraphics[width=\textwidth]{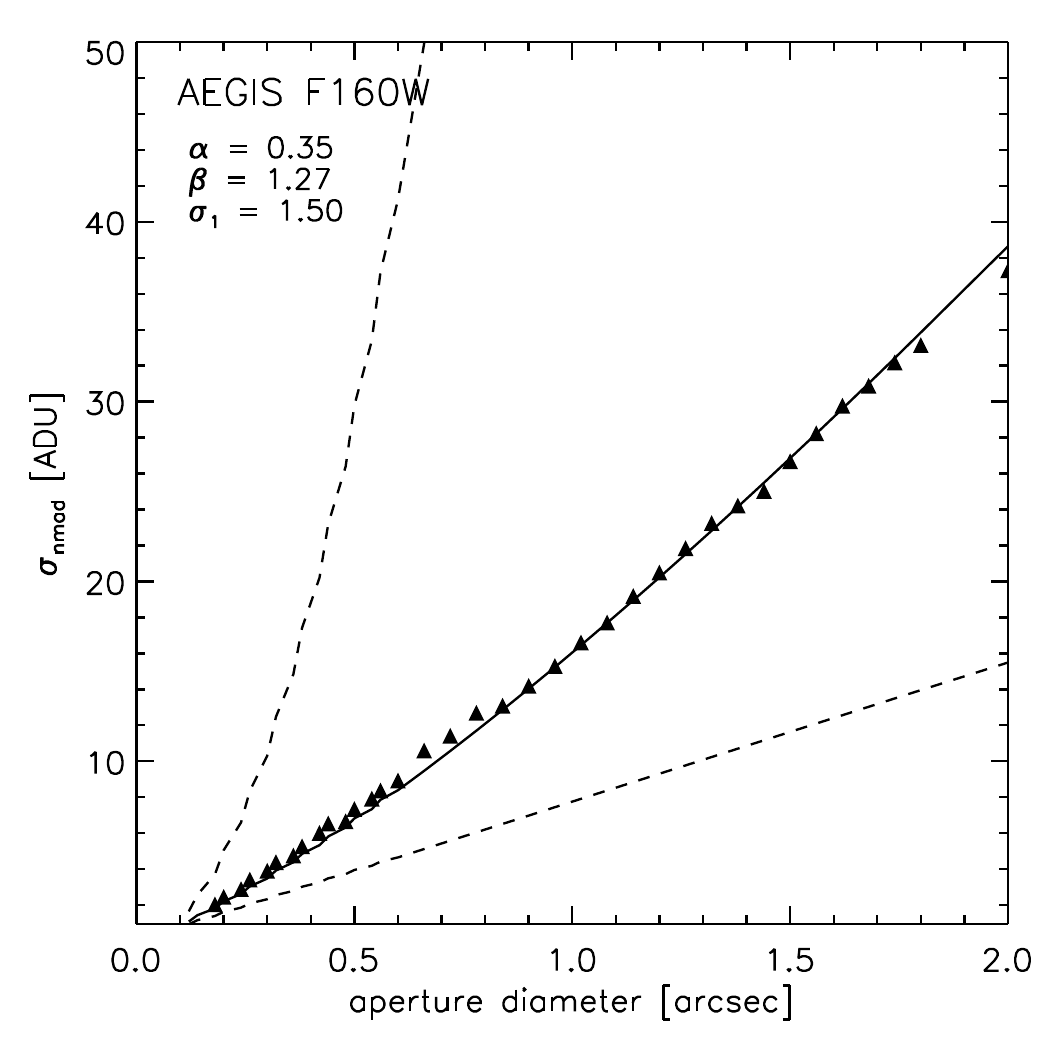}
\label{fig:subfig2}}
\end{minipage}
\caption{Histograms of summed counts in different aperture sizes from empty regions throughout the image (left-hand panel) and the resultant noise-scaling as a function of aperture (right-hand panel) for the AEGIS F160W image. The measured $\sigma_{nmad}$ are shown by the triangles. The solid line shows the power-law fit to the data, with the fit parameters given in the upper left. The dashed lines show the linear ($\propto N$) and $N^ 2$ scalings, which correspond to no correlation and perfect correlation between the pixels, respectively. }
\label{fig:emptyap_errors}
\end{figure*}

To estimate the error on the total magnitudes, we determine the background noise in an aperture the size of the circularized Kron radius for each object, and correct this to total in the same way as for the fluxes, described by Equation~\ref{eqn:totcor} above. To determine how the background noise scales with aperture size, we measure the distribution of counts in empty regions of increasing size within the noise-equalized F160W (F140W) image. For each aperture size we measure the flux in $\gt 2000$ apertures placed at random positions across the image. We exclude apertures that overlap with sources in the detection segmentation map. As an example, the left hand panel of Fig.~\ref{fig:emptyap_errors} shows the distribution of flux counts in empty apertures of 0.5, 0.7, 1, 1.2, 1.5 and 2$^{\prime\prime}$ diameter in the UDS noise-equalized F160W image. The other fields show similar distributions. Each histogram can be well-described by a Gaussian, with the width increasing as aperture size increases. The increase in standard deviation with linear aperture size $N = \sqrt A$, where $A$ is the area within the aperture, can be described as a power-law. A power-law index of 1 would indicate that the noise is uncorrelated, while if the pixels within the aperture were perfectly correlated, the background noise would scale as $N^2$. The right-hand panel of Fig.~\ref{fig:emptyap_errors} shows the measured standard deviation as a function of aperture size in the UDS noise-equalized F160W image. We fit a power-law of the form
\begin{equation}
\sigma = \sigma_{1} \alpha N^{\beta},
\end{equation}
where $\sigma_1$ is the standard deviation of the background pixels, fixed to a value of 1.5 here, $\alpha$ is the normalization and $ 1< \beta < 2$ \citep[see][]{Whitaker11}. The fitted parameters for each field are listed in Table~\ref{table:noiseprop}. The power-law fit is shown by the solid line in the figure. The bracketing $\beta = 1$ and $\beta = 2$ scalings are shown by the dashed lines. The final error in the catalog is divided by the square root of the weight at the position of the object, as described above. 
Errors in the aperture fluxes are similarly calculated by measuring the standard deviation of the flux 
distribution in apertures of $0\farcs7$ diameter on the noise-equalized \textit{HST} images. 

We note that the distributions for the largest apertures show that the background flux is
slightly negative far away from objects. This systematic error in the background is much smaller
than the random error. We did not correct for this error as it is very small compared to the
quoted errors and
as it is not clear whether the background
is similarly affected in regions where objects are detected.

\begin{table}[h]
\centering
\caption{Power-law parameters for empty aperture errors}\label{table:noiseprop}
\begin{tabular}{lll}
\hline \hline
\noalign{\smallskip}
Field & $\alpha$ & $\beta$ \\
\hline
\noalign{\smallskip}
AEGIS & 0.35 & 1.27 \\
COSMOS  & 0.35 & 1.28 \\
GOODS-N  & 0.3 & 1.31 \\ 
GOODS-S & 0.3 & 1.31  \\
UDS & 0.3 & 1.31\\
\hline
\noalign{\smallskip}

\end{tabular}
\vspace{+0.5cm}
\end{table}

\subsection{Low Resolution Data Photometry}
\label{sec:lowresphot}

The large differences between the PSF sizes in the \textit{HST} data and the ground-based and Spitzer data must be taken into account in order to obtain accurate color information, without degrading the \textit{HST} images to lower S/N through smoothing or losing the exquisite high resolution information they provide. We use the MOPHONGO code developed by one of us (I. Labb{\'e}) to perform the photometry on the ground-based and Spitzer IRAC images, as described in \citet{Labbe06, Wuyts07, Whitaker11}. 

The code uses a high-resolution image as a prior to estimate the contributions from neighboring blended sources in the lower resolution image. We use an average of the PSF-matched F125W, F140W and F160W images as the high-resolution prior. A shift map that captures small differences in the astrometry of the high resolution reference image and low resolution measurement image is created by cross-correlating the object positions in the two images. The position-dependent convolution kernel that maps the higher resolution PSF to the lower resolution PSF is determined by fitting a series of Gaussian-weighted Hermite polynomials to the Fourier transform of a number of point sources across each image. Point sources are selected in the same way as for the \textit{HST} PSF-matching described above. Poorly fit objects are rejected and a smoothed map of the appropriate coefficients is created in two iterations. The high resolution image is then convolved with the local kernel to obtain a model of the low resolution image, with the flux normalization of individual sources as a free parameter. Photometry in an aperture size appropriate to the size of the PSF is done on the original image, with a correction applied for contamination from neighboring sources around each object as determined from the model. Fluxes are further corrected to account for flux that falls outside of the aperture due to the large PSF size. The correction is given by the ratio of the flux enclosed in the photometric aperture in the high resolution image (before convolution) to the low-resolution model (after convolution). 

For the ground-based NIR and optical images we use aperture diameters that depend on the seeing. For images that have FWHM $ \leq 0\farcs8$ we use an aperture of 1$^{\prime\prime}$, for images with $0\farcs8 \lt$ FWHM $\leq 1\farcs0$ we use an aperture of $1\farcs2$, and for images with larger FWHM we use an aperture of $1\farcs5$. For the optical medium bands in GOODS-S we use a minimum aperture of $1^{\prime\prime}$ and an aperture of $3^{\prime\prime}$ for the few bands with seeing $\gt 1\farcs5$. For the IRAC bands we use an aperture diameter of $3^{\prime\prime}$. The apertures, FWHM and image zero points are listed in Tables~\ref{table:aegis_image_data} to \ref{table:uds_image_data}. 

The GOODS-N and GOODS-S fields have two epochs of IRAC 5.8 and 8$\mu$m data that overlap in the center of the field. For these fields we compute average of the two flux measurements for objects that have a weight greater than zero in both epochs, after correcting each flux to total as described below. The errors are added in quadrature and the weight given in the catalog is the average of the relative weights in the two epochs.

\subsection{Flux Corrections}
\label{sec:fluxcor}

We correct for Galactic extinction using the values given by the NASA Extragalactic Database extinction law calculator\footnote{\url{http://ned.ipac.caltech.edu/help/extinction_law_calc.html}} at the center of each field, based on the recalibration by \citet{Schlafly11} of the COBE/DIRBE and IRAS/ISSA dust maps \citep{Schlegel98}. The extinction estimates assume a \citet{Fitzpatrick99} reddening law with $R_V = 3.1$. We interpolate between the values provided for a small set of filters to determine the extinction at the central wavelength of each filter in our dataset. We do not account for the variable width of the filters or variations across the field. The Galactic extinction corrections are generally small ($\la 0.05$~mag). Tables~\ref{table:aegis_image_data}--\ref{table:uds_image_data} list the corrections applied for each band.

The fluxes provided in the catalog are total fluxes. We correct the color flux measured in each band to a total flux by multiplying by the ratio of the F160W (F140W) total flux to the F160W(F140W) flux measured in the appropriately sized aperture: 
\begin{equation}
f_{X, \rm{tot}} = f_X(r) \times \frac{f_{\mathrm{WFC3,tot}}}{f_{\mathrm{WFC3}}(r)},
\end{equation}
where $X$ represents each filter, $r$ is the appropriate aperture for that band as given in Tables~\ref{table:aegis_image_data}--\ref{table:uds_image_data}, and $f_{\mathrm{WFC3,tot}}$ is the total flux in the reference band as calculated by Equation \ref{eqn:totcor}. Note that differences between the PSFs have already been taken into account, so that $f_X(r)$ and $f_{\mathrm{WFC3}}(r)$ measure an equivalent fraction of the total light in different bands.  The aperture errors are similarly converted to a total error by multiplying by the same correction as the fluxes.

The F160W and F140W aperture flux measured within $0\farcs7$ as well as the total F160W and F140W flux are provided for each object in the catalog, allowing one to convert back to a consistent color measurement for any band.  The correction from AUTO to total is also provided as a column in the catalog. An addition correction is applied to all the catalog fluxes to account for zero point and template uncertainties, as listed in Tables~\ref{table:zp_offsets1} and \ref{table:zp_offsets2} . This is described in detail in Section~\ref{sec:eazy} below.

\subsection{Point Source Classification}
\label{sec:pointsource}

\begin{figure*}[ht]
\centering
\includegraphics{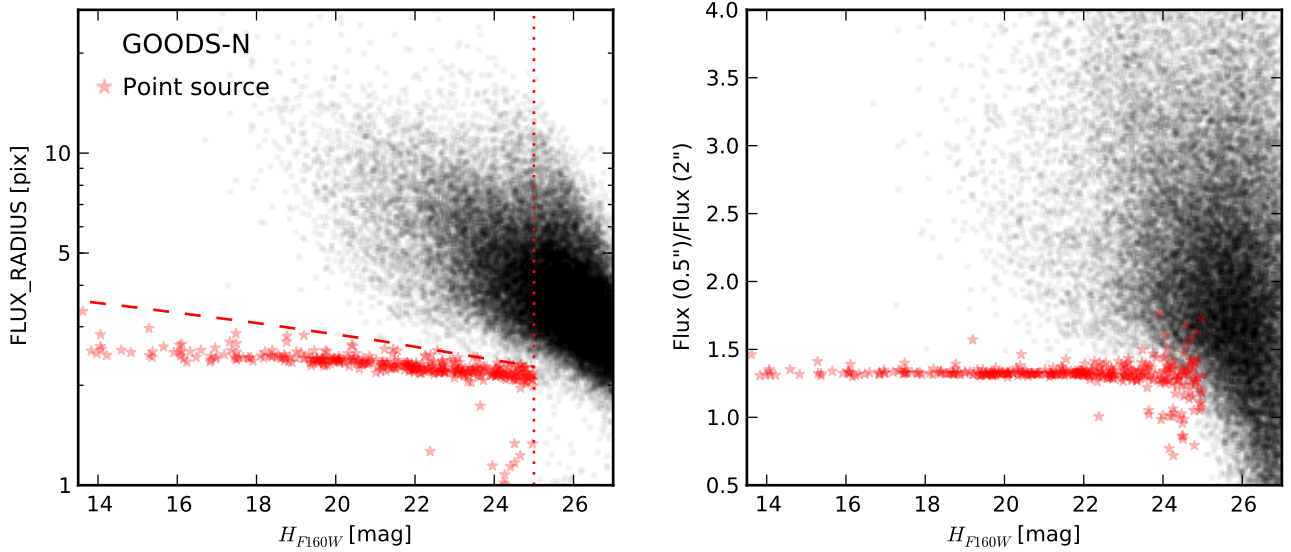}
\caption{Left panel: SExtractor's FLUX\_RADIUS against total F160W magnitude for the GOODS-N field.  Objects classified as point sources in the catalog are shown with red star symbols, galaxies and uncertain classifications (with $H_{F160W} > 25$~mag) as black symbols. The red dashed line, corresponding to the equation in the text, is used to make the selection. The red dotted line shows the magnitude limit of 25, beyond which the classification is deemed uncertain. The right panel demonstrates an alternate method of selecting point sources using the ratio of fluxes in a large and small aperture.  The tightness of the stellar sequence in this ratio at brighter magnitudes ($H_{F160W} \la 24$~mag) allows for a more stringent classification, but the separation becomes less clear than the flux radius selection at fainter magnitudes. The flux ratio was used to select stars for the PSF-matching and kernel fitting. The symbols in the right panel correspond to the selection shown in the left panel; in general, the two methods agree well.}
\label{fig:star_flag}
\vspace{+0.3cm}
\end{figure*}

Compact or unresolved sources form a tight sequence in size-magnitude space, with fairly constant, small sizes as a function of magnitude. In the left panel of Fig.~\ref{fig:star_flag} we show the SExtractor FLUX\_RADIUS against total F160W magnitude for the GOODS-N field. Point sources can be cleanly separated from extended sources down to $H_{F160W} \sim 25$~mag. We provide a point source flag in the catalog based on cuts in this space, as measured on the F160W images. Objects are classified as point sources (star\_flag = 1) if they have $H_{F160W} \leq 25$~mag and $\mathrm{FLUX\_RADIUS} < -0.115 H_{F160W} + 5.15$. 
These objects are shown as red stars in the figure. Note that the group of very compact objects lying well below the stellar sequence are mostly artifacts. Objects fainter than 25~mag (dotted red line) cannot be cleanly separated and are assigned a star\_flag of 2. All other objects are classified as extended, with star\_flag = 0. For the small fraction of objects with no coverage in F160W, we assign a star\_flag of 2.

We note that the ratio of fluxes in a large ($2\arcsec$) and small ($0\farcs5$) aperture plotted against magnitude provides a similar tight sequence of point sources for $H_{F160W} \la 24$~mag (right hand panel of Fig.~\ref{fig:star_flag}). A selection in this space was made for the PSF-matching. Both sequences proved to be useful diagnostics of the image quality, with large spread in the stellar sequence indicating that the centers of stars have been down weighted during cosmic ray rejection.

\subsection{`Use' Flag}

For convenience, we provide a flag with the catalog that allows a straightforward selection of galaxies that have photometry
of reasonably uniform quality. This 'use' flag (listed as 'use\_phot' in the catalog, to distinguish it from spectroscopic quality
flags) is set to 1 if the following criteria are met:
\begin{enumerate}
\item Not a star, or too faint for reliable star / galaxy separation: star\_flag\,=\,0 or  star\_flag\,=\,2
\item Not close to a bright star. The halos and diffraction spikes of bright stars can cause severe problems with the photometry,
  in particular in ground-based optical images. This criterion is implemented in two parts: not within 18\arcsec\ of a star with
  ${\rm F606W}<17$ and/or ${\rm F160W}<15$,
and not within $12\arcsec$ of  a star with ${\rm F606W}<19$ and/or
${\rm F160W}<17$. This near\_star flag is provided as a catalog column.
\item Well-exposed in the F125W and F160W bands. We require that a minimum of 2 individual exposures cover the object in both
F125W and F160W. This removes objects on the edges of the mosaics, and in gaps. The number of exposures for a given object is the
median number in a $0\farcs 7 \times 0\farcs 7$ ($12\times12$ pixel) box. The number of exposures in each of the WFC3 bands are provided as catalog columns.
\item A detection in F160W. We apply a very low S/N cut to limit the number of false positives, requiring f\_F160W\,/\,e\_F160W\,$> 3$.
\item A ``non-catastrophic'' photometric redshift fit (see Section~\ref{sec:photzs}). The criterion is $\chi_p < 1000$.
\item A ``non-catastrophic'' stellar population fit (see Section~\ref{sec:fast}). This criterion is $\log(M)>0$.
\end{enumerate}
This flag selects approximately 85\,\% of all objects in the catalogs. The flag is not very restrictive: for most science purposes
further cuts (particularly on magnitude or S/N ratio) are required. Furthermore, we caution that the flag is not 100\,\% successful in
weeding out problematic SEDs. The overall quality of the SEDs is higher for
galaxies with a higher S/N in the WFC3 bands. Although the use\_phot flag
only requires a S/N$>3$, we caution that the SEDs of galaxies with
S/N(F160W)$\lesssim 7$ will be quite noisy. As with all photometric catalogs, individual SEDs should be inspected when selecting objects for, say, spectroscopic follow-up
studies. For statistical studies of large samples the use\_phot flag should be sufficiently reliable, 
particularly when combined with a S/N or
magnitude criterion. 

\subsection{Catalog Format}

\begin{table*}[!th]
\centering
\caption{Catalog columns}\label{table:cat_cols}
\begin{tabular}{ll}
\hline \hline
\noalign{\smallskip}
Column name & Description \\
\hline
\noalign{\smallskip}
id & Unique identifier \\
x    &  X centroid in image coordinates\\
y  & Y centroid in image coordinates\\
ra  & RA J2000 (degrees)\\
dec &Dec J2000 (degrees)\\
faper\_{F160W}    &  F160W flux within a 0.7 arcsecond aperture\\
eaper\_{F160W}  &1 sigma error within a 0.7 arcsecond aperture\\
faper\_{F140W}    &  F140W flux within a 0.7 arcsecond aperture\\
eaper\_{F140W}  &1 sigma error within a 0.7 arcsecond aperture\\
f\_X   & Total flux for each filter X (zero point = 25)\\
e\_X  & 1 sigma error for each filter X (zero point = 25)\\
w\_X  & Weight relative to maximum exposure within image X (see text)\\
tot\_cor  &Inverse fraction of light enclosed at the circularized Kron radius\\
wmin\_ground  & Minimum weight for all ground-based photometry (excluding zero exposure)\\
wmin\_hst & Minimum weight for ACS and WFC3 bands (excluding zero exposure)\\ 
wmin\_wfc3 & Minimum weight for WFC3 bands (excluding zero exposure)\\ 
wmin\_irac & Minimum weight for IRAC bands (excluding zero exposure)\\
z\_spec  & Spectroscopic redshift, when available (see notes on quality in each field)\\
star\_flag & Point source=1, extended source=0 for objects with total $H_{F160W} \leq 25$~mag  \\
& All objects with $H_{F160W} > 25$~mag or no F160W coverage have star\_flag = 2\\
kron\_radius  & SExtractor KRON\_RADIUS (pixels)\\
a\_image & Semi-major axis (SExtractor A\_IMAGE, pixels)\\
b\_image  & Semi-minor axis (SExtractor B\_IMAGE, pixels)\\
theta\_J2000 & Position angle of the major axis (counter-clockwise, measured from East)\\ 
class\_star & Stellarity index (SExtractor CLASS\_STAR parameter)\\
flux\_radius & Circular aperture radius enclosing half the total flux  (SExtractor FLUX\_RADIUS parameter, pixels)\\
fwhm\_image  & FWHM from a Gaussian fit to the core (SExtractor FWHM parameter, pixels)\\
flags    & SExtractor extraction flags (SExtractor FLAGS parameter)\\
IRACx\_contam  & Ratio of contaminating flux from neighbors to the object's flux in each of the IRAC bands (x=1-4)\\
contam\_flag & A flag indicating if any of the photometry has a contamination ratio $\ge$50\% in any of the IRAC bands \\
&(1 if $\ge$50\% in at least 1 band, 0 = OK)\\
f140w\_flag & A flag indicating whether the corrections and structural parameters were derived from F140W rather than F160W\\
use\_phot & Flag indicating source is likely to be a galaxy with reliable measurements (see text) \\
near\_star & Flag indicating whether source is close to a star (1 = close to a bright star, 0 = OK)\\
nexp\_f125w & Median number of exposures in F125W within a $12\times12$ pixel box centered on the source\\
nexp\_f140w & Median number of exposures in F140W within a $12\times12$ pixel box centered on the source\\
nexp\_f160w & Median number of exposures in F160W within a $12\times12$ pixel box centered on the source\\
\noalign{\smallskip}
\hline
\noalign{\smallskip}
\multicolumn{2}{l}{X = filter name, as given in Table~\ref{table:ancil_data}, or IRAC1,IRAC2, IRAC3, IRAC4 representing the 3.6, 4.5, 5.8 and 8$\mu m$ IRAC bands}
\end{tabular}
\vspace{+0.5cm}

\end{table*}

We provide a full photometric catalog for each of the five 3D-HST fields, as well as a master catalog with a subset of parameters in common for objects from all five fields. The catalogs contain total flux measurements and structural parameters for 207967 objects in total - 41200, 33879, 38279, 50507, 44102 for AEGIS, COSMOS, GOODS-N, GOODS-S and the UDS, respectively. 

A description of the columns in each photometric catalog is given in Table~\ref{table:cat_cols}. All fluxes are normalized to an AB zero point of 25, such that 
\begin{equation}
\mathrm{mag}_{AB} = -2.5\times\log_{10}(F)+25.
\end{equation}

As described above, the catalogs contain the aperture flux in $0\farcs7$ and $1\sigma$ error for the F140W and F160W bands and the total fluxes and $1\sigma$ errors for every band listed in Table~\ref{table:ancil_data}. The structural parameters from SExtractor and the corrections to total magnitudes are derived from the F160W image where there is F160W coverage and F140W otherwise (only 0.3\% of objects). The `f140w\_flag' column indicates whether the F140W image was used rather than F160W (1 = F140W used, 0 = F160W used).

For the GOODS fields, where there is \textit{HST}/ACS data from both the GOODS survey and CANDELS, we append ``cand'' to the column names for the CANDELS data. The NIR column names correspond to the filters listed in Tables~\ref{table:aegis_image_data} to \ref{table:uds_image_data} and in some cases are appended with the survey name for clarity.  

We provide a weight column for each band to indicate the relative weight for each object compared to the maximum weight for that filter. In practice, the weight is calculated as the ratio of the weight at each object's position to the 95$^{th}$ percentile of the weight map. We smooth the weight map using a filter of 3 pixels, and use the 95th percentile rather than the absolute maximum, to avoid being affected by extreme values. Objects with weights greater than the 95$^{th}$ percentile weight have a value of 1 in the weight column. 

The catalog also contains a series of ``minimum weight'' columns (wmin\_ground, wmin\_hst, wmin\_wfc3 and wmin\_irac) which store the minimum of the relative weight values for the set of ground-based, \textit{HST}, \textit{HST/WFC3} and IRAC filters, respectively. These columns provide a useful way to select objects that have sufficient coverage in a particular set of filters. The ``nexp\_f125W'', ``nexp\_f140w'' and ``nexp\_f160W'' columns in the catalogs give the median number of exposures for each of the WFC3 bands in a $12\times12$ pixel ($0\farcs7 \times 0\farcs7$) box around each object.

\subsection{Completeness}
\label{sec:completeness}

\begin{figure}[!th]
\centering
\includegraphics[width = 0.48\textwidth]{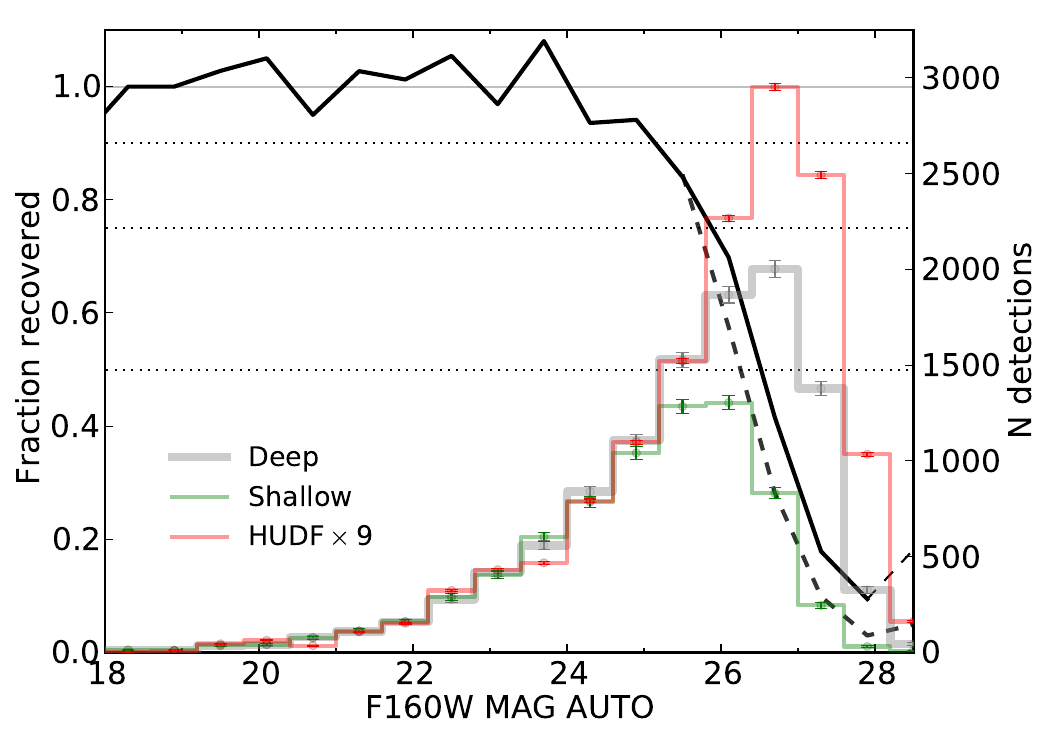} 
\caption{The completeness fraction as a function of F160W magnitude determined by comparing the detections in an image with the typical depth of the CANDELS/Wide images to the CANDELS/Deep image of the same area within GOODS-South and the HUDF. The deep image is twice as deep as the shallow image, with approximately four times the exposure time. The counts in the HUDF have been scaled by a factor of 9 to account for the difference in area. The solid black line shows the ratio of the counts in the shallow and deep areas. The fraction drops off faster relative to the HUDF, as seen from the dashed black line at $H_{160W} > 25$. The dotted lines show detection fractions of 90, 75 and 50\% with corresponding $H_{160W}$-AUTO magnitudes of 25.1, 25.9 and 26.5~mag relative to the deep image and 25.1, 25.8, and 26.2~mag relative to the HUDF. The gray, green and red histograms show the number of objects detected in the deep, shallow and HUDF images, respectively, with the error bars representing Poisson errors.} \label{fig:completeness}
\end{figure}

The depth of the images varies from field to field and across individual fields, as we show in the weight maps in Figures 2 -- 6. As a result, the completeness will depend on position, as well as varying for sources of different morphologies, magnitudes and sizes. For a discussion of the completeness within the deep and wide areas within the CANDELS/GOODS-S field and the simulated dependence of completeness on size and flux for de Vaucouleurs and exponential galaxy profiles, see \citet{Guo13}.

The nominal completeness within the CANDELS/Wide survey images can be estimated by comparing the number of detections in a shallow image with the typical exposure time of the CANDELS/Wide survey (2 orbit depth) to those in the CANDELS/Deep area in the GOODS-South field, which has approximately four times the exposure time. This is the method that was used by \citet{Tal14} to determine completeness as a function of various galaxy properties, such as redshift, mass etc. We follow the same procedure here to determine the completeness as a function of F160W magnitude. 

Deep and shallow F125W and F160W images of the central area of GOODS-S reduced in a similar way to that described in Sect.~\ref{sec:wfc3im} above were kindly provided by D. Magee. We created new deep and shallow detection images by coadding the noise-equalized F125W, F140W and F160W images in the same way as described in Sect.~\ref{sec:detection} and applied the same SExtractor parameters to detect sources. We adjusted the DEBLEND\_MINCONT parameter to 0.005 for the shallow image and 0.0001 for the deep image, as the detection of objects in the wings of nearby bright objects becomes problematic in the deeper image. We ran SExtractor in dual mode on each detection image, with our F160W image as the measurement image. The number of detections in the shallow image are compared to the number of detections in the deep image as a function of F160W AUTO magnitude in Fig.~\ref{fig:completeness} (green and gray histograms). The fraction of detections recovered in the shallow image compared to the deep image is shown by the thick black line. We find that the number of detections in the shallow image begins to deviate significantly from the deep image at an $H_{F160W}$ magnitude of 25, reaching 90, 75 and 50\% completeness levels at magnitudes of 25.1, 25.9 and 26.5~mag, respectively. Note that the decrease is fairly gradual at first, so the 75\% completeness level we quote is more reliable than the 90\% level. 

We also compare the number of objects detected in the shallow image to the number expected based on the detections in the HUDF (53 orbit depth), using the catalog of \citet{Lundgren14}. As the HUDF is much smaller in area than the images we use for these tests, we have scaled the counts in the HUDF by a factor of 9 to bring them into approximate agreement at the bright end. The scaled HUDF counts are shown by the red histogram in Fig.~\ref{fig:completeness}. We find good agreement between the scaled HUDF counts and the number of detections in the deep image down to $H_{F160W}\sim 25.8$~mag, but the HUDF is approximately 1 magnitude deeper, resulting in a faster drop in the fraction of objects detected in the shallow image. The fraction of objects recovered in the shallow image relative to the HUDF reaches 90, 75 and 50\% at magnitudes of 25.1, 25.8 and 26.2~mag, respectively. This is shown by the dashed black line in Fig.~\ref{fig:completeness}.  The weight of the shallow image used here corresponds to the shallowest portions of each of the full mosaics and this is therefore a conservative estimate of completeness. 
\subsection{Number Counts}
\label{sec:numbercounts}

\begin{figure*}[ht]
\includegraphics[width = 0.95\textwidth]{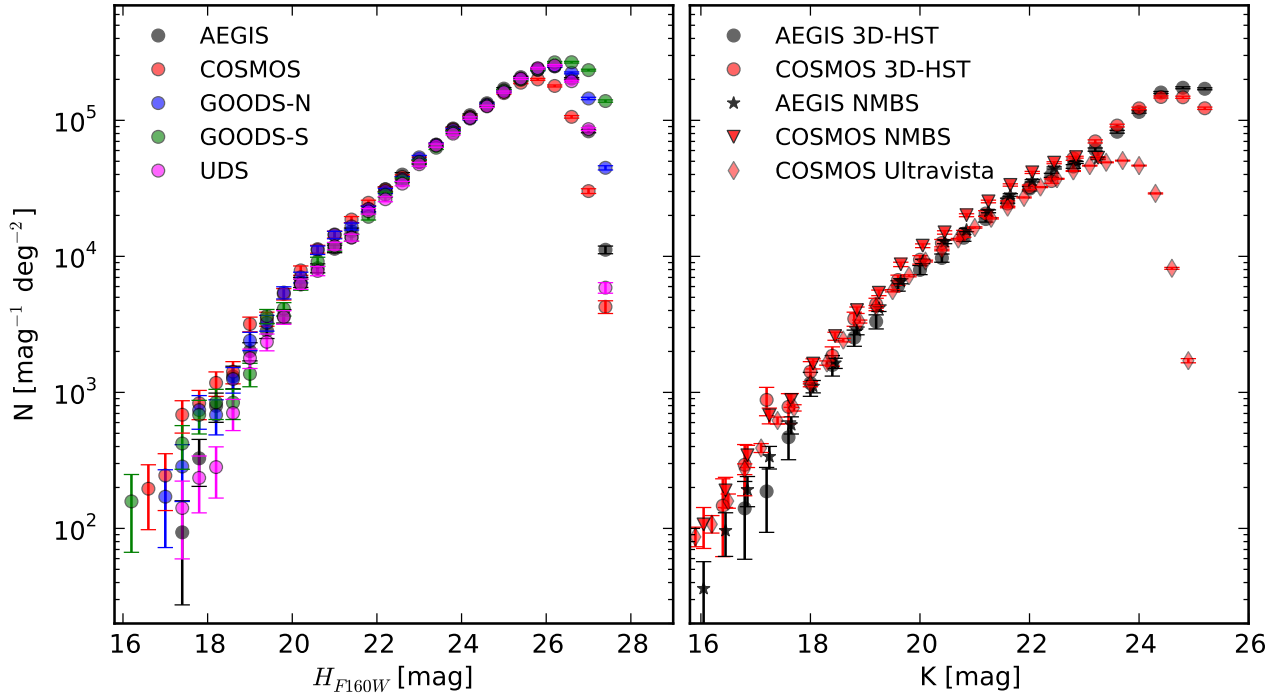}
\caption{Number counts per unit area.  The left hand panel shows the number counts of galaxies with Poisson errors in each of the five fields as a function $H_{F160W}$ total magnitude, with no correction for incompleteness. In the right hand panel, the COSMOS and AEGIS number counts in the $K$-band are compared to those from the NMBS survey \citep{Whitaker11} and UltraVISTA catalog from \citet{Muzzin13a}. The error bars on all the data points represent Poisson errors. The agreement between 3D-HST and the NMBS for the AEGIS field is excellent. In COSMOS we agree well with UltraVISTA over the full range where the ground-based survey is complete. The NMBS COSMOS number counts are somewhat higher, but agree well at the bright end, where all three surveys find an excess of bright galaxies compared to the AEGIS field.}\label{fig:numbercounts}
\end{figure*}

We determined the approximate effective survey area of each of the five
fields in the following way. For each of the WFC3 images we create a map of the number of exposures contributing to each pixel. We masked out the areas affected by bright stars, using the near\_star criteria described above. The useful science area (corresponding to the use\_phot flag) was then calculated by adding up the number of unmasked pixels with at least 2 exposures in both F125W and F160W. The maps are provided as part of the data release.
The area of each field is given in Table~\ref{table:fields}. The number density of galaxies (satisfying our ``use\_phot'' flag criterion) in the 3D-HST fields is shown in Fig.~\ref{fig:numbercounts}. The left hand panel shows the number counts as a function of total $H_{F160W}$ magnitude for each of the five fields, while the right hand panel compares the $K$-band number counts for AEGIS and COSMOS with those obtained by the NMBS \citep{Whitaker11} and the UltraVISTA catalog from \citet{Muzzin13a}. The error bars represent Poisson errors in both panels. The number counts are fairly consistent across the five fields. COSMOS shows a slight excess of objects compared to other fields, particularly at the bright end. The NMBS found a similar difference between COSMOS and the AEGIS field, as can be seen in the right hand panel. For bright sources, the agreement between the two surveys in the AEGIS field is excellent, while for COSMOS, 3D-HST and UltraVISTA show excellent agreement but slightly lower number counts than NMBS. 

\subsection{Photometry of Close Pairs}

We caution that the photometry of galaxies that are in close proximity
of one another may have systematic errors that are not properly accounted
for in the formal uncertainties. The ground-based and IRAC photometry
is performed after subtracting a model for neighboring sources
(see \S\,\ref{sec:lowresphot}), but the space-based photometry is
performed directly on PSF-matched data without explicitly accounting for the
flux of neighbors. SExtractor does attempt to mask and correct the aperture fluxes
symmetrically for regions affected by overlapping objects (with the
MASK\_TYPE parameter set to CORRECT). As described in \S\,\ref{sec:hstphot}
the photometric aperture has a diameter of $0\farcs 7$. We estimate
the fraction of potentially affected objects in the catalog by
determining the number of pairs with a distance smaller
than $0\farcs7$, i.e.,
with overlapping photometric apertures. This fraction ranges
from 3\,\% to 7\,\% in the five fields, with an average of 5\,\%.
If we assume that only the faintest object of the pair is
affected, we infer that 2\,\% -- 3\,\% of objects may have
problematic HST photometry due to the effects of neighbors.

\section{Quality and Consistency Tests}
\label{sec:tests}

We carried out a number of tests to assess the photometric quality of the catalogs. Here we limit the discussion to 
internal tests: we ask whether the colors and uncertainties are reasonable and also whether there are offsets between the
five fields. In Appendix~\ref{app:comparisons} we do external tests, comparing the photometry in our catalogs to other published surveys. In what follows, we assume that all objects defined as point-sources (see Section \ref{sec:pointsource}) are stars, which is likely to be true for the majority of objects. We note that QSOs may be labeled as stars in our catalogs.

\subsection{Colors}

\begin{figure*}[ht]
\includegraphics[width =1\textwidth]{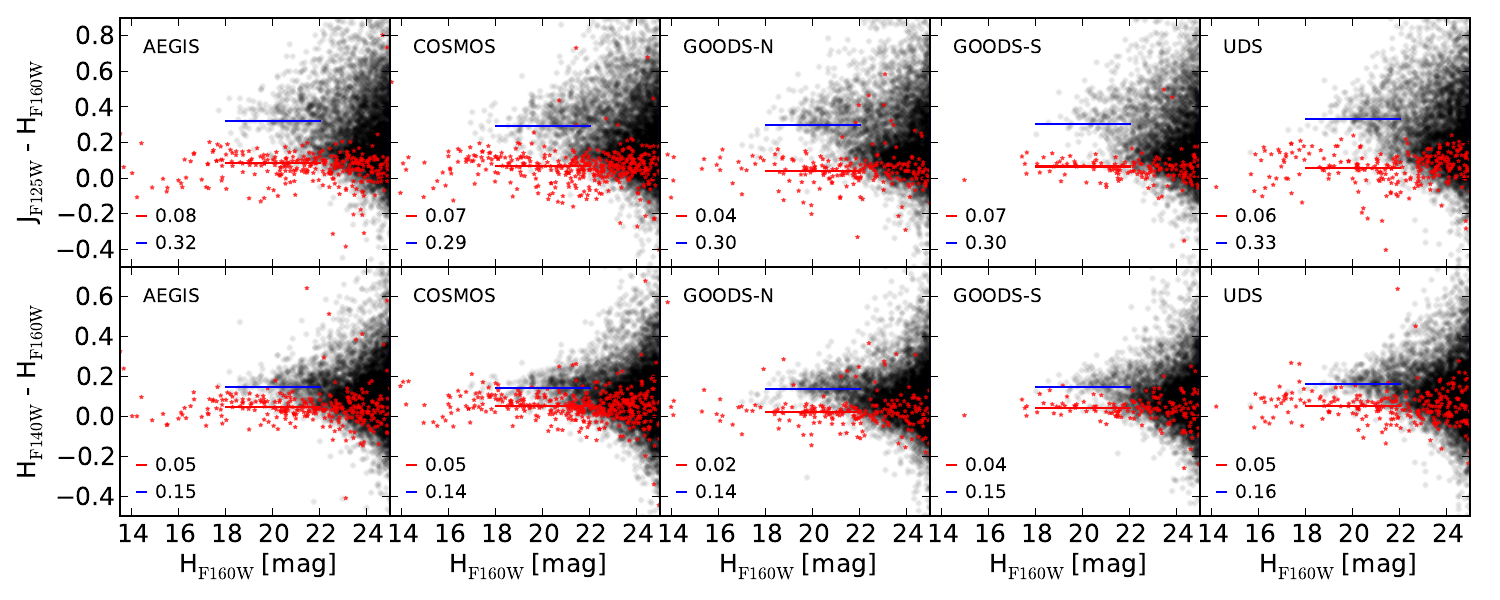}
\caption{$J_{F125W}$- $H_{F160W}$ and $H_{F140W}$- $H_{F160W}$ colors versus $H_{F160W}$ magnitude for each of the five fields. Point sources are shown in red and extended sources in black. The medians for point sources and extended sources in the range $18 < H_{F160W} < 22$ are labeled and shown by the red and blue lines, respectively. Objects with a SExtractor flag > 0 have been excluded. }\label{fig:colmag1}
\end{figure*}

In Fig.\ \ref{fig:colmag1} we compare the WFC3 colors of objects in the five fields. The top panels show the relation between $J_{F125W} -
H_{F160W}$ color and $H_{F160W}$ magnitude in each of the fields. The bottom panels show the relation between $H_{F140W}-H_{F160W}$ color and magnitude. Stars (objects with star\_flag = 1) are shown in
red. There is scatter in the colors of stars and galaxies, reflecting the fact that not all stars and galaxies have identical $J-H$ colors. In order to assess whether there are offsets between the fields we assume that the median observed $J-H$ colors do not have a
strong field dependence. The red and blue lines show the median color in the magnitude range $18<H_{F160W}<22$ for stars and
galaxies respectively. The median values are listed in the figure.

We find that the median WFC3 colors show very little field dependence. The rms field-to-field variation in the median $J_{F125W}-H_{F160W}$ color is 0.016 mag for both stars and galaxies. The difference between the highest and lowest field is 0.04 mag. The offsets of stars and galaxies are uncorrelated: subtracting the median color of the stars from the median color of the galaxies in each field does not reduce the field-to-field scatter. Taking the average of the median color of stars and the median color of galaxies does reduce the scatter, to an rms of only 0.01 mag and a maximum difference of 0.03 mag between the fields.  The $H_{F140W}- H_{F160W}$ colors show even less variation between fields than the $J_{F125W}-H_{F160W}$ colors: the rms field-to-field variation in the average color of stars (galaxies) is 0.013 (0.010)~mag, and the maximum difference is 0.03 mag.

\begin{figure*}[ht]
\includegraphics[width =1\textwidth]{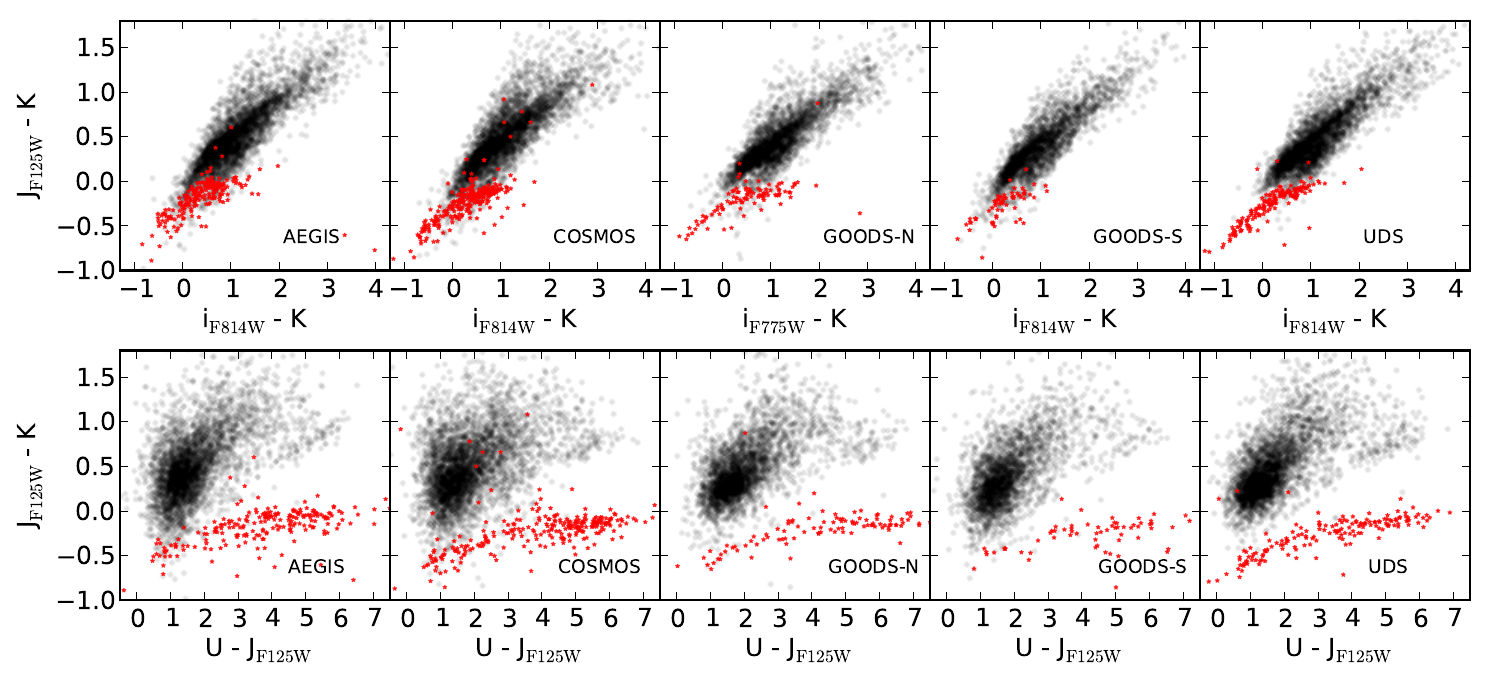}
\caption{$J_{F125W}  - K$ vs. $i - K$ and $J_{F125W} - K$ vs. $U - J_{F125W}$ color-color diagrams for objects with $K<24.5$~mag and $H_{F160W} < 25$~mag in each of the five fields. Point sources are shown in red and extended sources in black. Objects with a SExtractor flag > 0 have been excluded. Note that in the GOODS-N field, we use the F775W filter for the $i$-band, while the other fields use the F814W filter. The $U$-band and $K$-band filters also differ from field to field. The distributions are consistent with each other. The stellar locus can be clearly identified in color-color space.}\label{fig:obscols}
\end{figure*}

Figure \ref{fig:obscols} shows color-color relations in wavelength ranges that are particularly useful for separating stars from galaxies. There are subtle differences in the distributions of galaxies, partly due to non-uniformity in the ground-based filters that are used in the five fields. However, stars are well separated from galaxies in each field, particularly in the bottom panels of Fig.\ \ref{fig:obscols}. We infer that the star/galaxy separation is excellent, at least for objects with $K<24.5$.

\subsection{Total Fluxes}
\label{sec:totalfluxes}

As is usually the case,
the colors of objects in the catalogs are determined with higher accuracy than their total fluxes. Colors, and (more generally)
the shapes of the SEDs of objects, are measured using carefully matched apertures. However, total fluxes are based on measurements
in SExtractor's AUTO aperture, corrected on an object-by-object basis
for flux falling outside of this aperture
(see Section~\ref{sec:hstphot}). As described in Section~\ref{sec:fluxcor}, total fluxes were empirically determined for the $H_{F160W}$ band only;
all other bands were corrected to a total flux using the ratio of total flux to color aperture flux in the $H_{F160W}$ band.
As a result, the {\em shapes} of the SEDs in our catalog are based on psf-matched photometry using a reference aperture of $0\farcs 7$, and their {\em normalizations}
are based on the total $H_{F160W}$ flux. We note here
  that only the HST bands are true $0\farcs 7$ aperture
  measurements. The shapes of the SEDs of galaxies with strong color
  gradients outside of this aperture may have small systematic errors,
   in particular in the IRAC bands.

\begin{figure*}[!ht]
\includegraphics[width =1\textwidth]{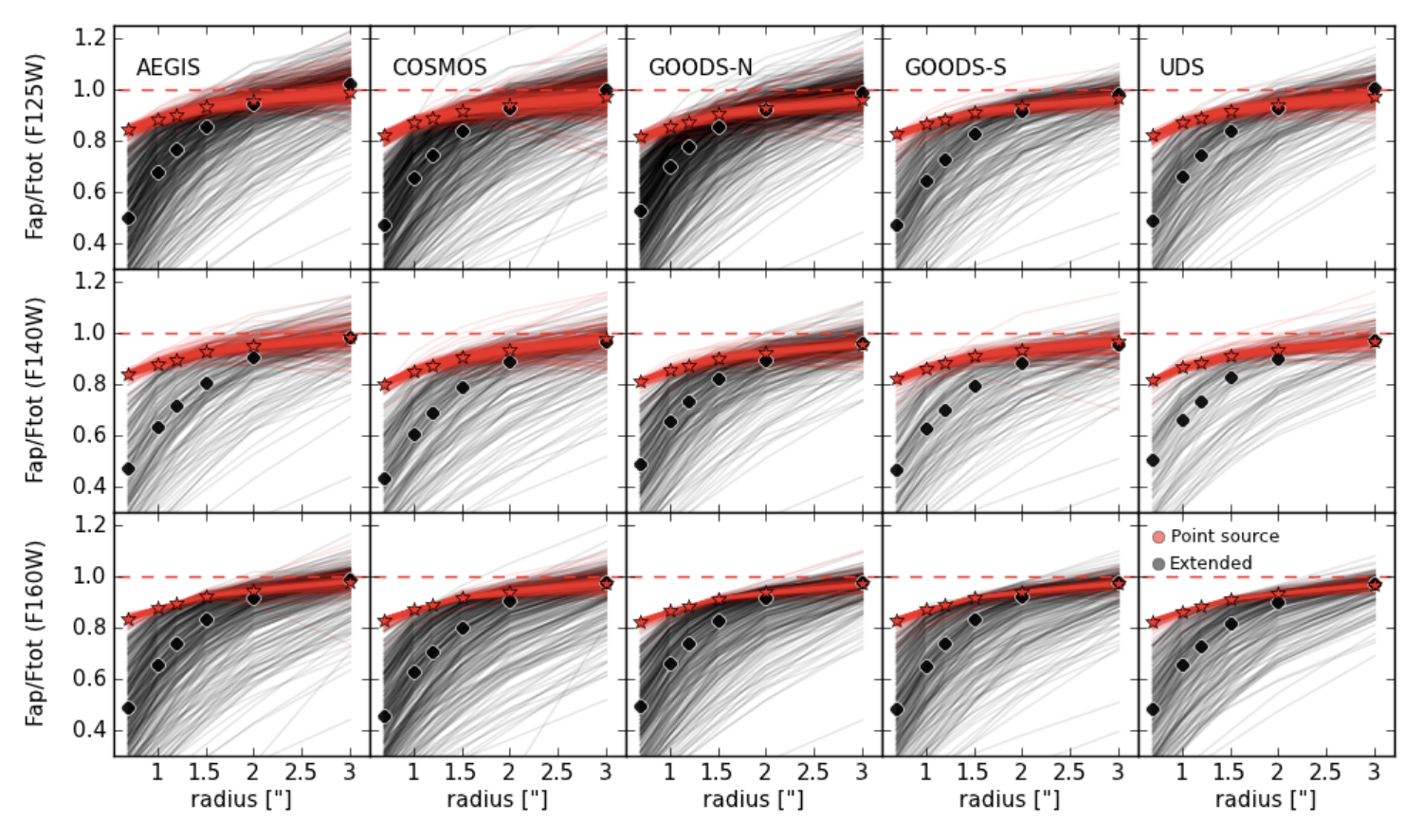} 
\caption{Ratio of aperture flux to total catalog flux as a function of aperture radius in each field and WFC3 band. Sources were selected to have S/N > 50. In each case few hundred extended sources were chosen randomly from the catalog with the requirement that they satisfy use=1 in addition to the S/N cut. Point sources are shown in red and extended sources in black. The median values for point sources and extended sources are shown by the large stars and filled circles, respectively. The agreement between the derived total catalog fluxes and the direct measurements of flux in $3^{\prime\prime}$ apertures are good. The measurements are consistent across the fields.}\label{fig:fapftot}
\end{figure*}

We test the accuracy of the total flux measurements in two ways. First, we measure fluxes in large apertures directly
from the WFC3 mosaics. Figure \ref{fig:fapftot} shows $F_{\rm ap}/F_{\rm tot}$,
the ratio of these aperture magnitudes and the total flux as listed in the
catalog, as a function of aperture size. Stars are shown as red lines, with median values indicated by open star symbols.
The growth curves show little scatter, and reach values that are within 0.05~mag of unity for an aperture
radius of $3\arcsec$. As our correction to total fluxes is partly based on the growth curves of stars this is not surprising;
nevertheless, this test empirically demonstrates that stellar photometry is reliable.
The grey curves and black points show growth curves of galaxies. There is a large variation in the curves, reflecting the
large variation in the apparent sizes of galaxies. Rather remarkably, the median growth curves again reach unity
(within 0.05~mag) at the largest aperture sizes, in all three filters and in all five fields. This  implies that our correction to total fluxes (and the PSF-correction for extended galaxies) is
correct to a few percent -- in the restricted sense that our catalog values correspond, in the median, to the measured flux in
a large aperture.

The second test is similar to the first in the sense that it also compares total fluxes in our catalog
to total fluxes that were measured in a different way. The alternative measurement here is not an aperture flux but
the total flux as implied by the best-fitting \citet{Sersic68} model. We used GALFIT \citep{Peng10} to fit these models
to galaxies in the catalog, using the same procedures as described in \citet{vanderWel12}. 
The GALFIT total fluxes are the best-fitting S\'ersic models integrated to $r=\infty$, and if galaxies are well-represented by
S\'ersic models out to large radii these fluxes are ``true'' total fluxes.

\begin{figure*}[!ht]
\includegraphics[width =1\textwidth]{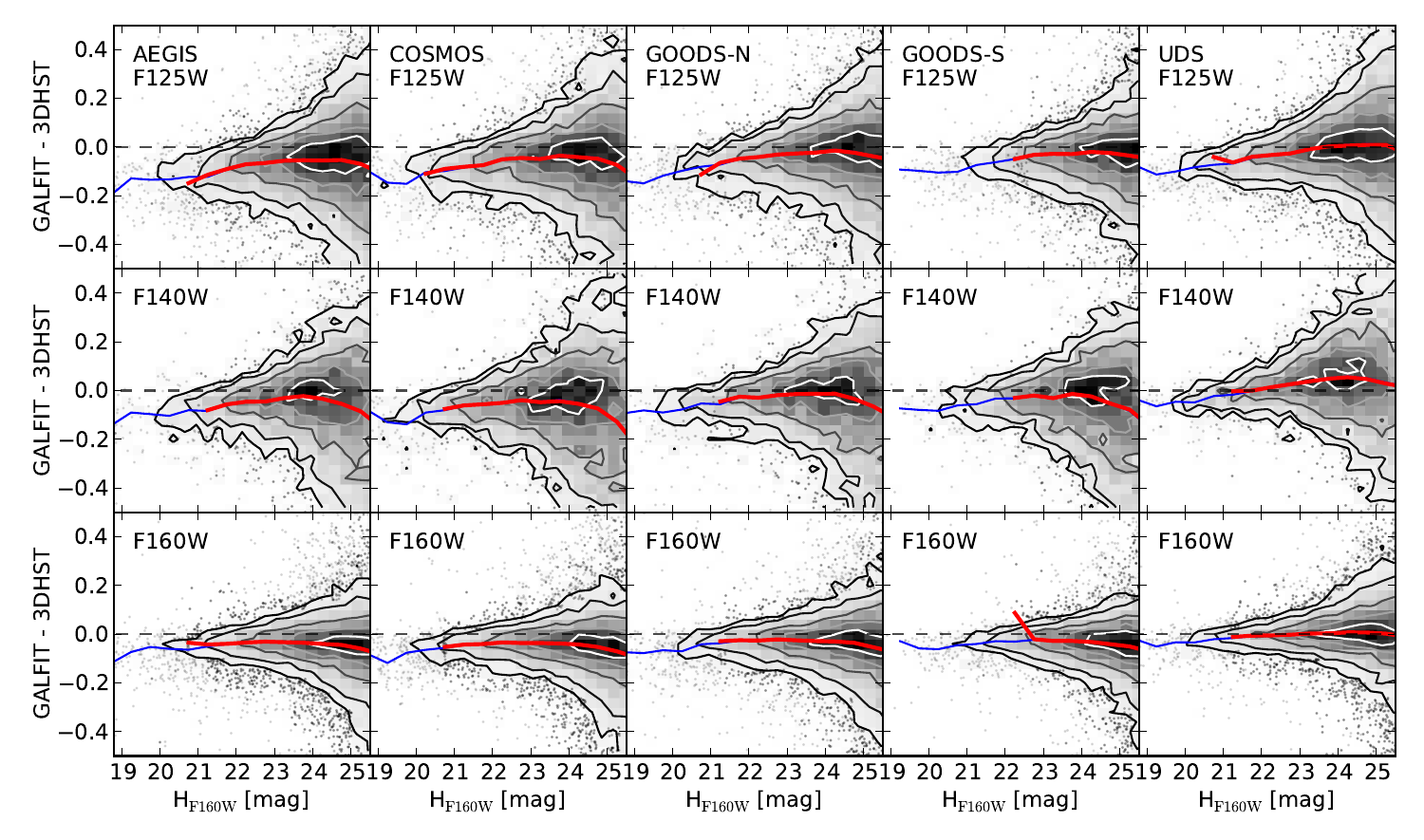} 
\caption{Comparison of the 3D-HST WFC3 total magnitudes to total magnitudes from morphological fits with GALFIT measured on the same images \citep[see][]{vanderWel12}. Objects with a good fit from GALFIT (GALFIT flag = 0) are shown dark gray, while objects with a suspicious fit (GALFIT flag = 1) are shown in light gray. Objects with a bad fit from GALFIT (GALFIT flag $> 1$), a 3D-HST use\_phot flag = 0 or SExtractor flags $\ge 2$ have been excluded from the comparison. The running medians for GALFIT flag=0 and flag=1 objects are shown in red and blue, respectively. The  contour levels are 5, 10, 25, 50 and 75\% of the maximum density.}\label{fig:galfitmags}
\end{figure*}

Figure \ref{fig:galfitmags} shows the difference between
total catalog magnitude and total GALFIT magnitude as a function of total catalog magnitude,
in the three WFC3 bands and for all five fields. Over the full magnitude range the median difference is $\le 0.06$~mag in each field. The median differences in the range $21 < H_{F160W} < 24$ are very small: $-0.03$, $-0.04$, $-0.03$, $-0.03$, and $0.00$ mag for $H_{F160W}$ in AEGIS, COSMOS, GOODS-N, GOODS-S, and UDS
respectively. The GALFIT magnitudes are slightly brighter, presumably because the S\'ersic profiles take into account
that the flux of galaxies exceeds that of point sources outside of the AUTO aperture.
The differences are also small in the bluer bands: the median difference for the five fields is $-0.04$ in $J_{F125W}$. This is remarkable given that galaxies have color gradients:
we use the $H_{F160W}$
total magnitudes to correct all other bands, and as galaxies generally become bluer with increasing radius one might
have expected that the $H_{F160W}$
correction underestimates the needed correction in bluer bands.
	
\begin{figure*}[!ht]
\includegraphics[width =1\textwidth]{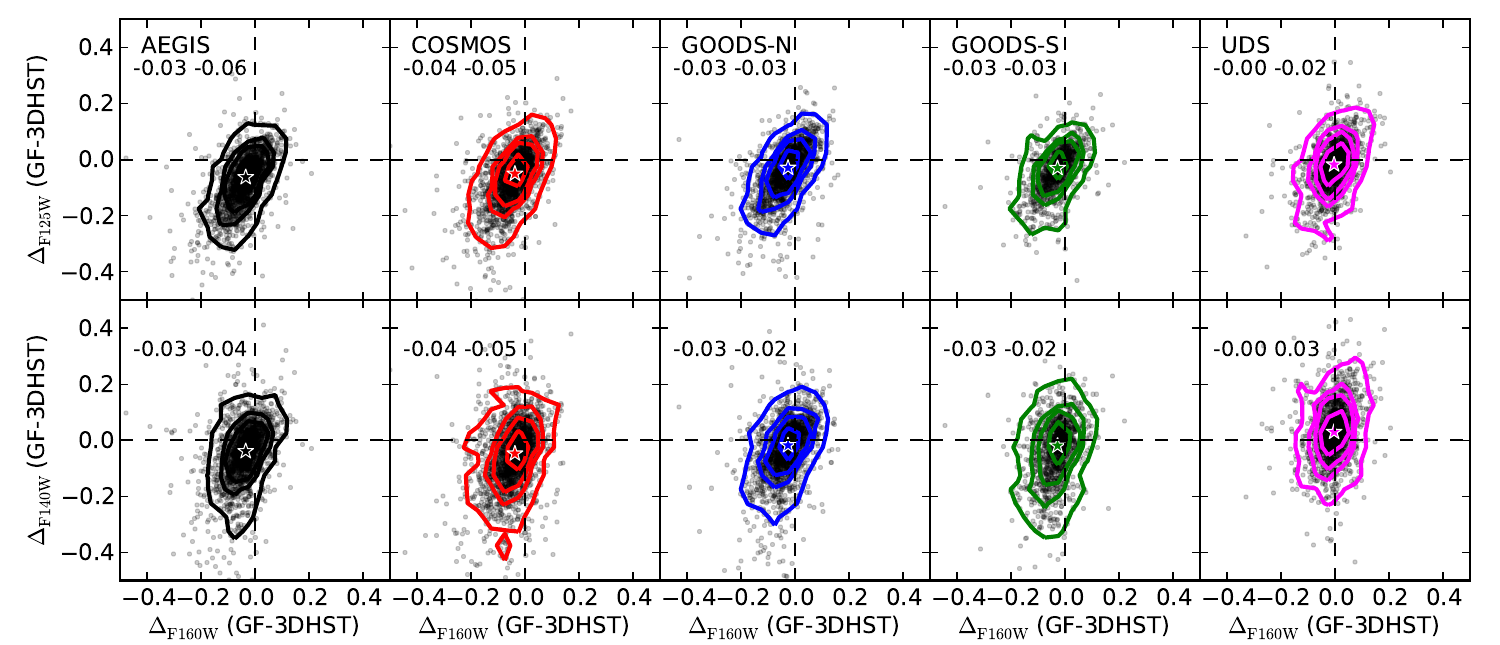} 
\caption{Difference between total magnitudes from GALFIT and 3D-HST in two WFC3 bands for  $21 < H_{F160W} < 24$. The upper (lower) panels compare the $J_{\mathrm{F125W}}$  ($H_{\mathrm{F140W}}$) and the $H_{\mathrm{F160W}}$ magnitude differences. Only objects with good GALFIT fits (GALFIT flag = 0) and 3DHST use\_phot = 1 are included. The medians are shown with a star symbol and labelled in the upper left corner of each plot. The contour levels represent 2, 10, 20 and 50\% of the maximum density. The differences are correlated and smallest in $H_{\mathrm{F160W}}$. }\label{fig:galfitdiff}
\end{figure*}

 In Fig. \ref{fig:galfitdiff} we compare the magnitude difference in two WFC3 bands (F125W vs. F160W in the upper panels, F140W vs. F160W in the lower panels) for galaxies with $21 < H_{F160W} < 24$~mag, SExtractor flags < 2, use\_phot = 1 and a good GALFIT fit (GALFIT flag = 0). Figure \ref{fig:galfitdiff} shows that the differences for individual galaxies are correlated, such that a relatively
high GALFIT flux in one band also implies a relatively high GALFIT flux in another band. This shows that the
observed differences for individual galaxies are not dominated by noise. It also suggests
that the differences are caused by the structure of the galaxies and not by color gradients, as color gradients would
not lead to correlations between offsets in $H_{F160W}$ and offsets in other bands.

\subsection{Error Estimates}

The errors in the photometry were determined by placing ``empty apertures'' in each of the images, and determining the width of the distribution of flux
measurements (see Section~\ref{sec:hstphot}). This method, described in more detail in \citet{Labbe03}, closely approximates the methodology that is used for
the actual flux measurements  and is insensitive to noise correlations (which affect
error estimates based on the observed pixel-to-pixel variation, such as the errors that are calculated by SExtractor).
As described in Section~\ref{sec:hstphot}, we ensured that the error for each
object is adjusted to take into account the photometric weight at its
position. 

 \begin{figure*}[!ht]
 \includegraphics[width =1\textwidth]{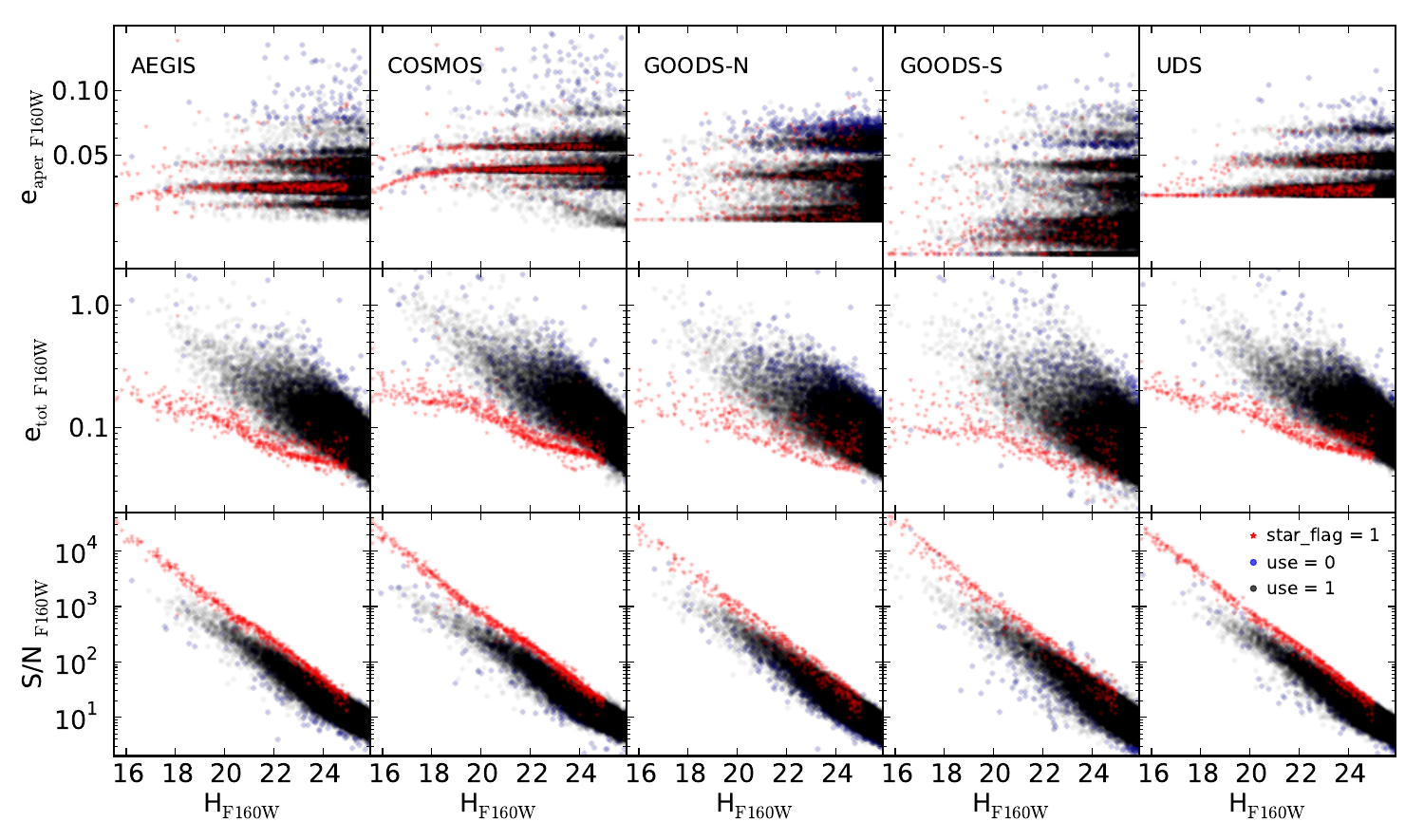} 
\caption{F160W error distributions in each of the five fields. The units are such that AB = $25 - \log ({\rm value})$. Objects with use\_phot = 0 are shown in blue if they are extended or in red if they are point sources (star\_flag = 1). Galaxies defined with use\_phot = 1 are shown in black. Upper panels: F160W errors within an aperture of $0\farcs7$ vs magnitude. The variable depths across each mosaic give rise to the discrete levels. GOODS-S is the deepest image with aperture errors reaching the lowest values, while COSMOS is the shallowest field on average. Middle panels: Total F160W error vs magnitude, as determined by scaling the noise within empty apertures to the aperture size of the circularized Kron radius of each object and making a correction to total as described in the text. Lowest panels: F160W (total) signal to noise vs. magnitude.  Point sources have the highest S/N at a given magnitude, while objects that should be excluded from analysis because they have low weight (and correspondingly large errors) form the lower envelope of the distribution. }\label{fig:f160errs}
\end{figure*}

 \begin{figure*}[!ht]
\includegraphics[width =1\textwidth]{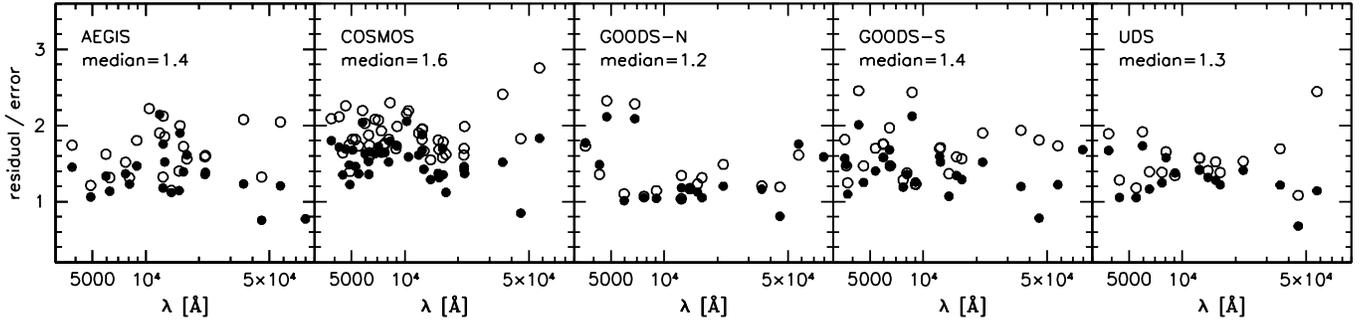} 
\caption{Comparison of the photometric uncertainties to the actual scatter in the data. For each filter in each field, the observed fluxes
were compared to the best-fitting EAZY model. The panels show the average residual from the fit divided by the expected residual based
on the photometric uncertainty. Open symbols show the catalog values, solid symbols
show the catalog values with the EAZY template error function added in quadrature. If the uncertainties are accurate the average should be 1. 
The uncertainties are slightly overestimated in most filters, but only by 20\,\% -- 50\,\%.}\label{fig:scatter_lam}
\end{figure*}

In Fig.~\ref{fig:f160errs} we show the catalog errors as a function of $H_{F160W}$ magnitude in each of the five fields. The top panels show the errors in our
standard photometric aperture of $0\farcs 7$. The scatter in the error at fixed magnitude is caused by the variation in the depth of the $H_{F160W}$
mosaics. The stripes reflect the fact that the weights, and hence the errors, largely reflect the number of exposures that went into a particular
position in the mosaic, and this number is an integer. Stars (red) fall in the same bands as galaxies, as their aperture fluxes are measured in the same
$0\farcs 7$ aperture. The distributions are not identical in each field, as the depths are not identical. The error distribution in GOODS-S extends to very
small errors, reflecting the great depth of the Ultra Deep Field data.

The middle panels show the errors in the ``total'' aperture. These errors are determined from the empty aperture errors using the power-law fit at the number of pixels in the circularized Kron aperture (see Fig.~\ref{fig:emptyap_errors} and \S\,~\ref{sec:hstphot}).  The stripes are blurred in these panels, as the scatter
in the error at fixed magnitude is now dominated by the variation in the Kron aperture size at fixed magnitude. The scatter in the Kron aperture size
reflects, in turn, the scatter in the sizes of galaxies at fixed magnitude. Stars are now clearly offset from galaxies: as the total
flux of stars is measured in a smaller aperture than the total flux of extended sources the errors in their total fluxes are smaller.
The scatter in the errors of stars is caused by the variation in weight: this also explains why some galaxies have smaller errors than some stars, particularly
in GOODS-S (which has the largest depth variation of any of the fields).

In the bottom panels  the S/N is plotted as a function of magnitude. The S/N was calculated by dividing the total $H_{F160W}$ flux by the error in the
total aperture. The relation of the S/N of stars with magnitude shows very little scatter, reflecting the small scatter in the errors of stars in the middle panels.
Galaxies show a large scatter. We note that the depths we, and others, quote for our data reflects the depth for point sources (red points). The errors
in the total magnitudes of galaxies are typically much larger, and this should be taken into account when
assessing the quoted depth of the $H_{F160W}$ images (see Section~\ref{sec:completeness}). By contrast,
the errors in the {\em colors} of galaxies, and the errors in the catalog for all
bands except $H_{F160W}$, are independent of the size of the object as they are based on the error in the $0\farcs 7$ aperture.

The previous tests show that the errors behave as expected, but they do not demonstrate that they 
are reasonably close to the true errors in the measurements. We tested this in the following way. For each photometric band in each field we selected
all objects with a S/N ratio of 15 {\em in that band}. Next, we subtracted the best-fitting EAZY template (see Section~\ref{sec:photzs}) from the observed flux, and
multiplied the residual by 15. 
Finally, we determined the biweight scatter in these distributions. If the errors are correct and the EAZY templates are perfect then we expect to find
that these distributions have a scatter of exactly 1. The results are shown in Fig.\ \ref{fig:scatter_lam}, for the catalog values (open symbols) and the
catalog values modified by the EAZY template error function (solid symbols).  We find that the scatter in these normalized residuals is close to 1 for
nearly all filters in all fields. The median deviations for the fields are between 1.2 and 1.5, implying that the errors are typically underestimated by 20\,\% --
40\,\%. This is a relatively small effect given that the observed residuals include all possible sources of error, including residual template mismatch.

\section{Redshifts, Rest-frame Colors and Stellar Population Parameters}
\label{sec:eazy}

We used the photometric catalogs to derive photometric redshifts, rest-frame colors, and stellar population parameters of the galaxies in the five fields.
As is well known, these derived parameters depend on the methodology that is used to derive them, and on
model assumptions \citep[see, e.g.,][]{Brammer08, Kriek09}.
A "default" set of parameters, described below, is provided with our photometric catalogs from the 3D-HST release pages. We expect to release future updates to these catalogs of derived parameters, in particular
versions in which the grism spectroscopic data are used to improve the redshift measurements.

\subsection{Spectroscopic Redshifts}
 \label{sec:speczs}
 
As a first step we searched the literature and other sources to find (ground-based)
spectroscopic redshifts of objects in the five fields. These redshifts are used to assess
the quality of photometric redshifts in \S\,\ref{sec:photzs},
at least for objects that are relatively bright at optical
wavelengths. Furthermore, when determining rest-frame colors
and stellar population parameters we always use
this spectroscopic redshift if it is available; otherwise we use the photometric redshift.
The spectroscopic redshifts in our catalogs are obtained by cross-matching the positions of objects within $0\farcs5$ to a number of publicly available catalogs.

In the AEGIS field there are 1139 spectroscopic redshifts obtained by matching to the DEEP2+3 catalogs \citep{Davis03, Newman13, Cooper12}.  Only objects with the highest quality flags (quality flag = 4) are included in the catalog. There are 1094 galaxies with use\_phot=1 and a spectroscopic redshift.

For the COSMOS field, we match to the zCOSMOS catalogs \citep{Lilly07}, finding 383 spectroscopic redshifts.  We additionally include 72 spectroscopic redshifts determined from MMT/Hectospec data \citep{Fabricant05, Mink07}\footnote{The MMT/Hectospec data includes about 8 hours in total, observed on November 23, 2011, February 2, 2013, March 31, 2013 and April 1, 2013 by M.Kriek.}, bringing the total to 455 spectroscopic redshifts. Here too, we keep only redshifts with an ``excellent'' quality flag. There are 420 galaxies with use\_phot=1 and a spectroscopic redshift.

There are 2081 spectroscopic redshifts included in the GOODS-N catalog. These were obtained by matching to the MODS catalog \citep{Kajisawa11}. The MODS redshifts are compiled from \citet{Yoshikawa10, Barger08, Reddy06, Treu05b, Wirth04, Cowie04, Cohen01, Cohen00, Dawson01}. No quality flags were provided, so there is a mix of reliable and less reliable redshifts in this field. There are 1837 galaxies with use\_phot=1 and a spectroscopic redshift.

In GOODS-S we find 2228 objects match to objects with spectroscopic redshifts in the FIREWORKS catalog \citep{Wuyts08}. We include the redshifts for the 1445 objects with a FIREWORKS quality flag of 1.0 in the catalog. There are 1284 galaxies with use\_phot=1 and a spectroscopic redshift.

There are 238 spectroscopic redshifts in the UDS catalog, 182 of which were obtained by matching to the compilation provided on the UDS Nottingham webpage \footnote{\url{http://www.nottingham.ac.uk/~ppzoa/UDS_redshifts_18Oct2010.fits}}. The redshifts are from a variety of sources, with some unpublished at the time the compilation was made \citep[][Akiyama et al. in prep.]{Yamada05, Simpson06, Geach07, vanBreukelen07, Ouchi08, Smail08, Ono10, Simpson12}. Redshifts with quality flags A (based on multiple reliable features), B (one reliable feature) or Z (flag not provided) are included in our catalog. Redshifts with quality flag C (one dubious feature) are not included. We also include 37 spectroscopic redshifts from IMACS/Magellan \citep[][and I. Momcheva, private communication]{Papovich2010}, 18 redshifts from \citet{Bezanson13} and 1 redshift from \citet{vandeSande13}. There are 178 galaxies with use\_phot=1 and a spectroscopic redshift. 

\subsection{Photometric Redshifts and Zero Point Corrections}
\label{sec:photzs}

\begin{figure*}[!ht]
\centering
\begin{minipage}[b]{0.32\linewidth}{
\includegraphics[width=\textwidth]{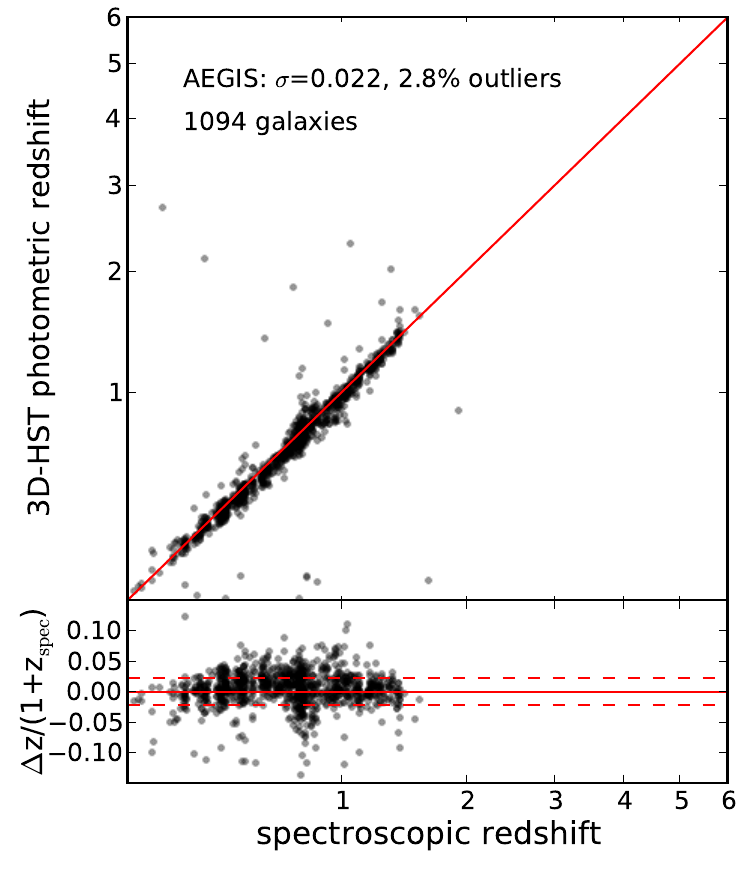}
}
\end{minipage}
\begin{minipage}[b]{0.32\linewidth}{
\includegraphics[width=\textwidth]{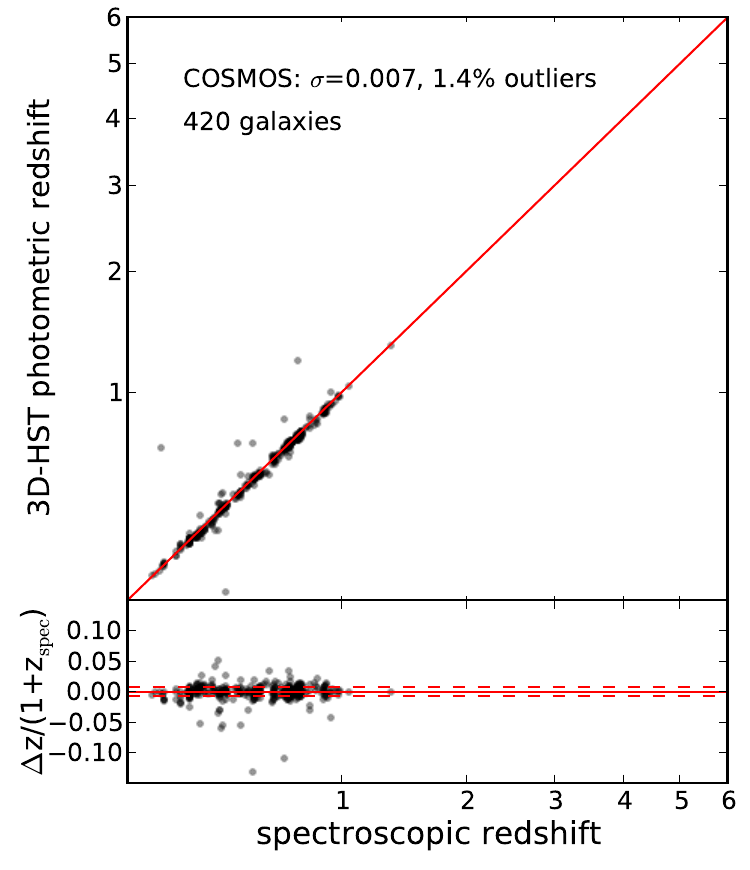}
}
\end{minipage}
\begin{minipage}[b]{0.32\linewidth}{
\includegraphics[width=\textwidth]{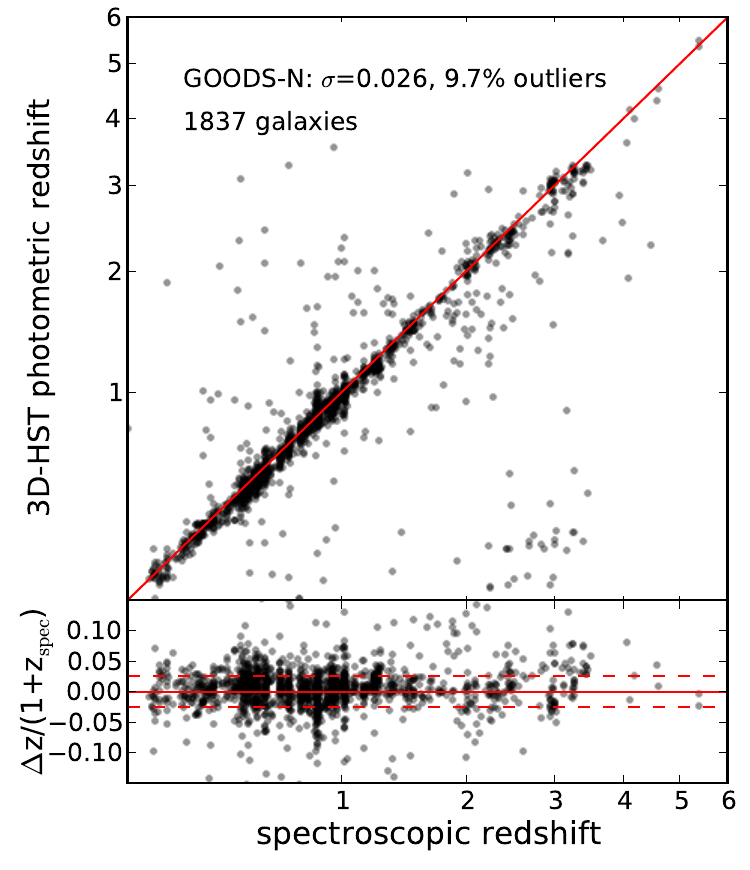}
}
\end{minipage}
\begin{minipage}[b]{0.32\linewidth}{
\includegraphics[width=\textwidth]{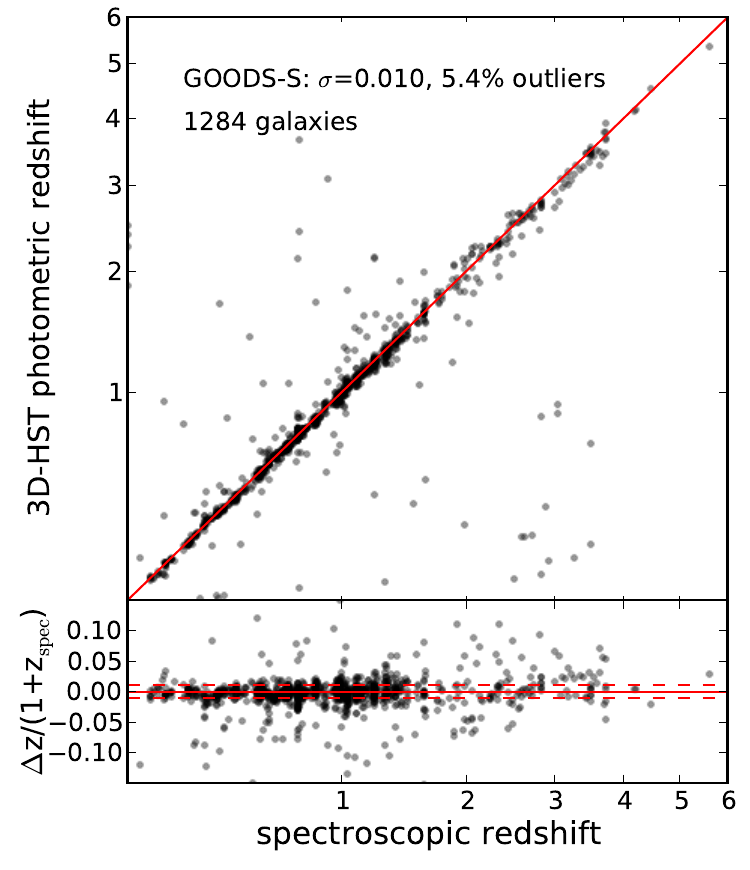}
}
\end{minipage}
\begin{minipage}[b]{0.32\linewidth}{
\includegraphics[width=\textwidth]{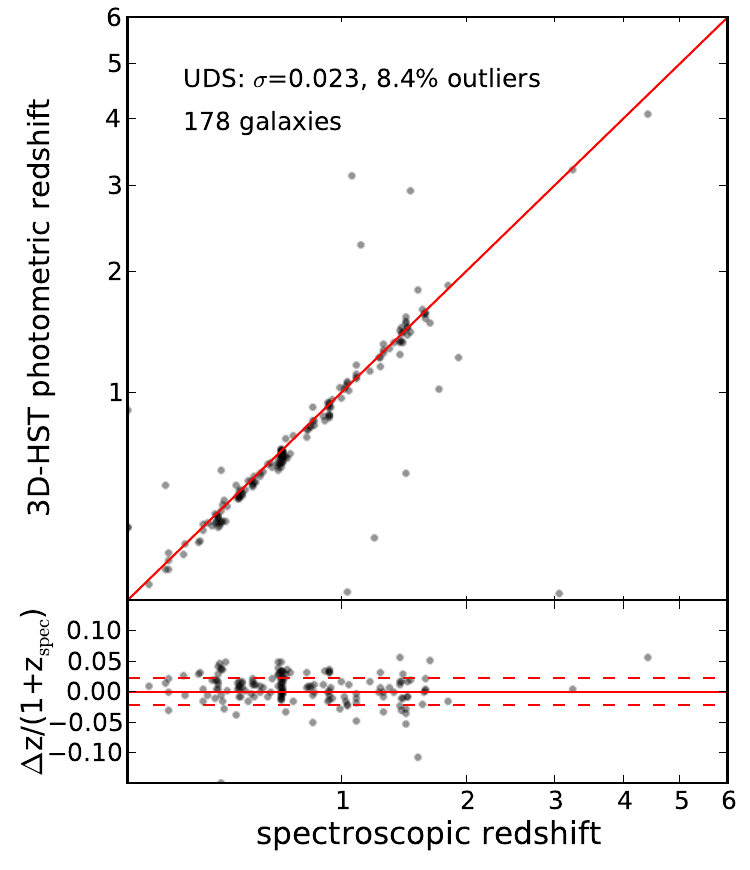}
}
\end{minipage}

\caption{Photometric redshifts versus spectroscopic redshifts from the literature in each of the five 3D-HST fields. The NMAD scatter $\sigma_{\mathrm{NMAD}}$, \% of objects with $|z_{\mathrm{phot}} - z_{\mathrm{spec}}|/(1+z_{\mathrm{spec}}) > 0.1$ and the number of galaxies in each comparison are shown in the upper left of the plot. The lower panels show the difference between the photometric and spectroscopic redshifts over $1+z_{\mathrm{spec}}$. The red dashed lines indicate $\pm \sigma_{\mathrm{NMAD}}$ in each case. }
\label{fig:speczphotz}
\end{figure*}

\begin{figure*}[ht]
\centering
\includegraphics[width=0.9\textwidth]{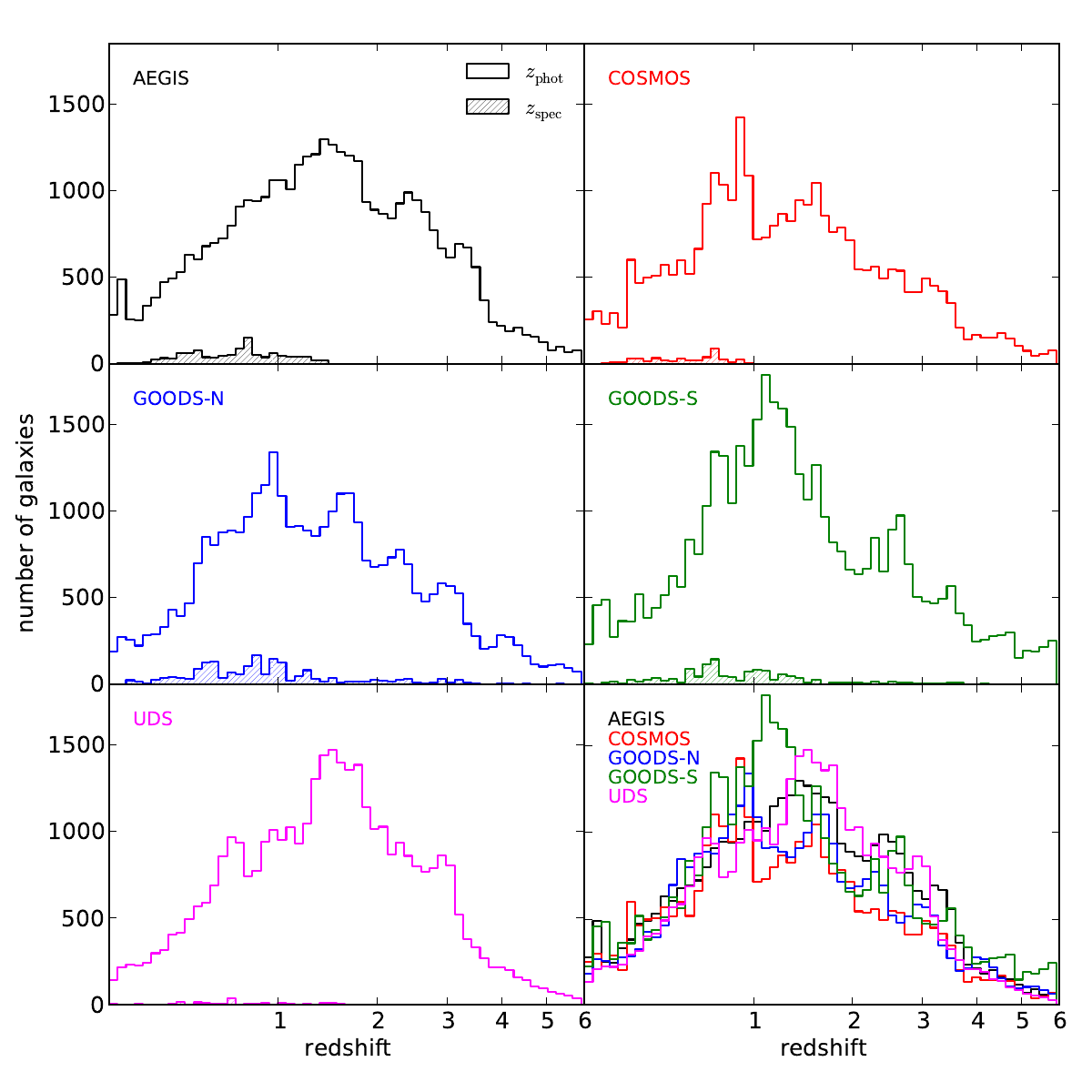} 
\caption{Redshift distribution in each of the five fields. The distribution of photometric and spectroscopic redshifts are shown as the open and hatched histograms, respectively. The final panel shows the distribution of photometric redshifts for all five fields, for comparison. \vspace{+10pt}}\label{fig:zdistr}
\end{figure*}
\vspace{+10pt}

\begin{figure}[!ht]
\includegraphics[width=0.48\textwidth]{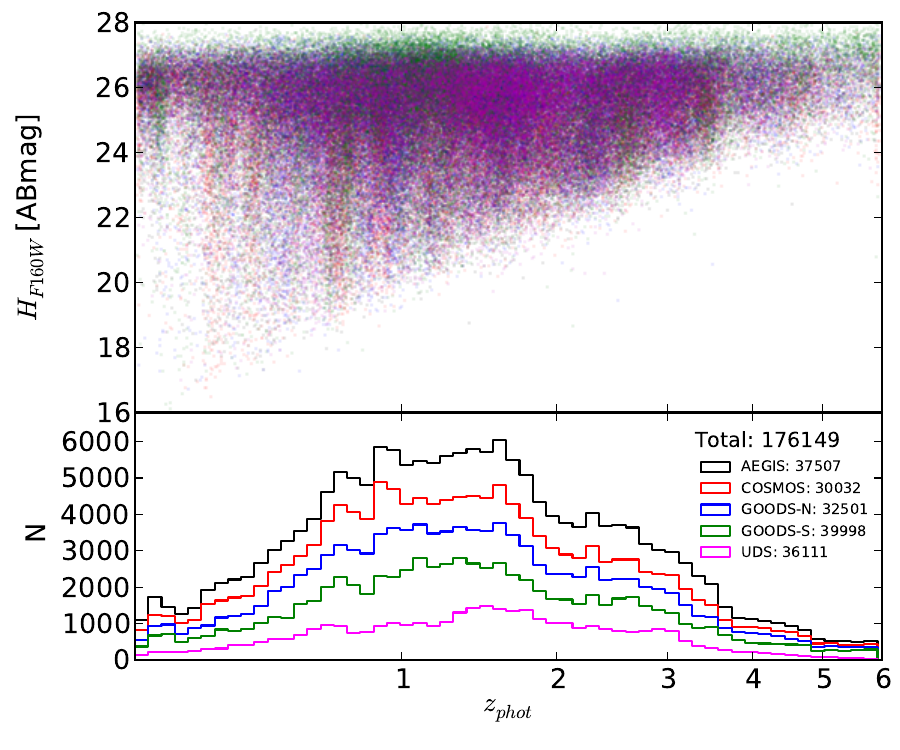}
\caption{The distribution of $H_{F160W}$ apparent magnitudes with photometric redshift (\texttt{z\_peak}), color-coded by field. The lower panel shows the number of galaxies as a function of \texttt{z\_peak} broken down into the contribution from each field. The histograms are successively added, with the black histogram giving the total distribution of all galaxies.}\label{fig:zphotmag}
\end{figure}

We determine photometric redshifts by fitting the SED of each object with a linear combination of seven galaxy templates, using the EAZY code
\citep{Brammer08}\footnote{\url{https://code.google.com/p/eazy-photoz/}}.
The template set is based on the default set described in \citet{Brammer08}. It contains five templates derived from a library of P\'EGASE stellar population synthesis models \citep{Fioc97}, a young, dusty template and an old, red galaxy template, as described in \citet{Whitaker11}. The old, red galaxy template is derived from the \citet{Maraston05} stellar population synthesis models with an age of 12.6 Gyr, a Kroupa initial mass function (IMF) and solar metallicity. We use the default template error function scaled by a factor of 0.5, which helps to account for systematic wavelength-dependent uncertainties in the templates, and a redshift prior based on the $K$-band apparent magnitudes.

We also modify both the templates and  the input photometry in the fitting procedure:  the templates are corrected for subtle differences between the observed SEDs of galaxies and the best-fitting
templates, and the photometry is corrected for empirically-determined zero point errors.
The methodology is discussed in the Appendix.
The zero point offsets for each field and band are provided in Table~\ref{table:zp_offsets1} and \ref{table:zp_offsets2}. The listed zero points have been applied to the catalogs and the corrected photometry used for the redshift fitting and stellar population parameters we present in the following sections. In what follows, galaxies are selected with a use\_phot flag of 1 and we use the spectroscopic redshift, where available, or the peak of the photometric redshift distribution (EAZY's \texttt{z\_peak}) as the galaxy redshift, unless specified otherwise.

Figure~\ref{fig:speczphotz} compares the photometric redshifts to spectroscopic redshifts from the literature. The number and quality of the spectroscopic redshifts in each field are heterogeneous as they are compiled from a number of different sources, as described in Section~\ref{sec:speczs}. In general we find excellent agreement with a normalized median absolute deviation $\sigma_{\mathrm{NMAD}} = 1.48 \times$~MAD/(1+z) of $\lt$ 2.7\% in all fields. The improvement in photometric redshifts that arises from including medium band data can be clearly seen by the reduction in scatter in the COSMOS and GOODS-S fields, reaching $\sigma_{NMAD} = 0.007$ and $\sigma_{NMAD} = 0.01$, respectively. Among objects with spectroscopic redshifts, there are few catastrophic failures: 3, 1, 10, 5 and 8\% for AEGIS, COSMOS, GOODS-N, GOODS-S and the UDS, respectively, where we define a catastrophic outlier as one with $|z_{\mathrm{phot}} - z_{\mathrm{spec}}|/(1+z_{\mathrm{spec}}) > 0.1$. We note that x-ray sources have not been excluded from these comparisons. In GOODS-S, 87 objects from the X-ray selected catalog of \citet{Szokoly04} are included. Removing these sources reduces the outlier fraction and scatter marginally. Ten objects from the XMM catalog of \citet{Ueda08} are included in the UDS spectroscopic redshift catalog, and four of these are outliers in the spectroscopic versus photometric redshift comparison. Given the small numbers of redshifts in the UDS, excluding these x-ray sources has a noticeable impact, reducing the outlier fraction to 6\%. In GOODS-N there are approximately 45 sources in common with the  Chandra 2Ms optically bright catalog from \citet{Alexander03}, however very few of these are outliers. The relatively high fraction of outliers in this field is likely to be caused by unreliable spectroscopic redshifts, as there are no quality flags made available in the merged source catalog. This can be contrasted with the low outlier fraction in the AEGIS field, where the spectroscopic redshifts are uniformly sourced from the DEEP2+3 surveys and stringent quality cuts have been applied. 

In Fig.~\ref{fig:zdistr} we show the photometric redshift distributions of galaxies (selected with use\_phot = 1) in each of the 3D-HST fields, using \texttt{z\_mc} from EAZY. \texttt{z\_mc} is a ``Monte Carlo'' redshift, randomly chosen from the EAZY probability distribution for each galaxy and is more appropriate for showing redshift \textit{distributions} incorporating the redshift uncertainties \citep{Wittman09}. The spectroscopic redshift distributions are shown by the hashed histograms. Overdensities such as those already known at $z=0.7$ and $z=1.1$ in GOODS-S \citep{Adami05} and the  $z=1.6$ cluster in the UDS \citep{Papovich2010} stand out clearly. These overdensities can also be seen in the distribution of apparent magnitudes with \texttt{z\_peak}, shown in Fig.~\ref{fig:zphotmag}, and in the mass distribution (see Fig.~\ref{fig:massz} below). In the lower panel of Fig.~\ref{fig:zphotmag} we show the number of galaxies as a function of \texttt{z\_peak}, adding the contributions from each consecutive field to the total number in that redshift bin. The upper (black) histogram then shows the total number in all five fields, while the difference between the upper histogram and the one below it shows the number of galaxies in AEGIS, and so on for each field.

\subsection{Rest-frame Colors}
\label{sec:rfcols}

The catalogs contain colors of galaxies in the observed frame, but to compare galaxies
at different redshifts rest-frame colors need to be used. These can be determined robustly as
we have a wide range of observed-frame photometry in each of the fields.
We use the EAZY templates and best-fitting redshift for each galaxy to determine its rest-frame luminosity in a series of filters, and determine rest-frame colors as the ratio of the luminosities in two filters.  More information on how the rest-frame colors are calculated is provided in \citet{Brammer11}, but we note that the calculation is now made for individual filters, rather than a set of two filters. That is, the templates are refit forcing $z=z_\mathrm{phot}$ and the flux in the rest-frame bandpass, $j$, is taken from the best-fit template considering only observed filters, $i$, where $|\lambda_\mathrm{obs,i} - \lambda_\mathrm{rest, j} (1+z) | < 1000 \AA$. We provide a catalog that contains the rest-frame luminosities in a variety
of commonly used filters.

\begin{figure*}[!ht]
\includegraphics[width=0.98\textwidth]{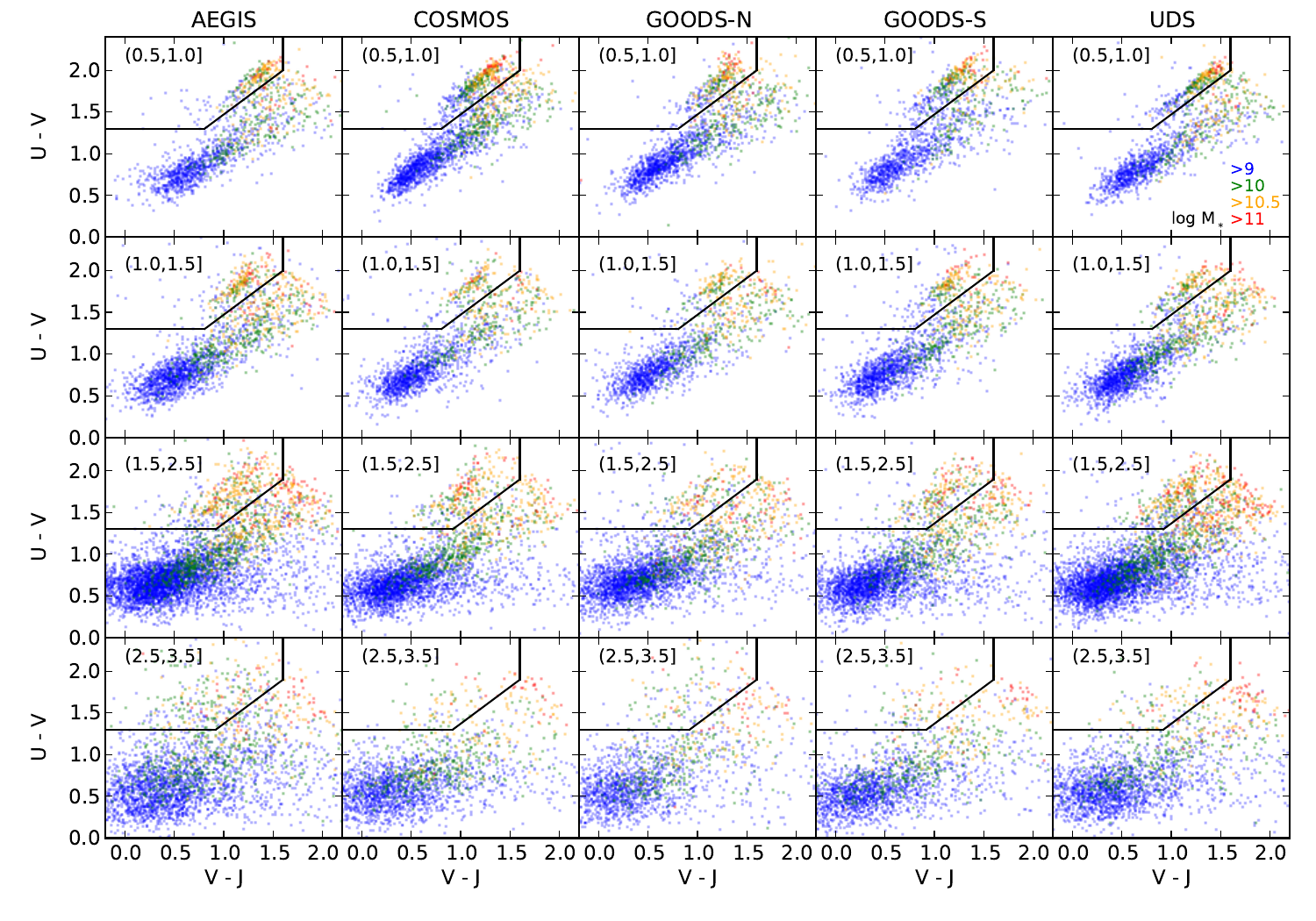}
\caption{Rest-frame $U-V$ vs $V-J$ colors. Each column presents the rest-frame colors for galaxies in one field, with redshift increasing from the upper to the lower panels. The redshift limits for each bin are given in the top left corner of each panel.  Galaxies are selected with a use\_phot flag = 1 and color-coded by mass, with the lowest mass galaxies in blue and the highest mass galaxies in red, as shown by the legend in the top right panel. The black lines indicate the selection box used to separate quiescent from star-forming galaxies, based on \citet{Williams09}.}\label{fig:rfuvj}
\end{figure*}

In order to assess the quality of the rest-frame photometry we show the "UVJ"
diagram of galaxies in Fig.\ \ref{fig:rfuvj}. This diagram shows the rest-frame $U-V$
color versus the rest-frame $V-J$ color. Each column represents one of the fields, with redshift increasing from top to bottom, as shown in the top left corner of each panel. Galaxies (selected with use\_phot = 1) are color-coded by mass, with the most massive galaxies in red ($\log M_* > 11$), galaxies with $10.5 < \log M_* <11$ in orange, $10 < \log M_* <10.5$ in green and $9 < \log M_* < 10$ in blue. The black lines mark the selection that is typically used to distinguish star-forming and quiescent galaxies \citep{Williams09, Whitaker11}. In this space, quiescent galaxies with low levels of star formation that are red in $U-V$ (upper left region) are separated from similarly red (in $U-V$), dusty star-forming galaxies, with the star-forming galaxies having redder $V-J$ colors.  A clear progression of increasing mass toward redder colors can be seen along the quiescent galaxy sequence, particularly in the lowest redshift bins. The majority of low mass galaxies lie in the star-forming ``blue cloud'' at all redshifts. In the highest redshift bin ($ 2.5 < z \le 3.5$) the most massive galaxies lie within the star-forming region and appear to be red due to higher levels of dust rather than older stellar populations. 

In future papers we will interpret the distribution of galaxies in this color-color plane; here we
simply note that
the $UVJ$ distributions are largely consistent from field to field.  The only exception is the quiescent galaxy sequence in the UDS at $z\le 1$, where the relation is tighter and the median somewhat bluer than the other fields. This is likely to be caused by the ground-based $u$-band data, where we found that a large zero point offset was required to make it consistent with the other bands (see Appendix~\ref{app:zps}).

\subsection{Stellar Population Parameters}
\label{sec:fast}

We use the FAST code \citep{Kriek09} to estimate the stellar masses, star formation rates, ages and and dust extinctions, given the spectroscopic redshift, where available, or the photometric redshift from EAZY  \texttt{z\_peak} otherwise. We use the \citet{BC03} stellar population synthesis model library with a \citet{Chabrier03} IMF and solar metallicity. We assume exponentially declining star formation histories with a minimum e-folding time of $\log_{10}(\tau/yr) = 7$, a minimum age of 40 Myr, $0 < A_V < 4$~mag and the \citet{Calzetti00} dust attenuation law. The stellar population parameters are provided in separate catalogs for each field. We stress that the star formation rates, dust absorption, and star formation histories
of the galaxies are uncertain when they are derived solely from optical -- near-IR photometry
(see, e.g., \citealt{Wuyts12}). By contrast, stellar masses and $M/L$ ratios
are relatively well-constrained as they mostly depend on the rest-frame optical colors of
the galaxies, and these
are well-covered by our photometry.

\begin{figure*}[!ht]
\includegraphics[width=0.98\textwidth]{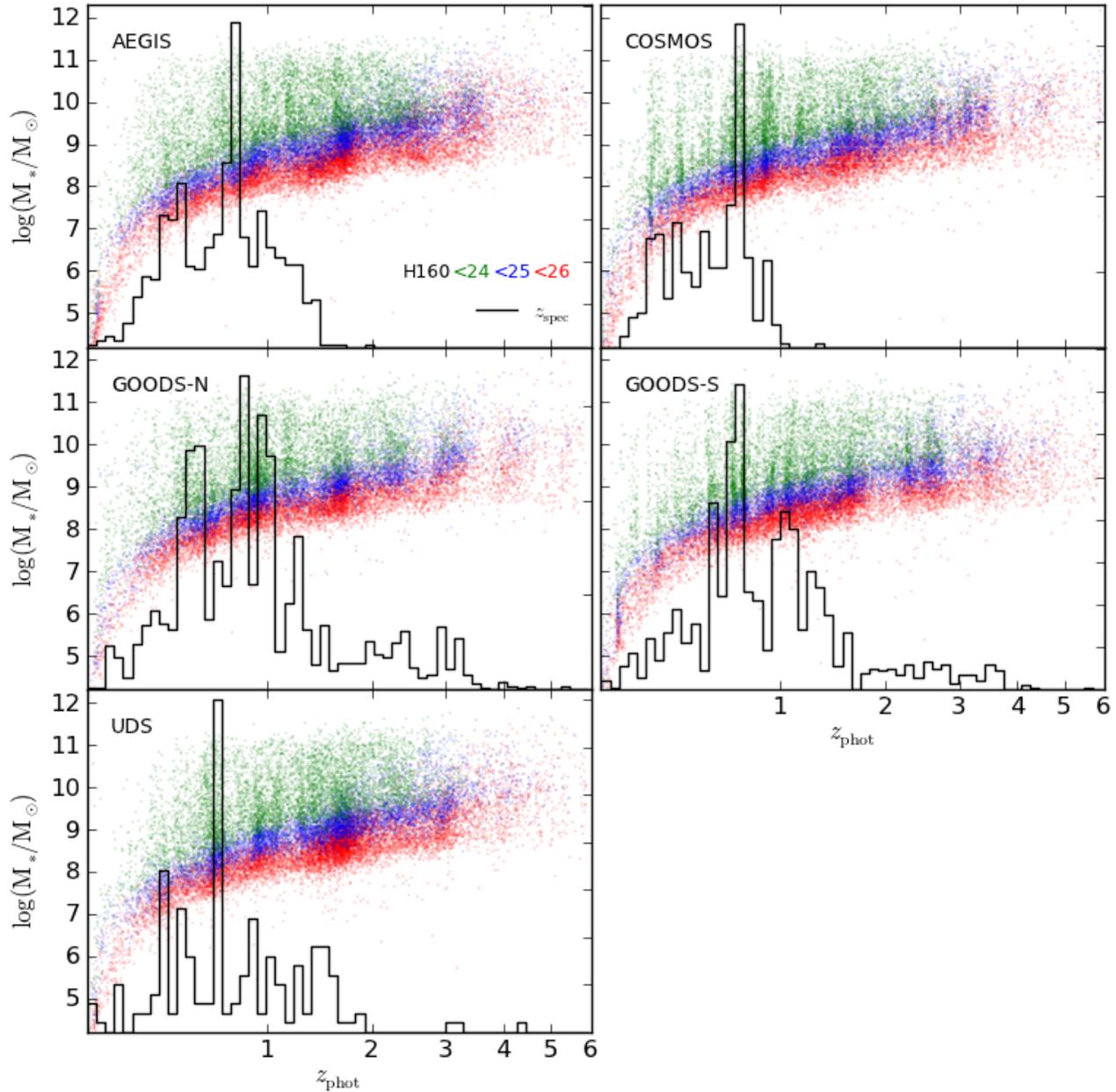} 

\caption{Stellar mass versus photometric redshift. The points are color-coded by magnitude such that galaxies with $H_{F160W} < 24$ are in green, $24 \le H_{F160W} < 25$ in blue and $25 \le H_{F160W} < 26$ in red. The gray histogram shows the distribution of spectroscopic redshifts from the literature, arbitrarily scaled. Many of the over densities in photometric redshift correspond to peaks in the spectroscopic redshift distribution. }\label{fig:massz}
\end{figure*}

In Fig.~\ref{fig:massz} we show the distribution of galaxy stellar masses with photometric redshift (\texttt{z\_peak}). The points are color-coded according to the galaxy's $H_{\mathrm{F160W}}$ magnitude, with the brightest galaxies in green ($H_{\mathrm{F160W}} < 24$), galaxies with $24 \le H_{F160W} < 25$ in blue and $25 \le H_{F160W} < 26$ in red. The histograms show the distribution of spectroscopic redshifts in each field (arbitrarily normalized). Many of the overdensities that can be seen in the photometric redshift distribution correspond to peaks in the distributions of already known spectroscopic redshifts, but extend to lower mass (fainter) galaxies than it is possible to measure spectroscopic redshifts for from the ground.

\section{Summary}

In this paper we have presented
the images and multi-wavelength photometric catalogs produced by the 3D-HST project for the five CANDELS/3D-HST extragalactic fields. The survey covers $\sim900$ arcmin$^2$ in the AEGIS, COSMOS, GOODS-North, GOODS-South and UDS fields with HST/WFC3 imaging and grism spectroscopy. The details of the WFC3 image reduction are given in \S\,~\ref{sec:wfc3reduction}. In addition to the new WFC3 data, we incorporated much of the available ground-based, Spitzer and HST/ACS data into the catalogs, using a total of 147 distinct data sets (see \S\,\ref{sec:otherdata} and Table~\ref{table:ancil_data}). We make all the images that have been used available on our website together with the catalogs. Each of the images is on the same astrometric system as the CANDELS WFC3 mosaics.

We have applied consistent methodology to produce multi-wavelength catalogs for all five of the fields.
The SExtractor software \citep{Bertin96} was used to detect sources on a noise-equalized combination of the F125W, F140W and F160W images. By using all three WFC3 bands, we exploited the maximum survey area and depth. 
As described in detail in \S\,\ref{sec:phot}, we measured the SEDs of objects
using
all the available ancillary data, carefully taking into account differences in image resolution (see \S\,\ref{sec:hstphot} and \ref{sec:lowresphot}). The results are consistent for all five of the fields and the total WFC3 magnitudes agree well with independently derived total magnitudes from morphological fitting (see \S\,\ref{sec:tests}). The resulting SEDs span from the $U$-band to 8$\mu$m and are of excellent quality, as demonstrated throughout the paper.  We used the EAZY code \citep{Brammer08} to fit photometric redshifts and reach an NMAD scatter between the photometric and spectroscopic redshifts of $\lt 2.7\%$ with fewer than 5\% significant outliers in all fields. In the two fields where there is good medium band coverage (COSMOS and GOODS-S), the scatter is $\le 1\%$. We provide rest-frame colors based on the best-fitting EAZY templates, as well as stellar masses and stellar population parameters for all the galaxies based on fits to their SEDs.

The CANDELS team has provided similar catalogs for two of the five fields discussed in this paper
\citep{Guo13, Galametz13}, and we can expect future CANDELS releases of the
other three fields. Our catalogs are complementary to these; we use slightly deeper detection
images and a larger number of photometric filters, but these differences are probably not critical
for most purposes. It will be very useful to have multiple "realizations" of the CANDELS datasets
in the public domain, using independent reductions and methodology. In an Appendix we show
band-by-band differences between the CANDELS catalogs and ours, and such comparisons
provide much-needed estimates of systematic uncertainties in the various catalogs.

As explained in the Introduction, in the context of the 3D-HST project, the work described in this
paper merely concludes the first phase of an even more ambitious undertaking. The most innovative
aspect of 3D-HST is the grism spectroscopy; future papers will describe these data and will
quantify how they improve the measurement of redshifts, masses, and other parameters.
\\
\\
\par
We are grateful to the many colleagues who have provided public data and catalogs in the
five fields described in this paper; high redshift galaxy science has thrived owing to this
gracious mindset and the TACs and Observatory Directors who have encouraged this. We would like to thank the referee for a careful reading of the paper and useful suggestions.
Formal acknowledgements follow below.
RS acknowledges the support of the South African National Research Foundation through the Professional Development Programme Postdoctoral Fellowship. 
KW acknowledges support by an appointment to the NASA
Postdoctoral Program at the Goddard Space Flight Center,
administered by Oak Ridge Associated Universities through a
contract with the National Aeronautics and Space Administration (NASA).
DM acknowledges the support of the Research Corporation for Science Advancements Cottrell Scholarship. BL acknowledges support from the NSF Astronomy and Astrophsics Fellowship grant AST-1202963. We acknowledge support from ERC Advanced Grant HIGHZ\#227749, and a NWO Spinoza Grant.
We thank the Lorentz Center for it support and hospitality.

This work is based on observations made with the NASA/ESA Hubble Space Telescope, obtained from the MAST Data Archive at the Space Telescope Science Institute, which is operated by the Association of Universities for Research in Astronomy, Inc., under NASA contract NAS 5-26555. Observations associated with the following GO and GTO programs were used: 12063, 12440, 12442, 12443, 12444, 12445, 12060, 12061, 12062, 12064 (PI: Faber); 12177 and 12328 (PI: van Dokkum); 12461 and 12099 (PI: Riess); 11600 (PI: Weiner); 9425 and 9583 (PI: Giavalisco); 12190 (PI: Koekemoer); 11359 and 11360 (PI: OÕConnell); 11563 (PI: Illingworth).
Based, in part, on data obtained at the W. M. Keck Observatory, which is operated as a scientific partnership among the California Institute of Technology, the University of California, and NASA and was made possible by the generous financial support of the W. M. Keck Foundation. 
This work makes use of data obtained as part of the ESO/GOODS survey. 
Observations have been carried out using the Very Large Telescope at the ESO Paranal Observatory under Programme ID 168.A-0485. 
Based on observations made with ESO Telescopes at the La Silla or Paranal Observatories under programme number 168.A-0485.
Based on data products from observations made with ESO Telescopes at the La Silla Paranal Observatory under ESO programme ID 179.A-2005 and
on data products produced by TERAPIX and the Cambridge Astronomy Survey Unit on behalf of the UltraVISTA consortium. 
Based on observations obtained with MegaPrime/MegaCam, a joint project of CFHT and CEA/IRFU, at the Canada-France-Hawaii Telescope (CFHT) which is operated by the National Research Council (NRC) of Canada, the Institut National des Science de l'Univers of the Centre National de la Recherche Scientifique (CNRS) of France, and the University of Hawaii. 
This work is based in part on data products produced at Terapix available at the Canadian Astronomy Data Centre as part of the Canada-France-Hawaii Telescope Legacy Survey, a collaborative project of NRC and CNRS.
This work makes use of data products produced at TERAPIX, the WIRDS consortium, and the Canadian Astronomy Data Centre.  
We thank H. Hildebrandt for providing the CARS-reduced CFHTLS images.  
This research has made use of the NASA/IPAC Infrared Science Archive, which is operated by the Jet Propulsion Laboratory, California Institute of Technology, under contract with the National Aeronautics and Space Administration.
This study makes use of data from AEGIS, a multiwavelength sky survey conducted with the Chandra, GALEX, Hubble, Keck, CFHT, MMT, Subaru, Palomar, Spitzer, VLA, and other telescopes and supported in part by the NSF, NASA, and the STFC.
This study makes use of data from the NEWFIRM Medium-Band Survey, a multi-wavelength survey conducted with the NEWFIRM instrument at the KPNO, supported in part by the NSF and NASA.
This paper uses data products produced by the OIR Telescope Data Center, supported by the Smithsonian Astrophysical Observatory. 
This work is based on observations made with the Spitzer Space Telescope, which is operated by the Jet Propulsion Laboratory, California Institute of Technology under contract with NASA. 
Based in part on data from the UKIDSS Ultra Deep Survey (UDS) and the Spitzer Public Legacy Survey of the UKIDSS Ultra Deep Survey (SpUDS), a Cycle 4 Spitzer Legacy program. 
\vspace{0.2cm}\\
Author contributions: RS and KW led this part of the 3D-HST project
and created the photometric catalogs described in this paper. RS, IM and PvD wrote the manuscript.
IM is the 3D-HST project lead, reduced all the WFC3 imaging in the CANDELS fields,
and was responsible for the integration
of the data products. GB and IL developed many of the algorithms and techniques
used in the creation of the catalogs. PvD is the PI of 3D-HST. The other authors provided help
at various stages with the project and/or commented on the manuscript.

\bibliographystyle{apj_url}
\bibliography{references}

\begin{appendix}

\section{Point Spread Functions}
\label{app:psfs}

\begin{figure*}[!ht]
\begin{minipage}[b]{0.5\linewidth}{
\includegraphics[width=\textwidth]{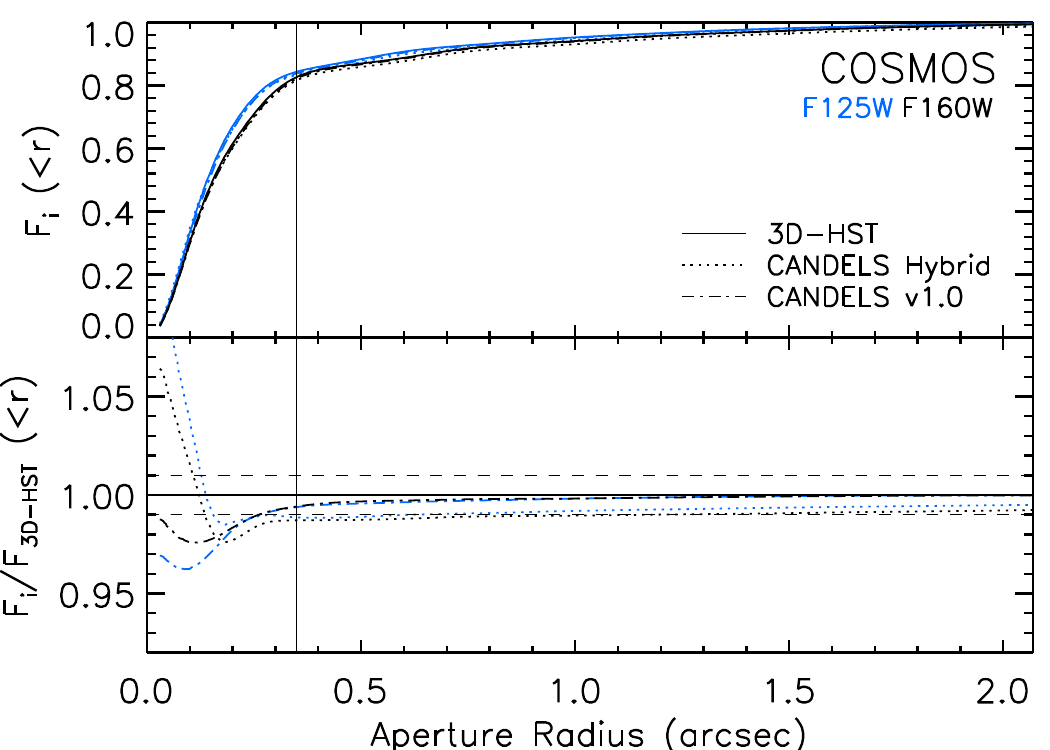}
\label{fig:subfig1}}
\end{minipage}
\begin{minipage}[b]{0.5\linewidth}{
\includegraphics[width=\textwidth]{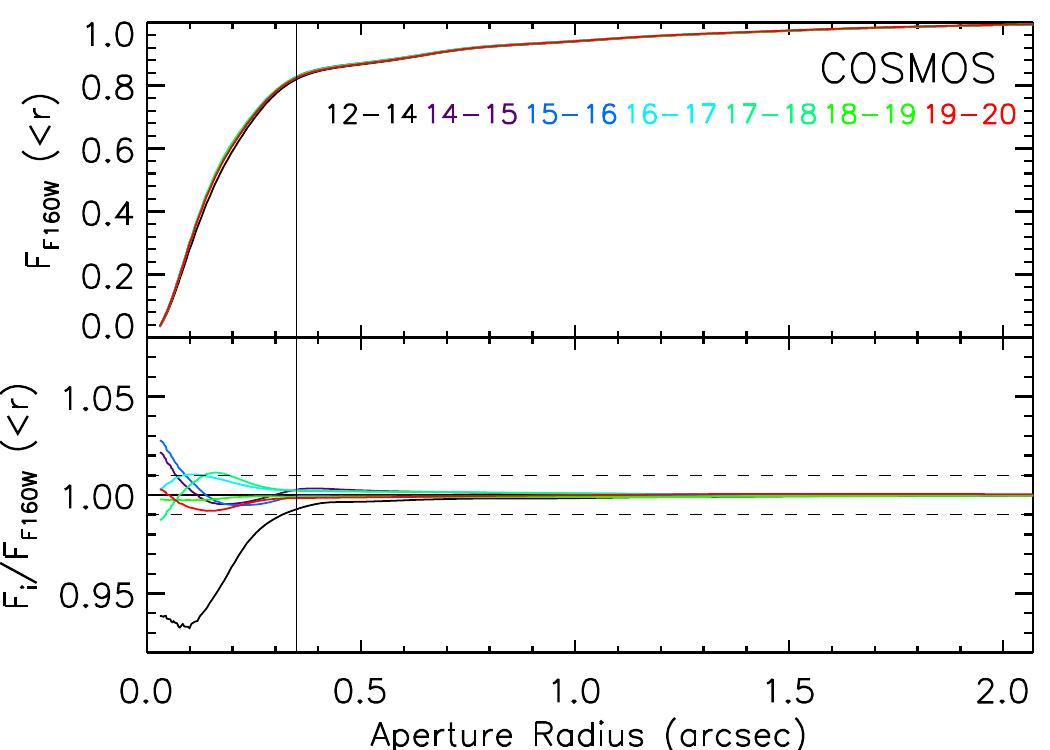}
\label{fig:subfig2}}
\end{minipage}
\caption{Left panels: A comparison of the F160W (black lines) and F125W (blue lines) growth curves from three PSFs for the COSMOS field. The solid lines show the 3D-HST PSFs used in this analysis, the dash-dotted line a PSF made in the same way using the CANDELS v1.0 mosaic, and the dotted line a "hybrid" PSF used for the GALFIT morphological measurements \citep{vanderWel12}. The hybrid PSF combines an artificial PSF created with Tiny Tim in the inner region with an empirical PSF from a stack of stars in the outer regions. Right panels: A comparison of the growth curves from PSFs created in the COSMOS field using stars in narrow magnitude bins (color-coded as shown in the legend) rather than the full magnitude range. There are very small differences between the PSFs. Saturation affects only the brightest stars, leading to a larger difference in the PSF of stars with $12 < H_{\mathrm{160W}} < 14$. At the aperture used for photometry ($0\farcs 35$) the differences between the PSFs in all the magnitude bins are smaller than 1\%.}
\label{fig:psf_comp}
\end{figure*}

In the left-hand panel of Fig.~\ref{fig:psf_comp} we compare the growth curves of the F125W and F160W PSFs obtained using the same method (described in \S\,\ref{sec:psfmatching}) from the 3D-HST (solid lines) and CANDELS v1.0 mosaics (dash-dot lines) to the hybrid PSF used for morphological fitting in \citet{vanderWel12} (dotted lines) in the COSMOS field. The three PSFs in each band are very similar over the full range of apertures, with differences of $\la 1\%$ at apertures $\gt 0\farcs5$. In the inner regions the hybrid PSF has larger flux than the 3D-HST PSF, indicating that we may still be missing the some flux due to the cosmic ray rejection. The CANDELS PSF has lower flux than the 3D-HST PSF in the cores. In the right hand panel we show the growth curves for PSFs created using stars in bins of 1 magnitude in the COSMOS field. There is excellent agreement across the full range in magnitude, with only the very brightest stars ($H_{F160W} < 14$) having significantly less flux within the central region. Tests on the other fields yield similar results.

\begin{figure*}[ht]
\centering
\begin{minipage}{0.48\linewidth}{
\includegraphics[width=\textwidth]{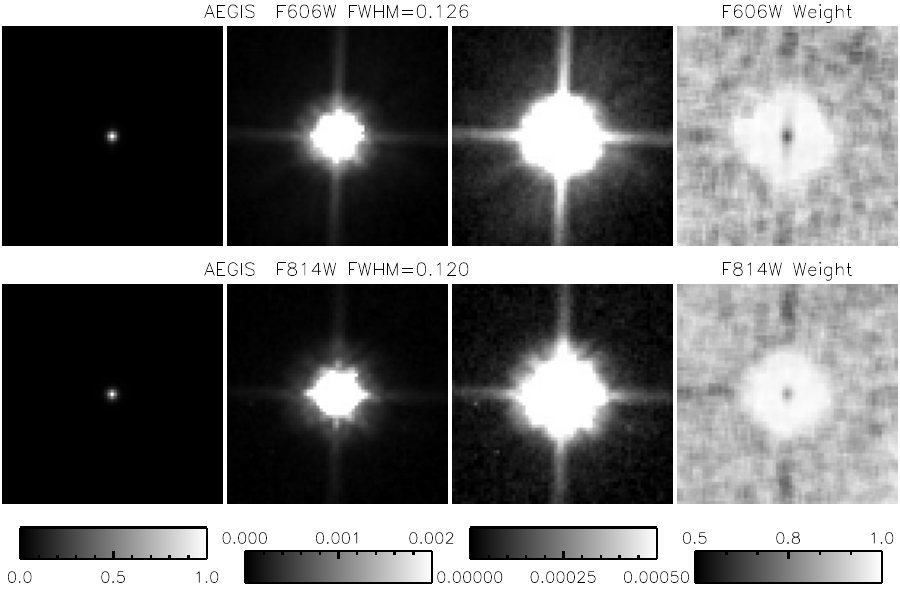}}
\end{minipage}
\begin{minipage}[c]{0.48\linewidth}{
\includegraphics[width=\textwidth]{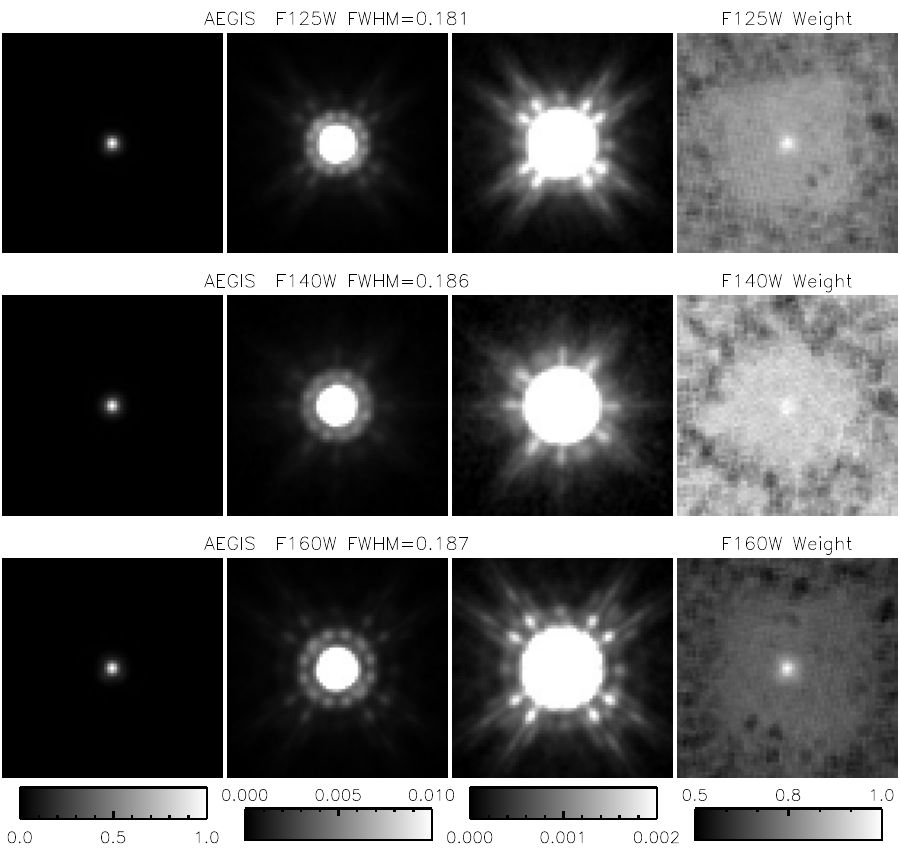}}
\end{minipage}
\caption{Point-spread functions (PSFs) for the ACS F606W and F814W bands (left 4 panels) and the WFC3 F125W, F140W, and F160W bands (right 4 panels) in the AEGIS field. The construction of the PSFs is described in \S\ref{sec:psfmatching}. Each image is $69\times69$ pixels or $4\farcs14\times4\farcs14$, i.e. it traces the PSF out to just over 2\arcsec radius. For each filter we show three stretch levels (panels 1 to 3 and 5 to 7) to expose the structure of the PSF: the core, the first Airy ring and the diffraction spikes. The images are normalized to a maximum value of one. The grayscale bars show the stretch for each panel. These are slightly different for ACS and WFC3 as a result of the different FWHMs (listed above the images). We also show the combined weight images for each PSF. The weight is largest in the center and lower at larger radii due to masking of neighboring objects. The ACS PSFs have lower weights in the central pixels because of cosmic ray rejection flagging the centers of stars.}
\label{fig:aegis_psfs}
\end{figure*}

\begin{figure*}[ht]
\centering
\begin{minipage}{0.48\linewidth}{
\includegraphics[width=\textwidth]{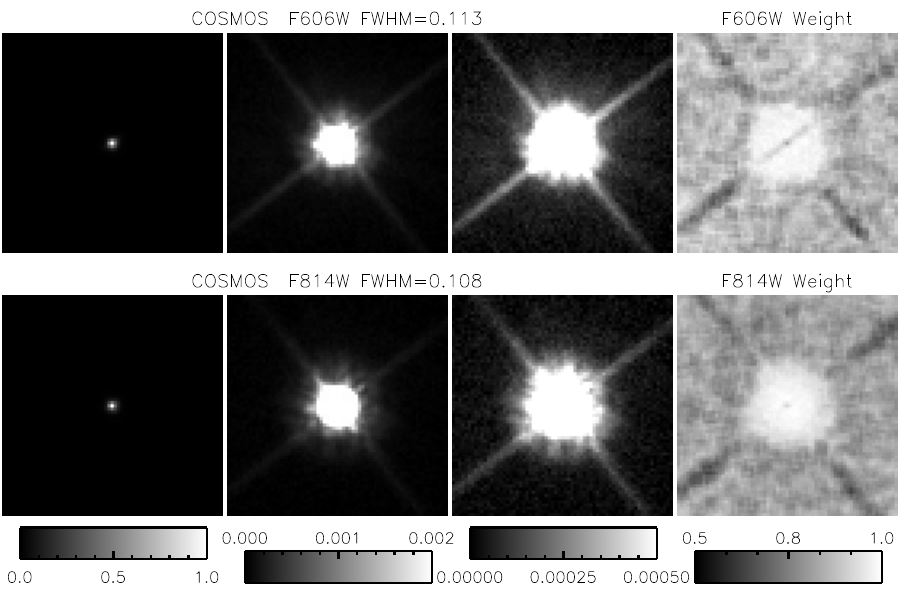}
\label{fig:subfig1}}
\end{minipage}
\begin{minipage}[c]{0.48\linewidth}{
\includegraphics[width=\textwidth]{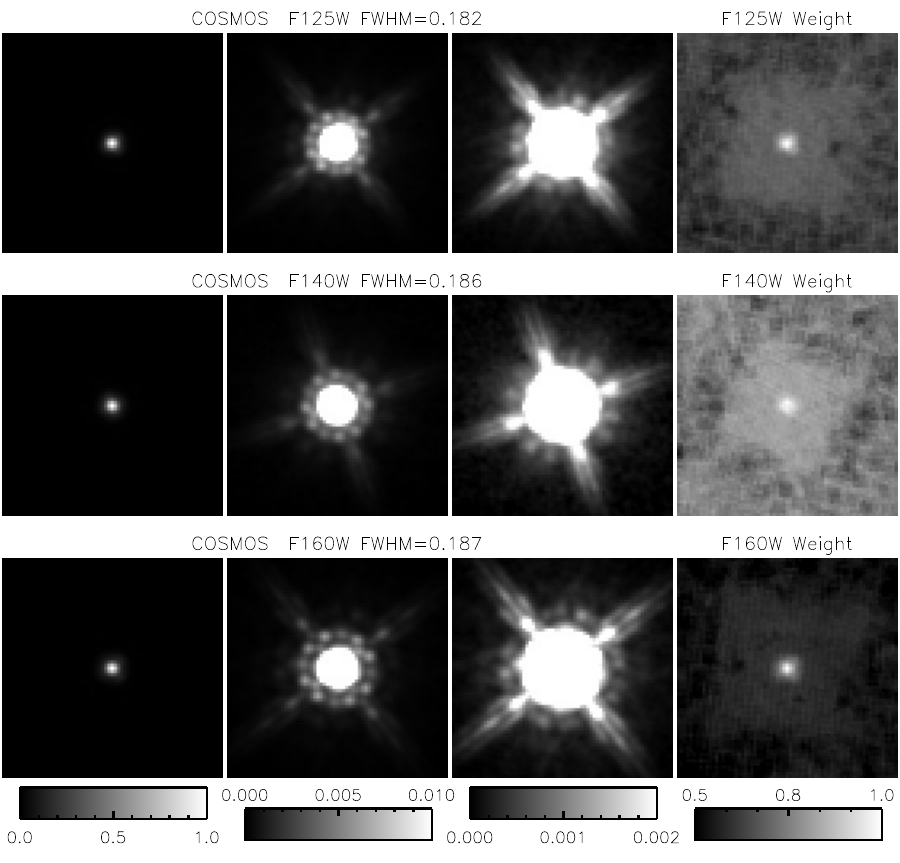}
\label{fig:subfig2}}
\end{minipage}
\caption{Same as Figure~\ref{fig:aegis_psfs} for the ACS and WFC3 PSFs in the COSMOS field.}
\end{figure*}

In Figures~\ref{fig:aegis_psfs} to \ref{fig:uds_psfs} we show the thumbnails of the PSFs and corresponding weight images for all the HST bands in all five fields.  Each image is $69\times69$ pixels or $4\farcs14\times4\farcs14$, i.e. it traces the PSF out to just over 2\arcsec radius. For each filter we show three stretch levels (panels 1 to 3 and 5 to 7) to expose the structure of the PSF: the core, the first Airy ring and the diffraction spikes. The images are normalized to a maximum value of one. The grayscale bars show the stretch for each panel. These are slightly different for ACS and WFC3 as a result of the different FWHMs (listed above the images). We also show the combined weight images for each PSF. The weight is largest in the center and lower at larger radii due to masking of neighboring objects. The ACS PSFs have lower weights in the central pixels because of cosmic ray rejection flagging the centers of stars.

\begin{figure*}[ht]
\centering
\begin{minipage}{0.48\linewidth}{
\includegraphics[width=\textwidth]{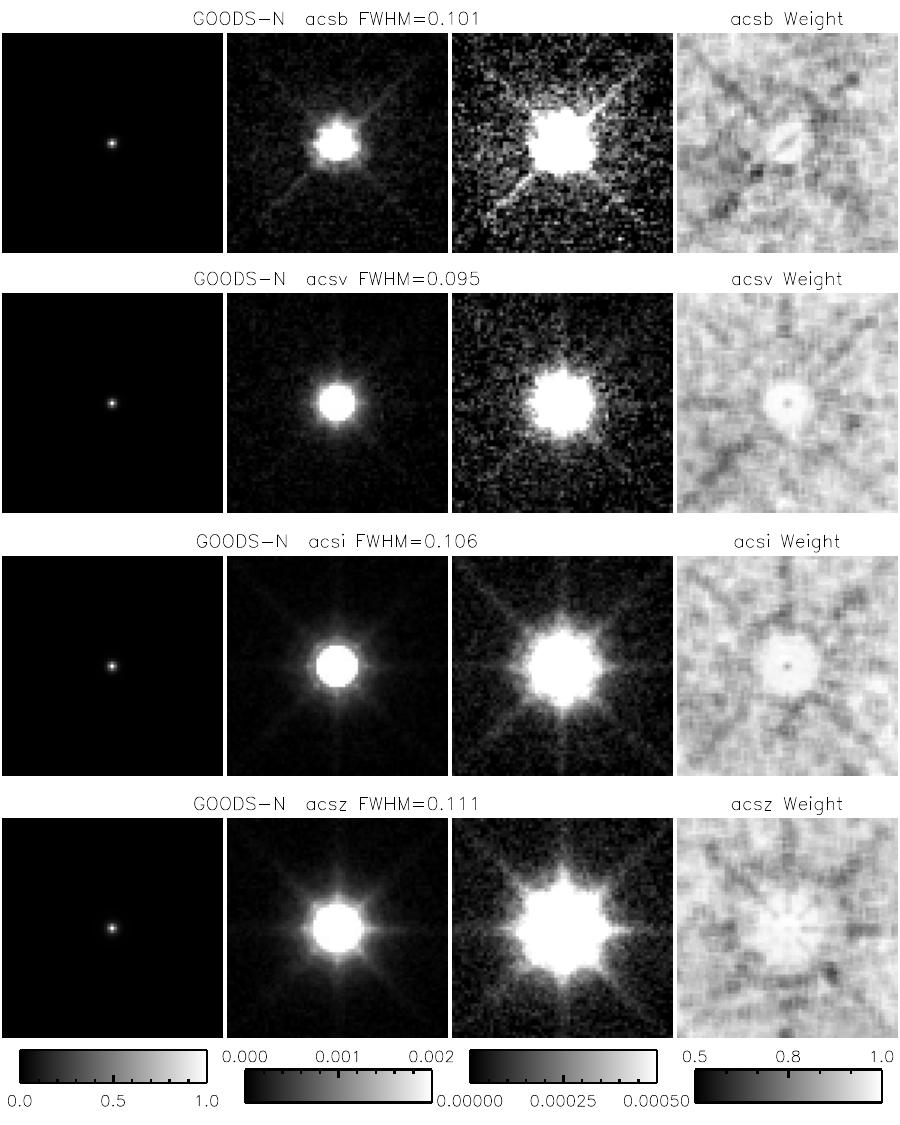}
\label{fig:subfig1}}
\end{minipage}
\begin{minipage}[c]{0.48\linewidth}{
\includegraphics[width=\textwidth]{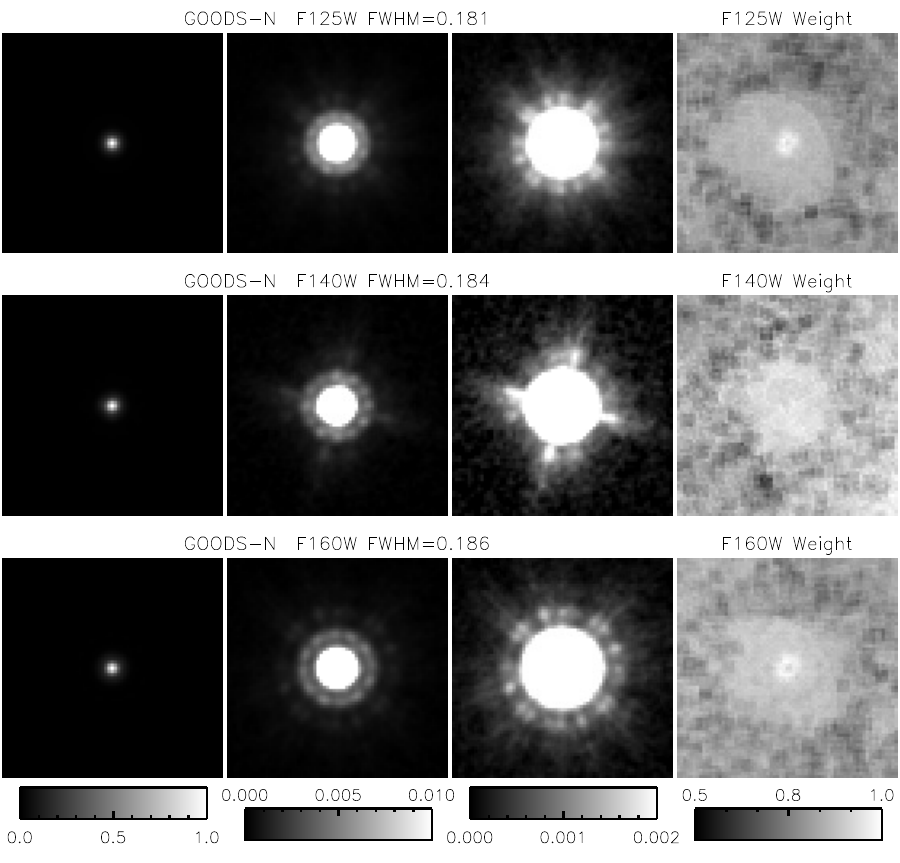}
\label{fig:subfig2}}
\end{minipage}
\caption{Same as Figure~\ref{fig:aegis_psfs} for the ACS and WFC3 PSFs in the GOODS-N field.}
\end{figure*}

\begin{figure*}[ht]
\centering
\begin{minipage}{0.48\linewidth}{
\includegraphics[width=\textwidth]{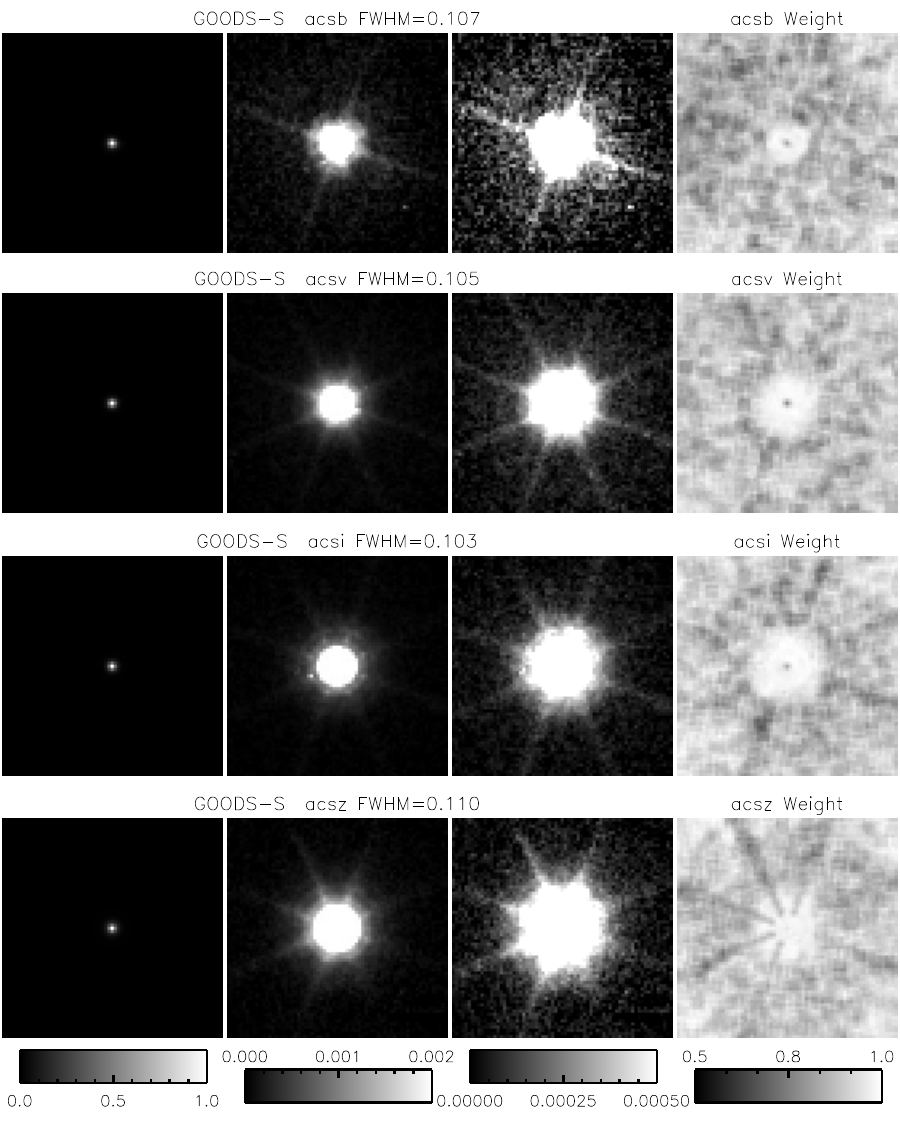}
\includegraphics[width=\textwidth]{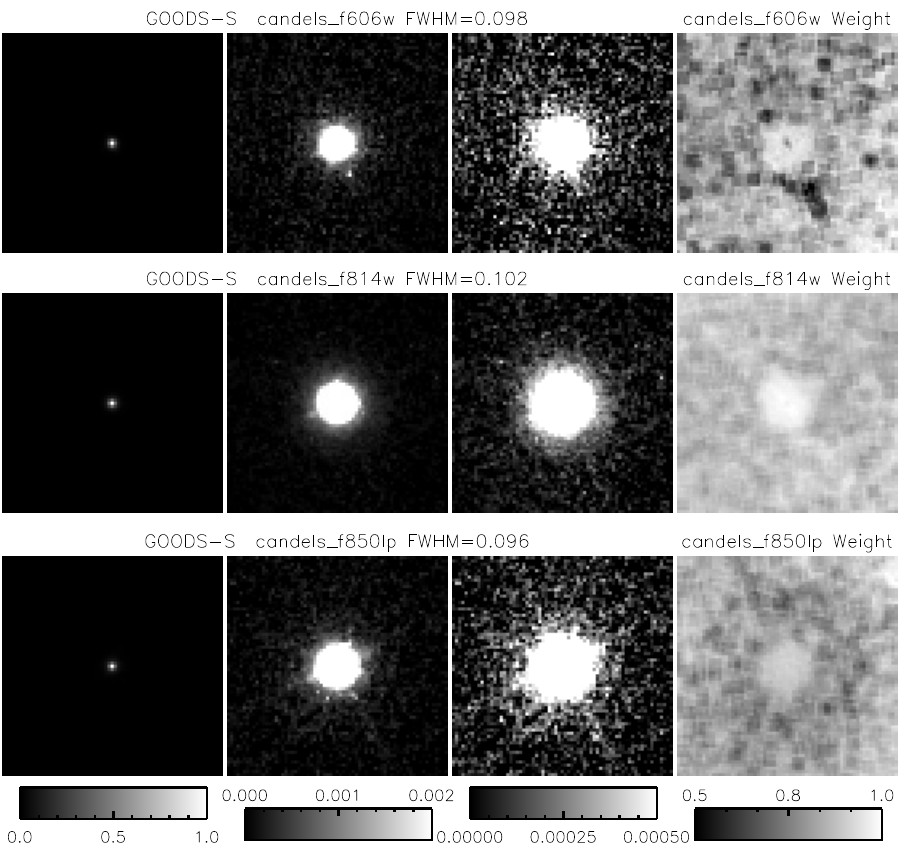}
\label{fig:subfig1}}
\end{minipage}
\begin{minipage}[c]{0.48\linewidth}{
\includegraphics[width=\textwidth]{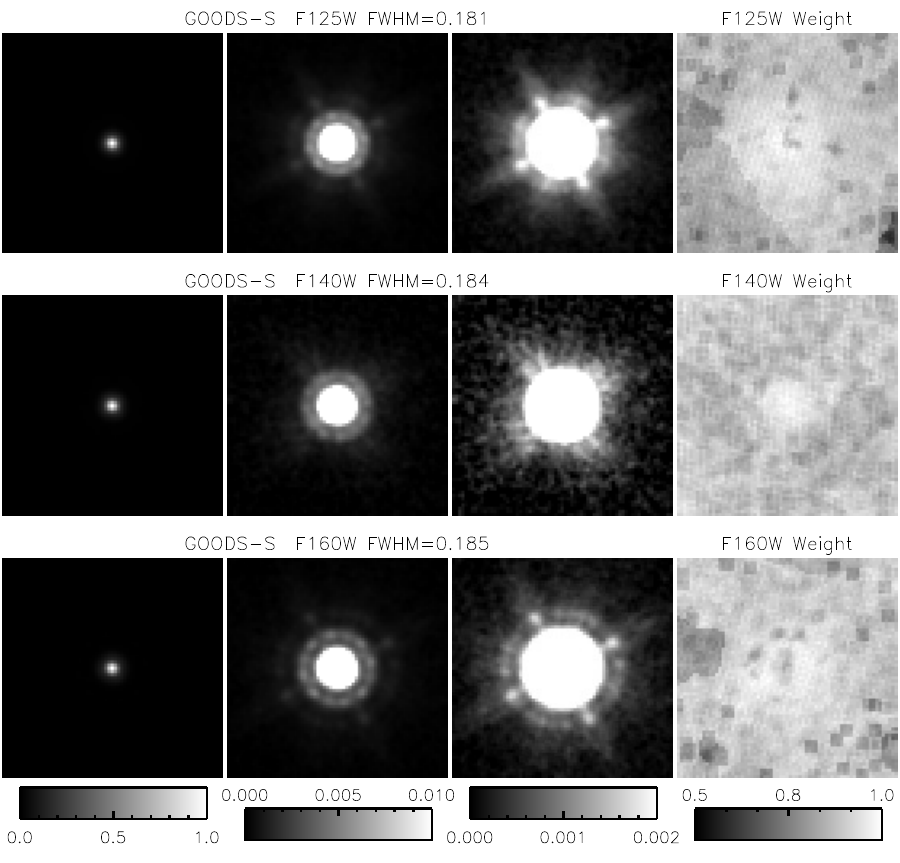}
\label{fig:subfig2}}
\end{minipage}
\caption{Same as Figure~\ref{fig:aegis_psfs} for the ACS and WFC3 PSFs in the GOODS-S field. We make use of two different sets of ACS images in this field. The PSFs for the images from the GOODS Survey are shown in the top four rows (left). The PSFs for the images from the CANDELS Survey are shown in the bottom three rows.}
\end{figure*}

\begin{figure*}[ht]
\centering
\begin{minipage}{0.48\linewidth}{
\includegraphics[width=\textwidth]{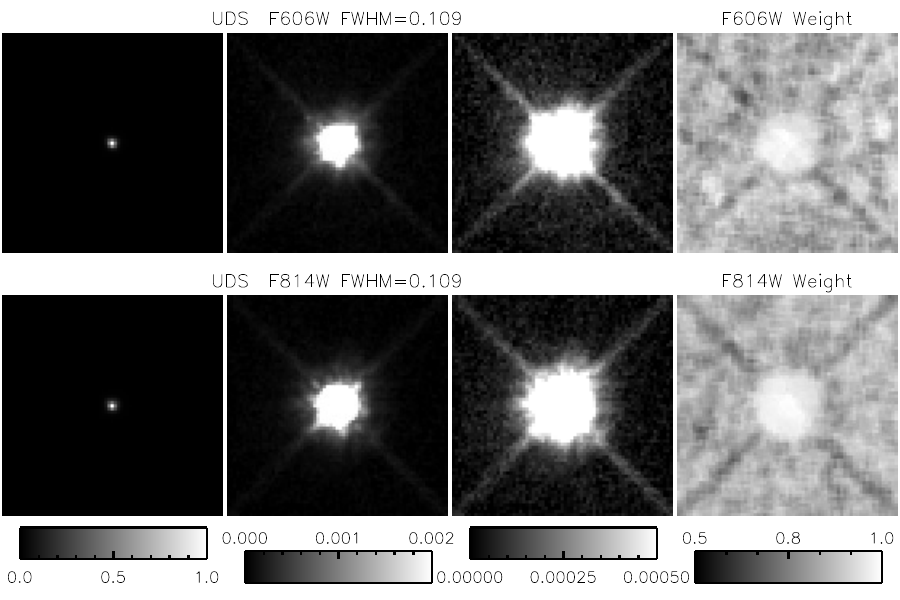}
\label{fig:subfig1}}
\end{minipage}
\begin{minipage}[c]{0.48\linewidth}{
\includegraphics[width=\textwidth]{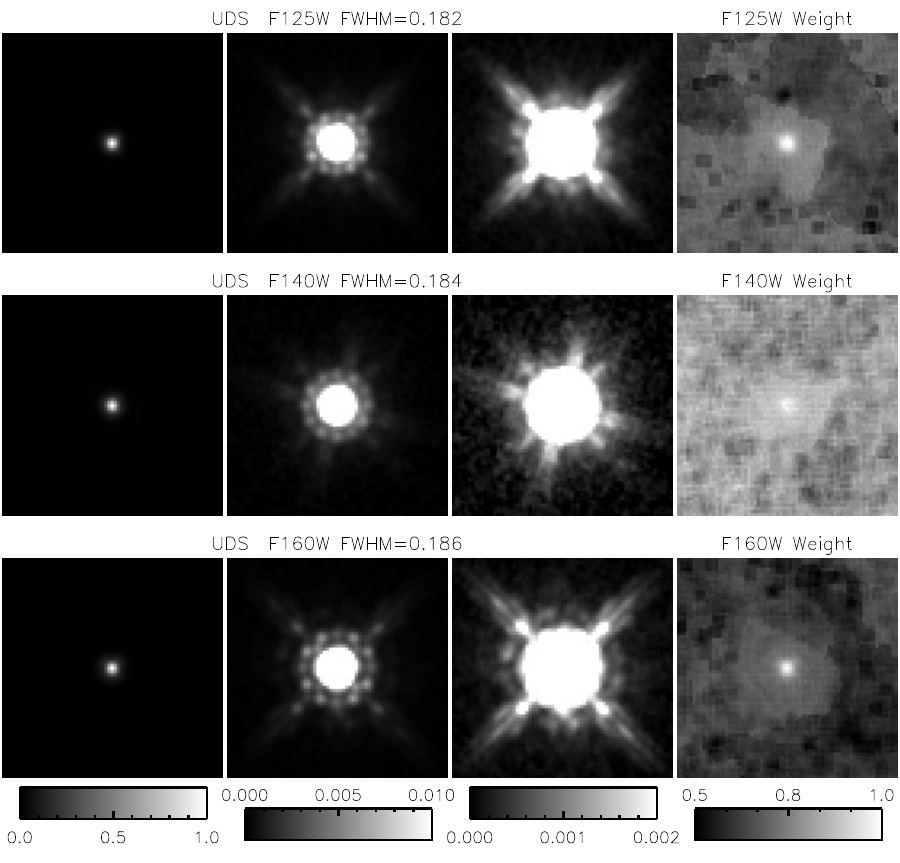}
\label{fig:subfig2}}
\end{minipage}
\caption{Same as Figure~\ref{fig:uds_psfs} for the ACS and WFC3 PSFs in the UDS field.}\label{fig:uds_psfs}
\end{figure*}

\section{Zero Point Offsets From SED-fitting}
\label{app:zps}

When fitting photometric redshifts, we apply two corrections: we modify the templates and we modify the
photometric zero points. These two corrections are separable, as the template correction is derived
in the rest-frame and the zero point correction is derived in the observed frame. As we have many
objects and many filters we can robustly determine the required corrections. In detail, we
shift
the observed and best-fit SEDs to the rest-frame and examine differences between the two for a large number of objects as a function of wavelength. We then improve the templates to include subtle features that are not included in the models, such as the broad dust absorption feature at 2175\AA. Since the galaxies span a wide range of redshifts, shifting to the rest-frame ensures that each part of the spectrum is sampled by a number of different photometric bands. This allows one to disentangle template effects from systematic offsets between the photometric bands. After adjusting the templates we fit for a photometric zero point offset for each filter from the residuals in an iterative fashion.  We largely follow the procedure described in \citet{Whitaker11} and other works, but with two significant differences. We now include all objects in the fit rather than just galaxies that have spectroscopic redshifts. By including all objects, we avoid the bias toward lower redshift
star-forming galaxies which dominate the sample with spectroscopic redshifts and are able to obtain good estimates of the zero point offsets even in fields with limited numbers of spectroscopic redshifts. Additionally we follow a two-step process to account for the relatively large zero point uncertainties in the ground-based data compared to the well-calibrated \textit{HST} filters. In the first iteration we allow the \textit{HST} bands to vary with respect to each other, keeping $H_{F160W}$ as a fixed reference point. The \textit{HST} band zero points are then fixed and the ground-based and IRAC band zero points are allowed to vary until convergence is reached.  Convergence is defined as having the largest change in any band other than $U$ or IRAC be less than 0.5\%, and this condition is usually met after only three or four iterations of the procedure.

We list the offsets applied to the fluxes in each field in Tables~\ref{table:zp_offsets1} and \ref{table:zp_offsets2}. The zero point offsets for the \textit{HST} bands are generally small, of the order of 0.01~mag. It is hardest to separate template and photometric zero point errors where there are no bracketing filters at the bluest and reddest ends of the spectra, sampled by the $U$-band and IRAC 8~$\mu$m, respectively. The COSMOS CFHTLS $u$-band zero point is known to be highly uncertain \citep{Erben09, Whitaker11} as is the UDS $u'$-band (R. Quadri, private communication). As a result, the zero point offsets applied in the $U$-band can be as large as 0.25~mag. The IRAC offsets are $\leq 0.15$~mag in the 8~$\mu$m band and generally much smaller in the other channels. The GOODS-N optical Subaru data are offset from the other optical data by $\sim$0.2 mag. Some of the optical medium bands with large FWHM in GOODS-S were found to have particularly large offsets and were excluded from further analysis (IA464, IA484, IA709, IA827). These bands are not included in the released catalogs.

In the COSMOS field there appears to be a systematic difference between the ground-based and the space-based NIR data, with an average offset of 0.1~mag. Similar offsets are found in the AEGIS J-bands.  The $K$-bands from the NMBS, WIRDS and UltraVISTA are known to have differences in the calibrated zero point (see the discussion in the Appendix of \citealt{Muzzin13a}). In both NMBS and UltraVISTA, the difference between the measured fluxes for NMBS $K$ and WIRDS $K_s$ was found to be $\sim0.03$~mag. In the NMBS COSMOS catalog a zero point offset of 0.05~mag is applied to the WIRDS $K_s$-band to bring it into agreement with the NMBS $K$-band. \citet{Muzzin13a} find a difference a difference of 0.08~mag between the NMBS $K$ and UltraVISTA $K_s$, such that the NMBS fluxes are brighter, and choose to correct the UltraVISTA fluxes to be consistent with the NMBS. We find very similar differences between the zero point corrections for the three $K$-bands ($\Delta{K_{\mathrm{NMBS}} - K_{s \mathrm{, WIRDS}}} = 0.05$~mag, $\Delta{K_{\mathrm{NMBS}} - K_{s \mathrm{, UVISTA}}} = 0.07$~mag). We have shifted all three bands fainter to agree with the (fainter) $HST$ reference system, however. 

\begin{table}[ht]
\centering
\caption{Zero point offsets applied to the v4.1 catalogs}
\label{table:zp_offsets1}
\begin{tabular}{llcc c llcc}
\cline{1-4} \cline{6-9}
\cline{1-4} \cline{6-9}
\noalign{\smallskip}
Field & Band & Flux correction & Magnitude offset &  & Field & Band & Flux correction & Magnitude offset  \\
\cline{1-4}\cline{6-9}
\noalign{\smallskip}
AEGIS  
 &    U& 1.2258&-0.22 & & COSMOS &    U& 1.1624&-0.16\\
 &    G& 1.0076&-0.01 & &  &   B& 0.9747& 0.03\\
& F606W& 0.9999& 0.00 & & &    G& 0.9436& 0.06\\
&     R& 0.9594& 0.05 & &   &  V& 0.8055& 0.23\\
&     I& 0.9065& 0.11 & & & F606W& 1.0109&-0.01\\
& F814W& 0.9353& 0.07 & & &    Rp& 0.8305& 0.20\\
&    Z& 0.8980& 0.12 & &    & R& 0.9239& 0.09\\
&    J1& 0.8568& 0.17& & &    Ip& 1.2181&0.12\\
&    J2& 0.8749& 0.15 & & &    I& 0.8799& 0.14\\
&    J3& 0.8863& 0.13 & & & F814W& 0.9658& 0.04\\
&     J& 1.0575&-0.06 & & &  Z& 0.8644& 0.16\\ 
& F125W& 0.9966& 0.00 & & &  Zp& 0.8232& 0.21\\
&    H1& 0.8899& 0.13 & & & UVISTA Y& 0.9152& 0.10\\
&    H2& 0.9196& 0.09 & & &    J1& 0.8461& 0.18\\
&     H& 1.0028& 0.00 & & &    J2& 0.8593& 0.16\\
& F140W& 0.9967& 0.00 & & &   J3& 0.8834& 0.13\\
& F160W& 1.0000& 0.00 & & &     J& 0.9290& 0.08\\
&     K& 0.9572& 0.05 & & & UVISTA J& 0.9284& 0.08\\
&   Ks& 0.9817& 0.02 & & & F125W& 1.0129&-0.01\\
& IRAC1& 1.0142&-0.02 & & &    H1& 0.8766& 0.14\\
& IRAC2& 0.9918& 0.01 & & &   H2& 0.8866& 0.13\\
& IRAC3& 1.0453&-0.05 & & &  H& 0.9372& 0.07\\
& IRAC4& 1.0234&-0.03 & & & UVISTA H& 0.9860& 0.02\\

\cline{1-4} 

GOODS-N 
& U & 0.8324 & 0.20 & & & F140W& 1.0344&-0.04\\
& B & 0.7878 & 0.26 & & & F160W& 1.0000& 0.00\\
& F435W & 1.0237 & -0.03 & & &    K& 0.8918& 0.12\\
& G & 0.9716 & 0.03 & & &    Ks& 0.9398& 0.07\\
& V & 0.7966 & 0.25 & & & UVISTA Ks& 0.9580& 0.05\\
& F606W & 1.0000 & 0.00 & & & IRAC1& 1.0170&-0.02\\
& Rs & 0.9497 & 0.06 && & IRAC2& 0.9724& 0.03\\
& R & 0.7244 & 0.35 & & & IRAC3& 0.9581& 0.05\\
& I & 0.7702 & 0.28 & & & IRAC4& 0.8917& 0.12\\
& F775W & 0.9899 & 0.01 & & &IA427& 1.0314&-0.03\\
& Z & 0.8524 & 0.17 & & & IA464& 1.0382&-0.04\\
& F850LP & 0.9877 & 0.01 & & & IA484& 0.9929& 0.01\\
& F125W & 1.0099 & -0.01 & & & IA505& 0.9497& 0.06\\
& J & 0.9050 & 0.11& & & IA527& 0.9419& 0.06\\
& H & 0.9924 & 0.01 & & & IA574& 0.9695& 0.03\\
& F140W & 1.0129 & -0.01 & & & IA624& 0.7890& 0.26\\
& F160W & 1.0000 & 0.00 & & & IA679& 0.6997& 0.39\\
& Ks & 1.0009 & 0.00 & & &IA709& 0.9086& 0.10\\
& IRAC1 & 0.9889 & 0.01 & & & IA738& 0.9501& 0.06\\
& IRAC2 & 0.9984 & 0.00 && & IA767& 0.8625& 0.16\\
& IRAC3 & 1.0891 & -0.09 & & & IA827& 0.9712& 0.03\\
& IRAC4 & 1.0799 & -0.08\\
\noalign{\smallskip}
\hline
\noalign{\smallskip}

\multicolumn{9}{l}{Notes. Corrected\_AB $=25 -2.5 \log_{10}(\mathrm{Flux} \times$ Flux correction) or Corrected\_AB $= 25-2.5 \log_{10}\mathrm{(Flux)}+$magnitude offset. } 
\end{tabular}

\vspace{+0.5cm}
\end{table}

\begin{table}[ht]
\centering
\caption{Zero point offsets applied to the v4.1 catalogs continued}
\label{table:zp_offsets2}

\begin{tabular}{llcc c llcc}
\cline{1-4} \cline{6-9}
\cline{1-4} \cline{6-9}
\noalign{\smallskip}
Field & Band & Flux correction & Magnitude offset &  & Field & Band & Flux correction & Magnitude offset  \\
\cline{1-4}\cline{6-9}
\noalign{\smallskip}
GOODS-S 
& U & 1.0846 & -0.09 & & UDS & u & 1.2635 & -0.25\\
& U38 & 1.2200 & -0.22 & & & B & 0.9756 & 0.03\\
& B & 1.0055 & -0.01 & & & V & 0.9472 & 0.06\\
& F435W & 1.0819 & -0.09 & & & F606W & 1.0079 & -0.01\\
& V & 0.9787 & 0.02 & & & R & 0.8442 & 0.18\\
& F606Wcand & 1.0038 & 0.00 & & & i & 0.7961 & 0.25\\
& F606W & 1.0033 & 0.00 & & & F814W & 0.9342 & 0.07\\
& R & 1.0185 & -0.02 && & z & 0.8448 & 0.18\\
& Rc & 0.9367 & 0.07 & & & F125W & 1.0048 & -0.01\\
& F775W & 0.9845 & 0.02 & & & J & 1.0096 & -0.01\\
& I & 0.9910 & 0.01 & & & H & 1.0549 & -0.06\\
& F814Wcand & 0.9919 & 0.01 & & & F140W & 1.0257 & -0.03\\
& F850LP & 0.9838 & 0.02 & & & F160W & 1.0000 & 0.00\\
& F850LPcand & 1.0022 & 0.00 & & & K & 1.0614 & -0.06\\
& F125W & 1.0028 & -0.00 & & & IRAC1 & 1.0426 & -0.05\\
& J & 0.9975 & 0.00 & & & IRAC2  & 1.0041 & 0.00\\
& tenisJ & 0.8736 & 0.15 & & & IRAC3 & 1.1500 & -0.15\\
& H & 1.0752 & -0.08 & & & IRAC4 & 1.1500 & -0.15\\
& F140W & 1.0072 & -0.01\\
& F160W & 1.0000 & 0.00\\
& tenisK & 0.7644 & 0.29\\
& Ks & 1.0360 & -0.04\\
& IRAC1 & 1.0226 & -0.02\\
& IRAC2 & 1.0124 & -0.01\\
& IRAC3 & 0.9693 & 0.03\\
& IRAC4 & 0.9451 & 0.06\\
& IA427 & 0.9815 & 0.02\\
& IA445 & 0.9839 & 0.02\\
& IA505 & 0.9861 & 0.02\\
& IA527 & 0.9519 & 0.05\\
& IA550 & 0.9632 & 0.04\\
& IA574 & 1.0385 & -0.04\\
& IA598 & 0.9468 & 0.06\\
& IA624 & 0.8999 & 0.11\\
& IA651 & 0.9798 & 0.02\\
& IA679 & 0.9991 & 0.00\\
& IA738 & 0.9228 & 0.09\\
& IA767 & 0.9092 & 0.10\\
& IA797 & 0.9102 & 0.10\\
& IA856 & 0.8628 & 0.16\\
\noalign{\smallskip}
\hline
\noalign{\smallskip}
\multicolumn{9}{l}{Notes. Corrected\_AB $=25 -2.5 \log_{10}(\mathrm{Flux} \times$ Flux correction) or Corrected\_AB $= 25-2.5 \log_{10}\mathrm{(Flux)}+$magnitude offset. } 

\end{tabular}
\vspace{+0.5cm}
\end{table}

\section{Comparisons to Other Catalogs}
\label{app:comparisons}

\begin{figure*}[h]
\includegraphics[width=0.98\textwidth]{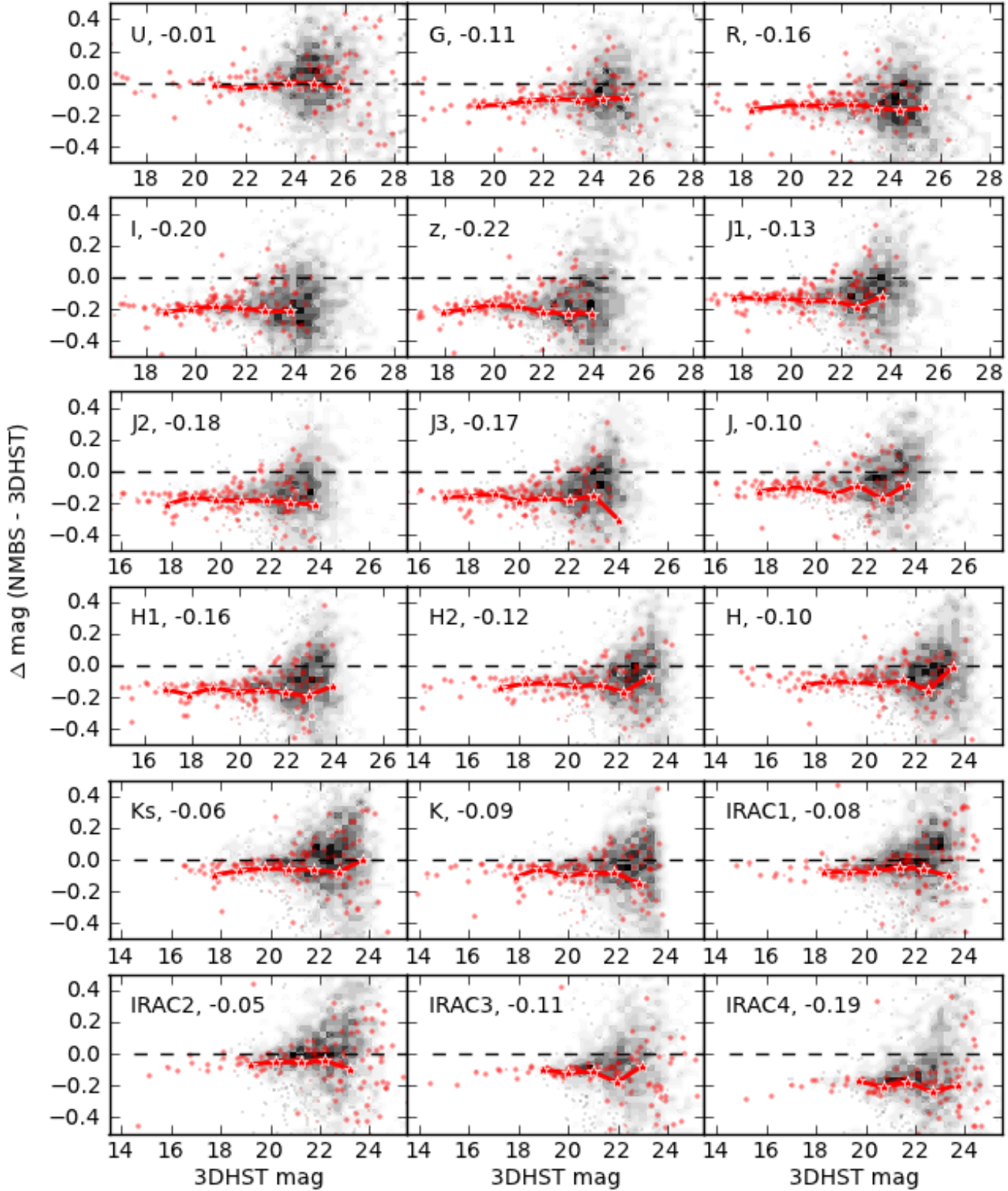}
\caption{Comparison of the AEGIS catalog to the NMBS \citet{Whitaker11} catalog. We compare our zero point-corrected total fluxes to the total fluxes from NMBS. The difference in magnitudes vs. 3D-HST magnitude is shown for each band in common in the two catalogs. The density of galaxies, selected to have use\_phot  = 1, is shown by the shaded contours, with objects outside of the lowest contour (2\% of the maximum density) shown as individual gray points. Point sources with star\_flag = 1 are shown in red.  Objects with SExtractor flags $<2$ ($\ge 2$, ie. blended or otherwise problematic) are shown with large (small) points. The median magnitude difference for all stars in bins of 1 magnitude is shown by the red solid line and large red star symbols. } \label{fig:nmbscomp}
\end{figure*}

\begin{figure*}[h]
\includegraphics[width=0.98\textwidth]{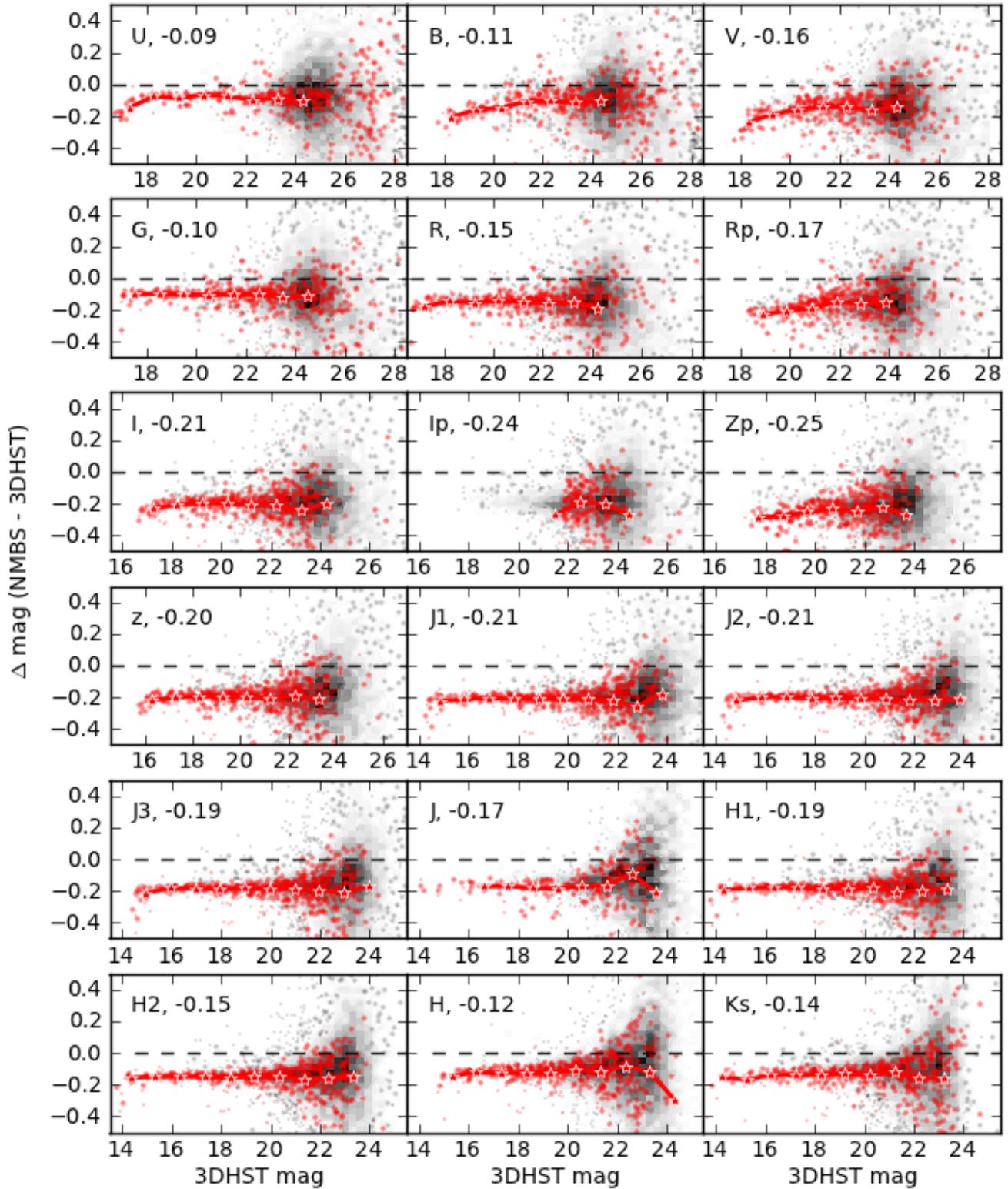}
\caption{Comparison of the COSMOS catalog to the NMBS \citet{Whitaker11} catalog. Symbols are the same as in Fig.~\ref{fig:nmbscomp}. The offsets are constant with magnitude and can largely be explained by the zero point corrections we have made to adjust the ground-based data to the HST system. } \label{fig:nmbs1comp}
\end{figure*}

\begin{figure*}[h]
\includegraphics[width=0.98\textwidth]{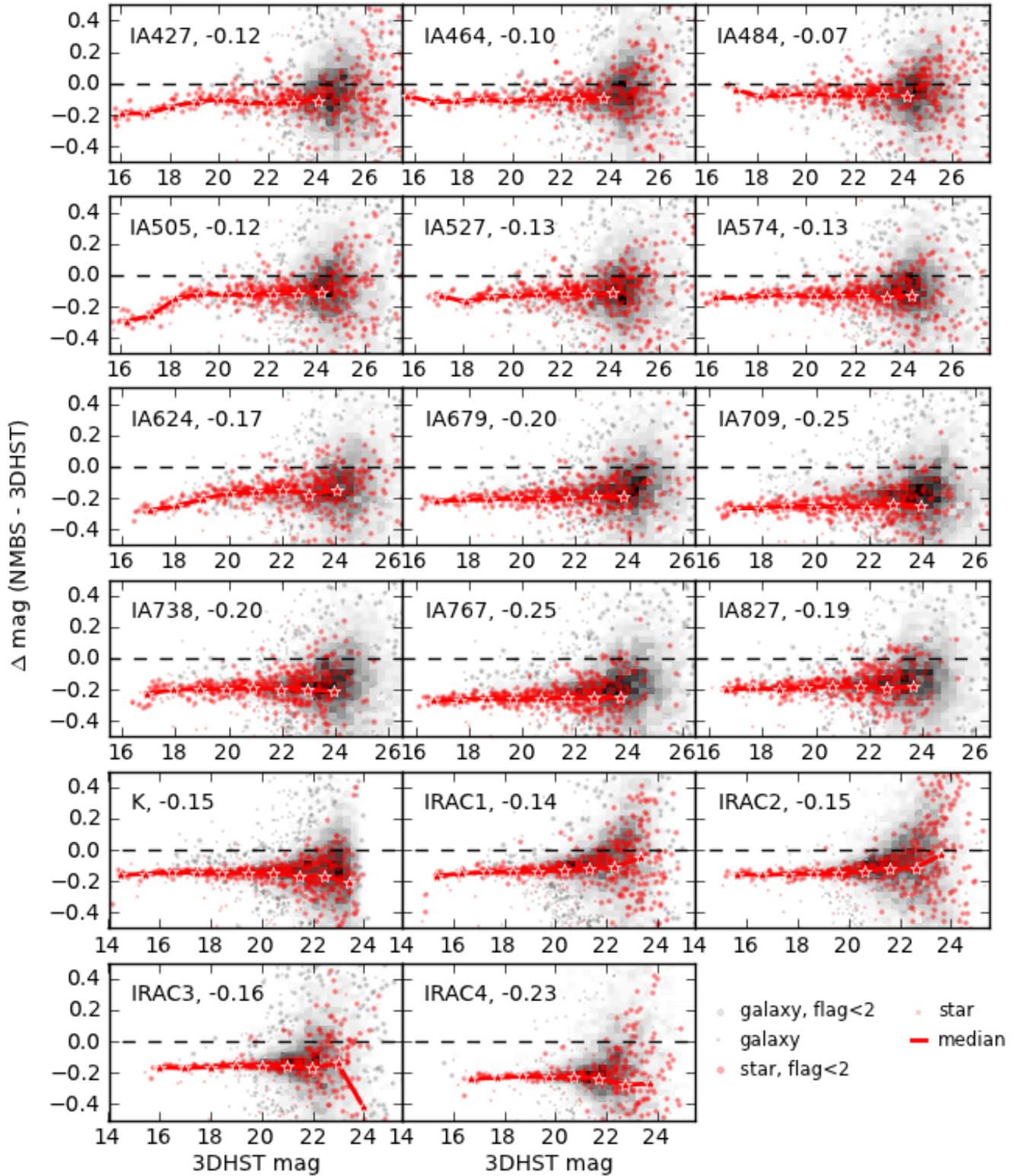}
\caption{Comparison of the COSMOS catalog to the NMBS \citet{Whitaker11} catalog continued. Symbols are the same as in Fig.~\ref{fig:nmbscomp}. } \label{fig:nmbs2comp}
\end{figure*}

\begin{figure*}[h]
\includegraphics[width=0.98\textwidth]{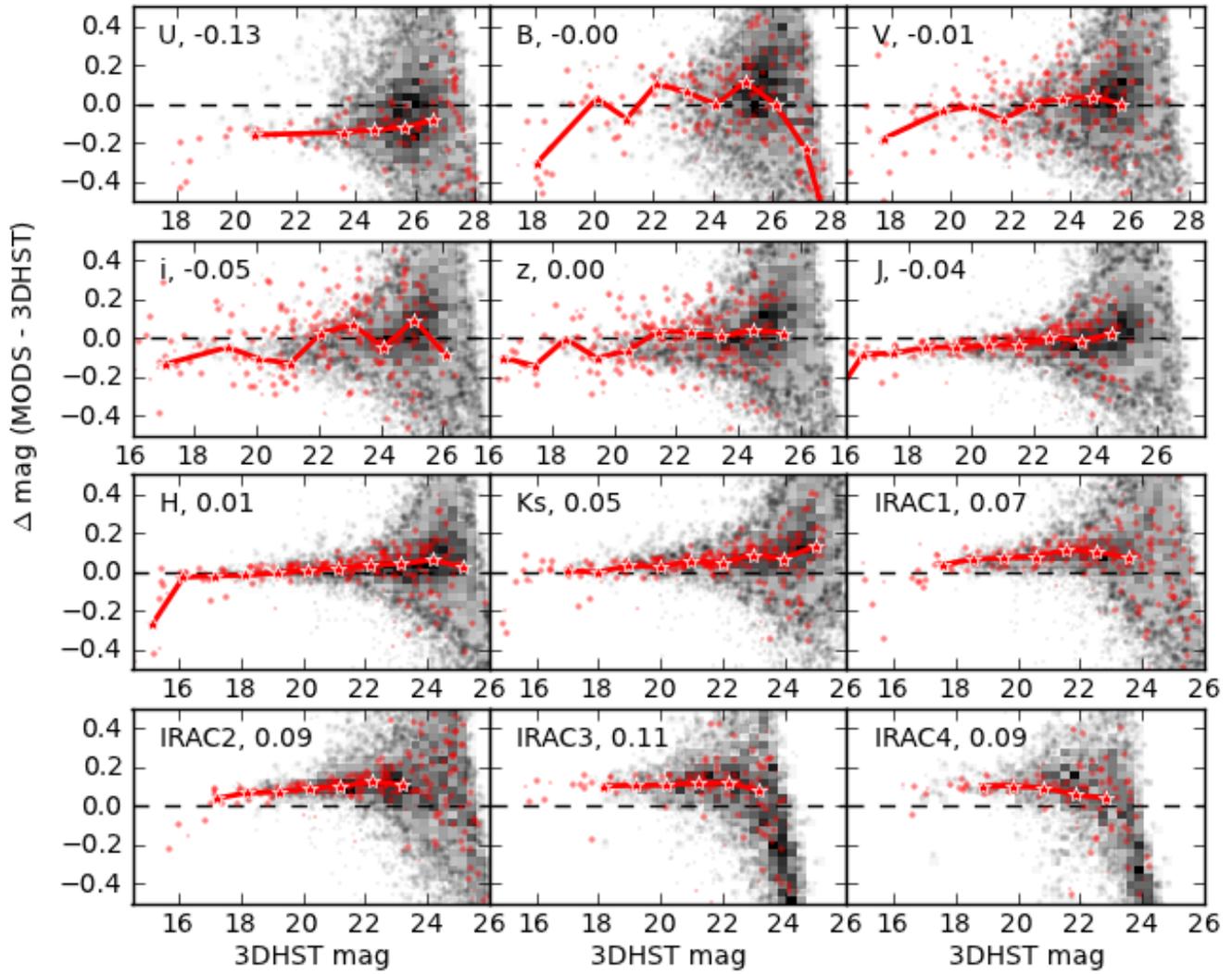}
\caption{Comparison of the GOODS-N catalog to the Moircs Deep Survey catalog \citep{Kajisawa11}. Symbols are the same as in Fig.~\ref{fig:nmbscomp}. } \label{fig:modscomp}
\end{figure*}

\begin{figure*}[h]
\includegraphics[width=0.94\textwidth]{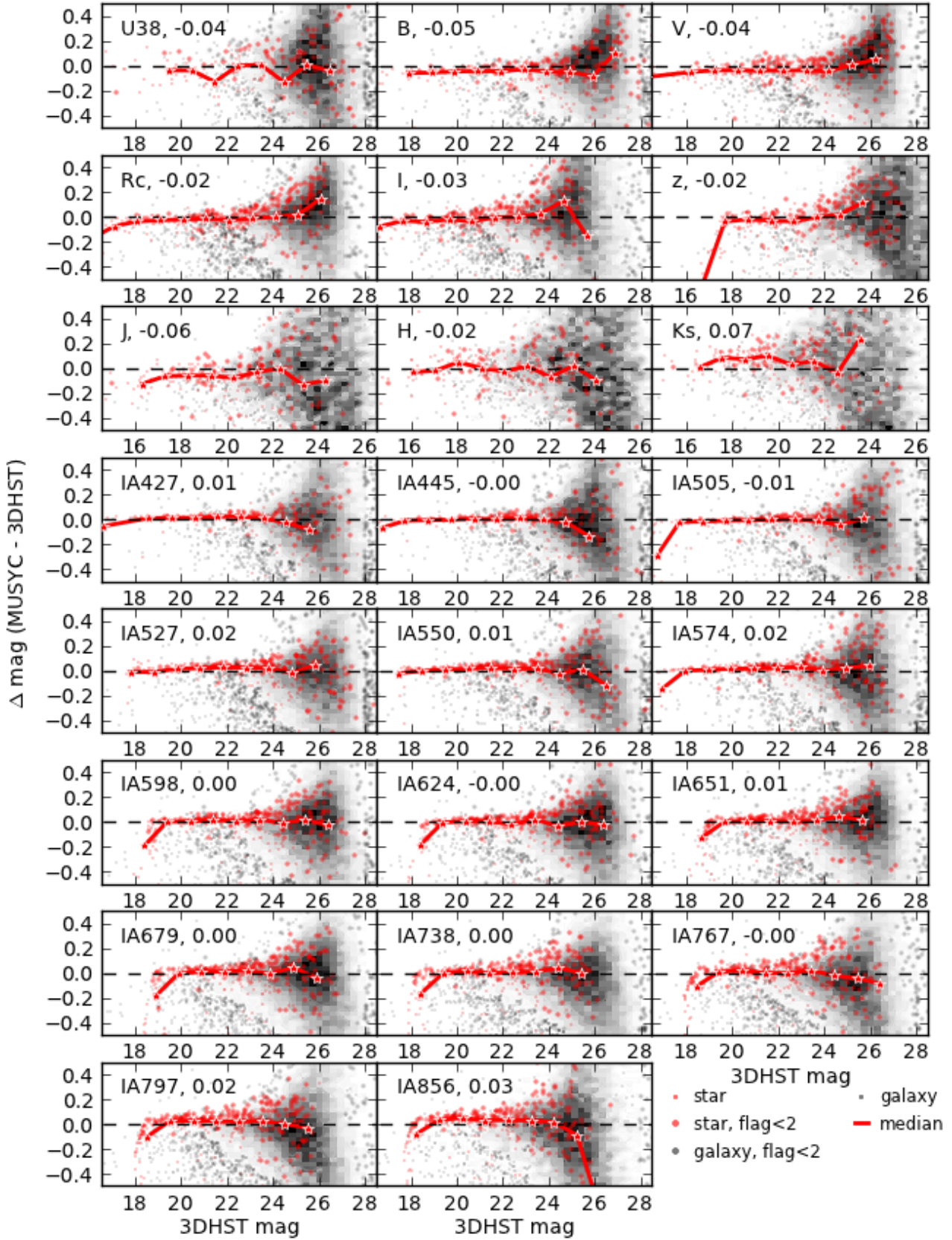}
\caption{Comparison of the GOODS-S catalog to the MUSYC catalog \citep{Cardamone10}. We have applied the extinction and color corrections provided with the MUSYC release to their fluxes and compare to our total zero point-corrected fluxes. The agreement for stars is good in all the medium bands.} \label{fig:musyccomp}
\end{figure*}

\begin{figure*}[h]
\includegraphics[width=0.98\textwidth]{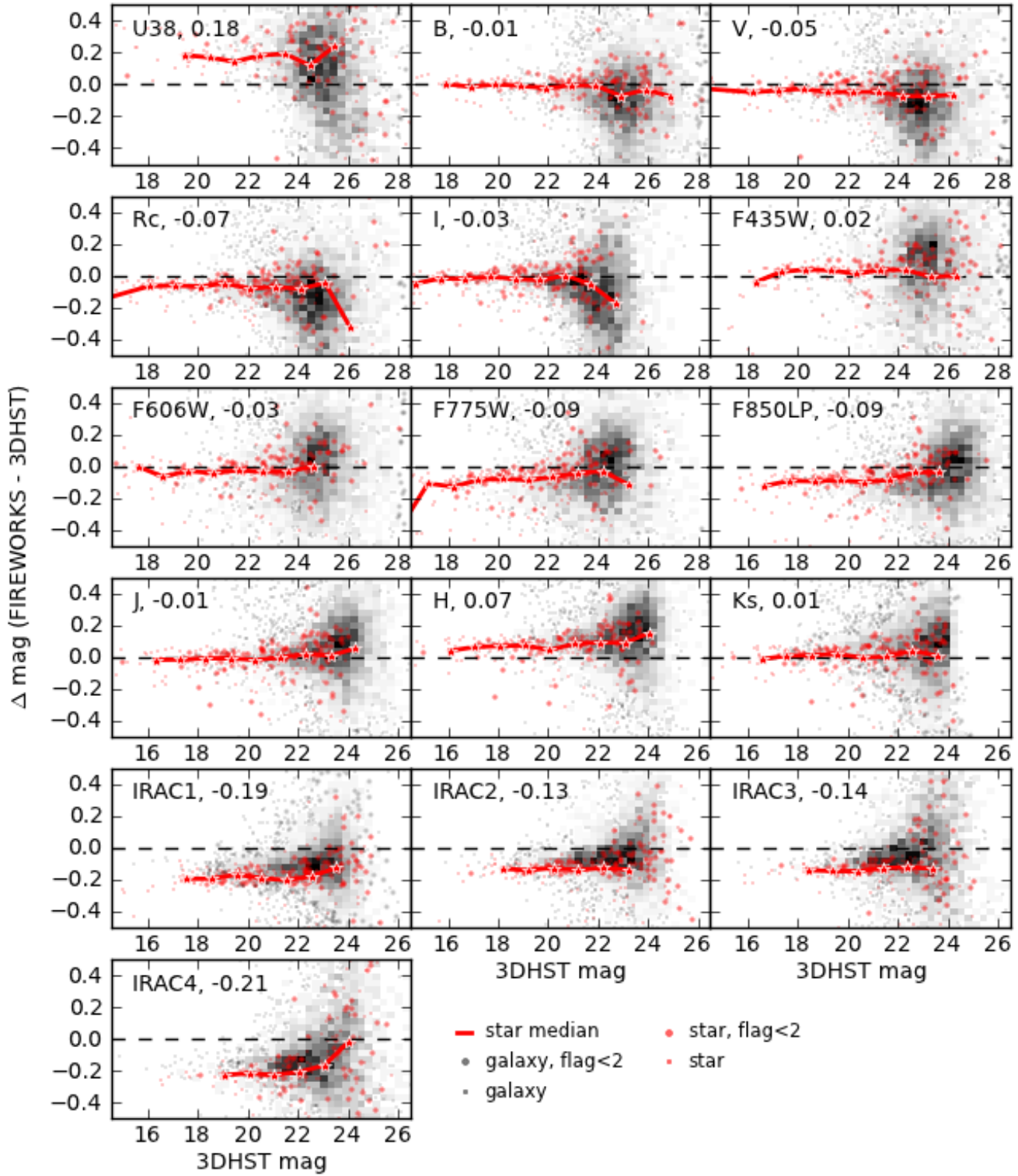}
\caption{Comparison of the GOODS-S catalog to the FIREWORKS catalog \citep{Wuyts08}. We compare total (rather than color aperture) fluxes from both catalogs. Symbols are the same as in Fig.~\ref{fig:nmbscomp}.The U-band is brighter by $\sim0.18$~mag in 3DHST, while the FIREWORKS IRAC magnitudes are brighter by $\sim0.1$--0.2~mag. There are no significant trends with magnitude. } \label{fig:fwcomp}
\end{figure*}

\begin{figure*}[h]
\includegraphics[width=0.94\textwidth]{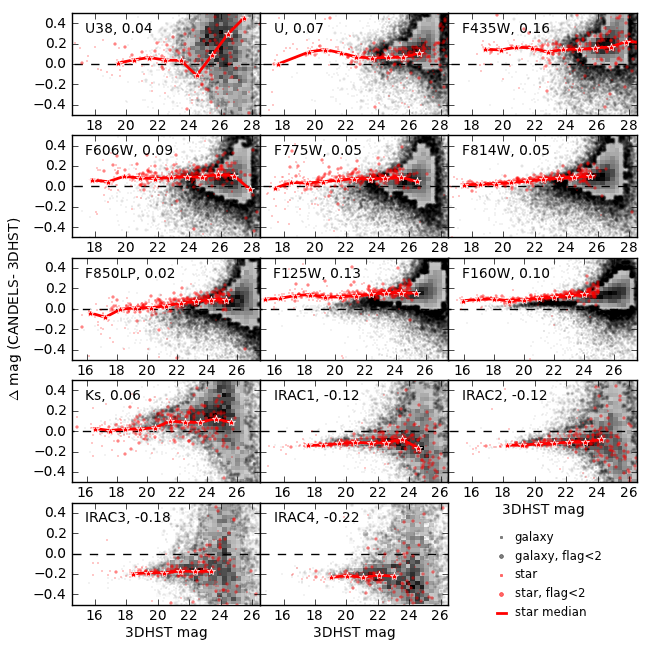}
\caption{Comparison of the GOODS-S catalog total fluxes to the CANDELS catalog total fluxes \citep{Guo13}. Symbols are the same as in Fig.~\ref{fig:nmbscomp}. The fluxes in the CANDELS catalog have not been corrected for the amount of light falling outside of the Kron radius, which accounts for the brighter total magnitudes in 3D-HST. } \label{fig:guocomp}
\end{figure*}

\begin{figure*}[h]
\includegraphics[width=0.94\textwidth]{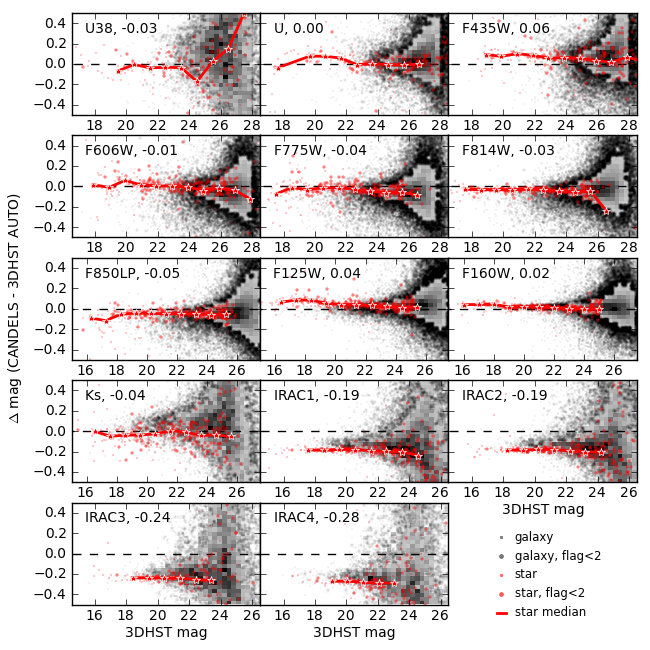}
\caption{Comparison of the GOODS-S catalog AUTO fluxes to the CANDELS catalog total fluxes \citep{Guo13}. Symbols are the same as in Fig.~\ref{fig:nmbscomp}. } \label{fig:gsauto}
\end{figure*}

\begin{figure*}[h]
\includegraphics[width=0.98\textwidth]{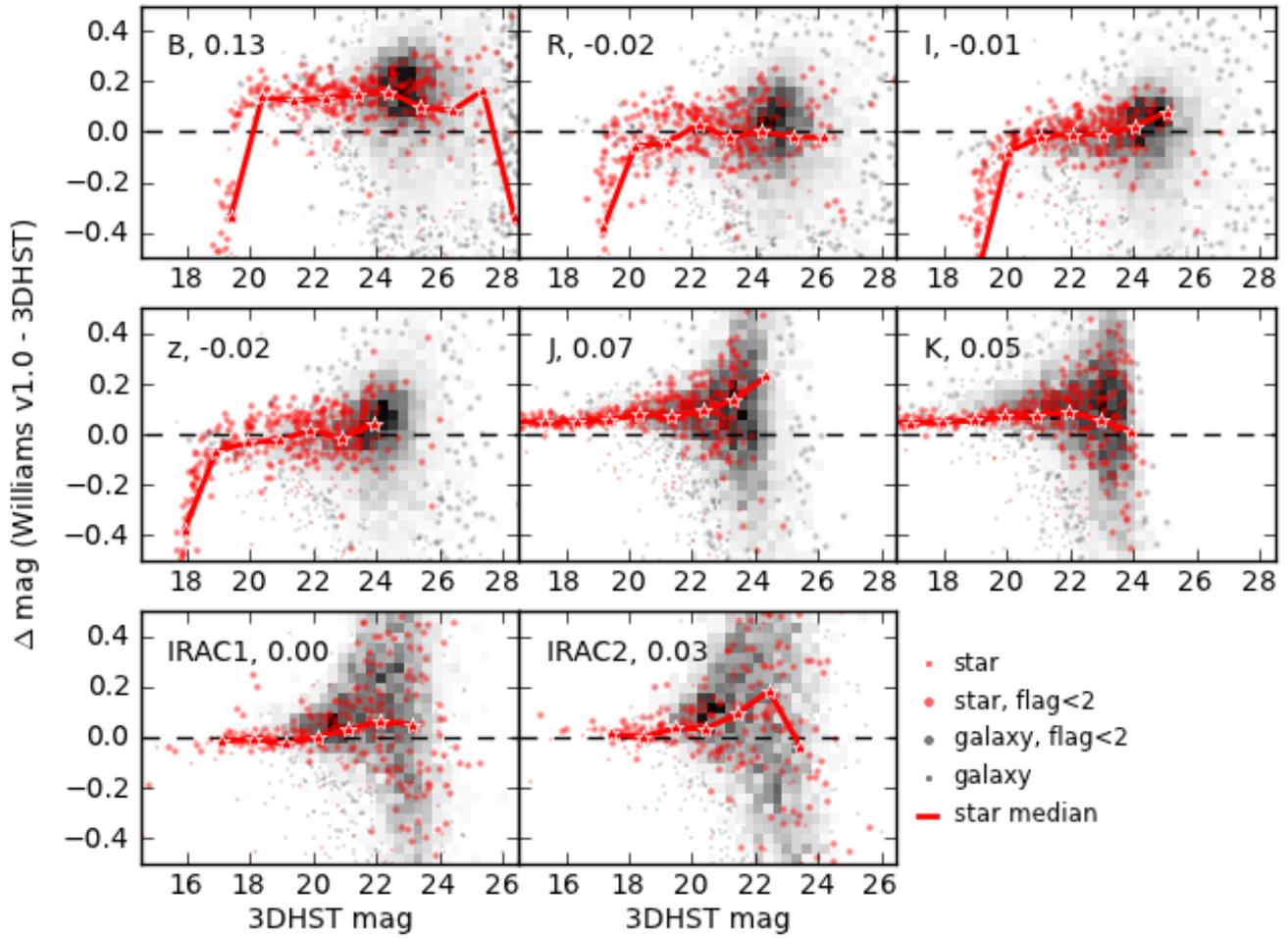}
\caption{Comparison of the UDS catalog to the \citet{Williams09} catalog. We compare total (rather than color aperture) fluxes from both catalogs. Symbols are the same as in Fig.~\ref{fig:nmbscomp}. } \label{fig:rwcomp1}
\end{figure*}

\begin{figure*}[h]
\includegraphics[width=0.98\textwidth]{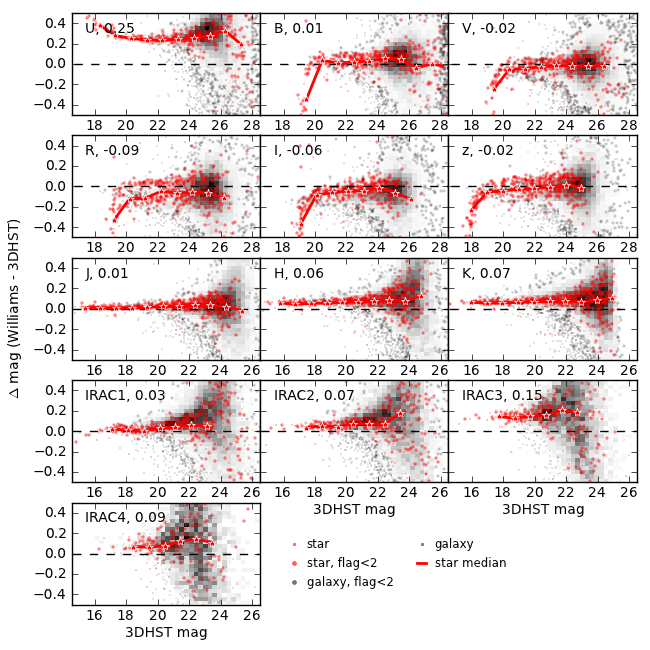}
\caption{Comparison of the UDS catalog to an updated version of the \citet{Williams09} catalog (see \citealt{Quadri12}). We compare total (rather than color aperture) fluxes from both catalogs. Symbols are the same as in Fig.~\ref{fig:nmbscomp}. Two distinct tracks can be seen for stars in the $R$-band, probably caused by a difference in size of the PSF in two regions of the image that is not taken into account by one of the PSF-matching methods. The  $z$-band may be similarly affected. The large scatter at the bright end in the other ground-based optical data suggests that there may also be (more subtle) variations in the PSF in these images. } \label{fig:rwcomp2}
\end{figure*}

\begin{figure*}[h]
\includegraphics[width=0.98\textwidth]{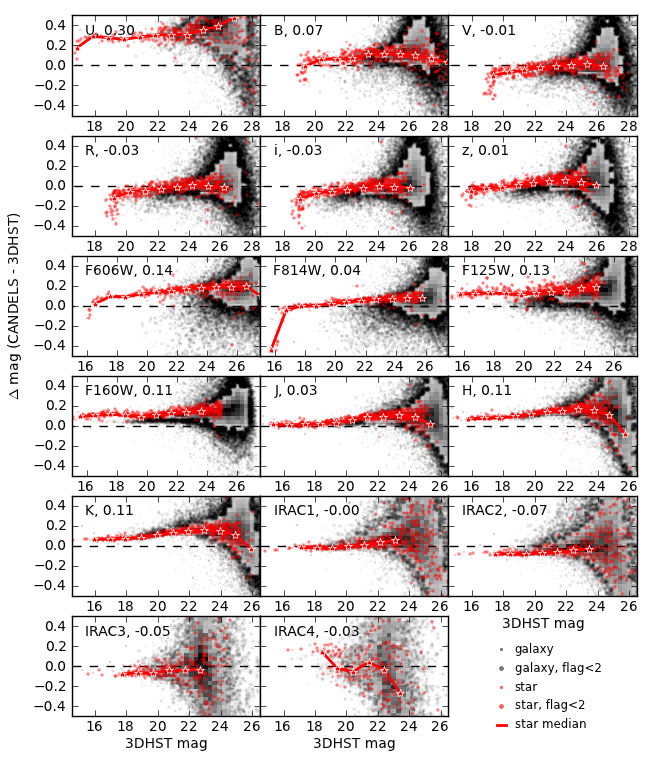}
\caption{Comparison of the UDS catalog total fluxes to the CANDELS \citet{Galametz13} catalog total fluxes. Symbols are the same as in Fig.~\ref{fig:nmbscomp}. The fluxes in the CANDELS catalog have not been corrected for the amount of light falling outside of the Kron radius, which accounts for the brighter total magnitudes in 3D-HST. } \label{fig:galametzcomp}
\end{figure*}

In Figures~\ref{fig:nmbscomp} to \ref{fig:galametzcomp} we show comparisons of each of our catalogs to other publicly available catalogs. We compare the AEGIS and COSMOS catalogs to their counterparts from the NEWFIRM Medium Band Survey \citep{Whitaker11}. The GOODS-N catalog is compared to the Moircs Deep Survey (MODS) catalog from \citet{Kajisawa11}. There are a number of catalogs covering the GOODS-S field. Here we compare to the recent CANDELS catalog \citep{Guo13}, the FIREWORKS catalog \citet{Wuyts08} and the MUSYC survey \citep{Cardamone10}, from which the medium band data are drawn. We compare the 3D-HST UDS catalog to the catalog published by \citet{Williams09}, an updated version of the same catalog, and the recent CANDELS catalog \citep{Galametz13}.

A direct comparison of the aperture fluxes measured in two independent
surveys is a useful diagnostic of problems with the photometry and
catalog processing, particularly when the same images have been used
and similar methods applied. However, where there are unresolved
differences, it is not clear which catalog is more accurate. In this
Appendix we aim to inform the reader what the differences between the
3D-HST and other available catalogs are, rather than commenting on the
quality of either catalog. We therefore present comparisons of the
default fluxes that would be used by anyone accessing the catalogs,
rather than the fluxes from any intermediate stage, which may be more
directly comparable. We note, for instance,
that the fluxes in the CANDELS catalogs are more
equivalent to our AUTO fluxes than our (default) total
fluxes. Also, in the comparison to the NMBS
we find better agreement before applying any zero point
corrections than after, due to our choice of
the HST filters as a reference for the zero point fitting. Where
possible, we explain the offsets between the catalogs in the text.
 
We compare the 3D-HST catalog total fluxes, which have been adjusted
for zero point offsets, as described in Appendix~\ref{app:zps}, to the
total fluxes provided in each of the other catalogs. In some cases
zero point offsets were calculated and applied in a similar way in the
comparison catalogs. In the catalogs where color aperture fluxes and a
correction to total in the detection band are provided, we convert the
colors to total fluxes. In each panel we plot the difference between
the total flux from the comparison catalog and the total flux in the
3D-HST catalog as a function of magnitude in that band from the 3D-HST
catalog. We cross-match objects within $1\arcsec$ when the comparison
catalog uses a ground-based detection image, and $0\farcs5$ for the
two WFC3-detected CANDELS catalogs. The filter and median difference
for stars is shown in the top left hand corner of each panel. The
density of galaxies, selected with use\_phot = 1 is shown in the gray
scale, with outlying objects shown as individual gray points. Stars
(with star\_flag = 1) are shown in red. Objects that are not blended,
with a SExtractor flag of $< 2$, are shown by the larger
points. Objects that have a flag $\ge 2$ are shown with small
points. The median values for stars in bins of 1 magnitude are shown
by the large red star symbols and the red solid line.
 
The AEGIS catalog is compared to the NMBS AEGIS v5.1 catalog \citep{Whitaker11} in Fig.~\ref{fig:nmbscomp}. The methods used for photometry in the two surveys are very similar and zero point corrections have been applied to both catalogs. The U-band measurements from the two surveys are in excellent agreement, although there is large scatter, and the same zero point offset of -0.22~mag was found in both cases. In the other bands, 3D-HST is fainter than the NMBS, with median offsets of 0.1 to 0.2 mag, and no significant trends with magnitude. The largest difference is 0.2~mag in IRAC channel 4 (8$\mu$m). The offsets in the NIR can largely be explained by the zero point corrections applied to our catalogs in order to bring the ground-based data into agreement with the HST data. The (unadjusted) fluxes measured on the NMBS narrow-band ($J1$, $J2$, $J3$, $H1$, $H2$) and $K_s$-band images are in excellent agreement with the NMBS catalog fluxes, however we find that a shift of $\sim$0.15 mag is necessary in order to bring them into agreement with the HST filters and the other ground-based NIR bands. 

In Figures~\ref{fig:nmbs1comp} and \ref{fig:nmbs2comp} we compare the
COSMOS catalog to the NMBS v5.1 catalog \citep{Whitaker11}. Zero point
corrections have been applied to both catalogs. Here the offsets are
larger on average but again constant with magnitude for most bands.
The bands found to have the most uncertain zero points in NMBS ($U$
and $Ip$) have the largest offsets. For three of the medium bands and
some of the optical bands, there is a turn-down towards fainter
magnitudes in 3D-HST for stars at bright end of the stellar
sequence. These may be saturated stars for which the larger aperture
($1\farcs5$) used in NMBS captures more of the light. The offsets in
the NIR can largely be explained by the zero point corrections applied
to both catalogs. If we remove the corrections from both and compare
the measured photometry directly, we find differences of $\Delta$(NMBS
- 3D-HST) = -0.02 to -0.09~mag for the bands between $z$ and $K$ in
wavelength. Taking the NMBS $K$-band difference of -0.03~mag as a
reference point would lead to differences between the bands of
-0.06~mag (in the $J$-band) to 0.01~mag (in $H2$), both qualitatively
and quantitatively similar to the actual zero point corrections
applied within the NMBS catalog. We have also verified that the fluxes
measured in large apertures on the NIR NMBS images agree well with the
total fluxes in the catalog and that the \textit{HST} - ground based
colors measured in large apertures are consistent with the catalog
colors. We are therefore confident that it is the zero points of the
images themselves that are uncertain at this level rather than an
error introduced by the aperture or total corrections applied in the
analysis.

Figure~\ref{fig:modscomp} compares the GOODS-N catalog to the MODS Wide catalog \citep{Kajisawa11}. The $1\farcs2$ aperture fluxes in the MODS catalog are converted to total using the ratio of the $K_s$-band MAG\_AUTO\_COR and aperture fluxes. There is large scatter between the fluxes in the Subaru optical bands, but fairly good agreement in the median. The NIR measurements agree well, with smaller scatter. There is a slight trend with magnitude, such that 3D-HST is brighter than MODS for fainter objects. 
The IRAC fluxes are brighter in 3D-HST, with an offset of $\sim0.1$~mag in all four channels.

In Fig.~\ref{fig:musyccomp} we compare the total zero point corrected fluxes from the MUSYC Subaru v1.0 Catalog \citep{Cardamone10} to the 3D-HST fluxes. We have applied the recommended corrections to go from the MUSYC color aperture fluxes to galactic extinction and zero point corrected total magnitudes. The agreement for all bands is excellent. Outlying galaxies tend to be fainter in 3D-HST than MUSYC, however, with more objects are scattered below zero than above. There are also some differences for bright stars, with 3D-HST measuring lower fluxes at the very bright end than MUSYC.

Figure~\ref{fig:fwcomp} compares the FIREWORKS catalog \citep{Wuyts08} to the 3D-HST GOODS-S catalog. We convert the color aperture fluxes in FIREWORKS to total using the ratio of the total to aperture flux in the $K_s$-band. The agreement is generally good in the optical and NIR. The $U$-band measurements are brighter by 0.18~mag in 3D-HST, while the ACS F775W and F850LP bands are fainter by 0.09~mag. In the IRAC bands there are offsets of 0.13 to 0.21~mag, with FIREWORKS brighter than 3D-HST.

In Fig.~\ref{fig:guocomp} we compare the 3D-HST GOODS-S catalog to the recent CANDELS catalog by \citet{Guo13}. The CANDELS fluxes in all the HST images are measured in an aperture given by the isophotal area of each source in the F160W image. These `color' fluxes are then converted to total fluxes using the ratio of the AUTO and isophotal fluxes in F160W. The resulting total fluxes do not include a correction for the flux outside of the Kron radius and are fainter than the 3D-HST total fluxes by $\sim0.05$--0.15~mag as a result. The offsets in the optical and NIR bands and the difference in the median for stars and galaxies are largely removed by comparing the 3D-HST AUTO fluxes to the CANDELS catalog fluxes. This comparison is shown in Fig.~\ref{fig:gsauto}. The IRAC fluxes measured by CANDELS using the TFIT software \citep{Laidler07}, which applies similar techniques for the photometry of low-resolution images with a high-resolution prior, are 0.12--0.22~mag brighter than the 3D-HST total fluxes.

Figure~\ref{fig:rwcomp1} compares the 3D-HST UDS catalog to the UDS v1.0 catalog from \citet{Williams09}. Figure~\ref{fig:rwcomp2} shows the comparison to an updated version of the same catalog, with significant updates as described by \citet{Quadri12}. The public catalog  contains only 8 bands, with the newer catalog adding the $U$, $V$, and $H$-bands, and IRAC 5.8$\mu$m and 8$\mu$m. The agreement with both catalogs is generally good, with large scatter in the $R$ and $z$-band, that separate more clearly into two tracks in the second $R$-band comparison. The $R$-band offsets show a spatial dependence, indicating that there may be a systematic variation in the PSF or astrometry between two of the original SXDS fields that has not been accounted for in the mosaicking or by the photometry in one of the catalogs. Bright stars are systematically fainter in 3D-HST. There is a dependence on magnitude in the $I$, $J$ and IRAC bands in Fig.~\ref{fig:rwcomp1} that is no longer visible in Fig.~\ref{fig:rwcomp2}. The $U$-band, for which the zero point is known to be uncertain, is offset brighter by 0.25~mag in 3D-HST.

\begin{figure*}[h]
\includegraphics[width=0.98\textwidth]{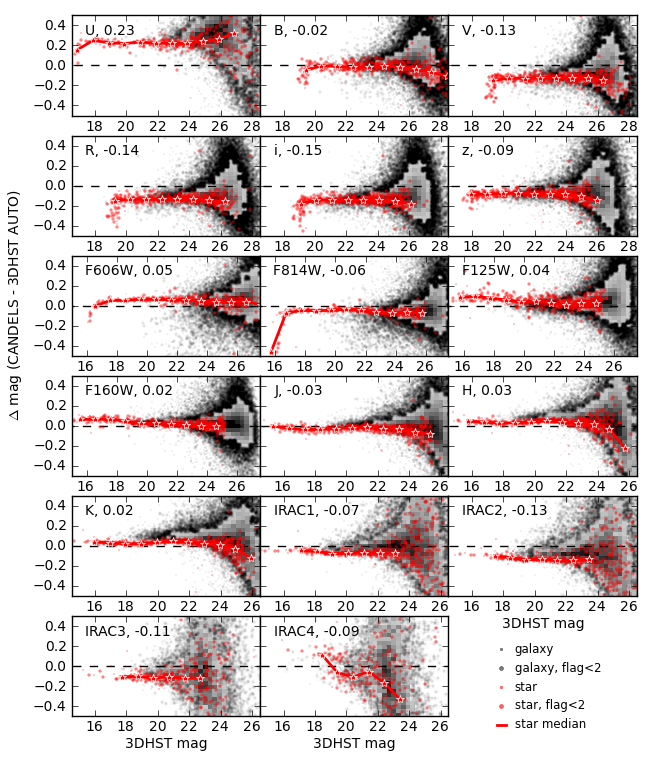}
\caption{Comparison of the UDS catalog AUTO fluxes to the CANDELS \citet{Galametz13} catalog total fluxes. Symbols are the same as in Fig.~\ref{fig:nmbscomp}.  } \label{fig:udsauto}
\end{figure*}

Figure~\ref{fig:galametzcomp} compares the 3D-HST UDS catalog to the CANDELS catalog from \citet{Galametz13}. The methods applied for the CANDELS UDS catalog are very similar to those used for the \citet{Guo13} catalog described above, and similar trends are seen in the comparisons. When we account for the correction from AUTO to total fluxes, which is not applied in the \citet{Galametz13} catalog, the offsets in the ACS and WFC3 bands largely disappear, as shown in Fig.~\ref{fig:udsauto}. The median difference for stars decreases from 0.11 to 0.02~mag in F160W, for example. With the exception of the $B$-band and NIR, the total fluxes measured on the low-resolution images agree better without adjusting for this difference, suggesting that the `dilation' method to grow the isophotal areas of objects before applying TFIT returns fluxes that are closer to the total flux measured by 3D-HST. We again find that the $U$-band is brighter in 3D-HST. There is large scatter in the IRAC4 (8$\mu$m) band.

\end{appendix}

\end{document}